\documentclass[aps, prd, showpacs, nofootinbib, preprintnumbers,floatfix,letterpaper,longbibliography,superscriptaddress, reprint]{revtex4-1}

\usepackage[utf8]{inputenc}
\usepackage{bm}
\usepackage{latexsym,amssymb,amsmath,amsfonts}
\usepackage{graphicx}
\usepackage{longtable}
\usepackage{footnotebackref}
\usepackage{physics}
\usepackage[usenames]{xcolor}
\usepackage{soul}
\usepackage{subfigure}
\usepackage{makecell}
\usepackage{amssymb}
\usepackage{amsmath}
\usepackage{rotating}
\usepackage{color}
\usepackage{epsf}
\usepackage{fancyhdr}
\usepackage{ifthen}
\usepackage{slashed}
\usepackage{xspace}
\usepackage{tabularx}

\def\OGW{\Omega_\text{gw}}

\def \hpx{{\tt HEALPix} }

\begin{document}

\title{Upper limits on persistent gravitational waves using folded data and the full covariance matrix from Advanced LIGO’s first two observing runs}
\author{Deepali Agarwal}
\email{deepali@iucaa.in}
\affiliation{Inter-University Centre for Astronomy and Astrophysics (IUCAA), Pune 411007, India}

\author{Jishnu Suresh}
\email{jishnu@icrr.u-tokyo.ac.jp}
\affiliation{Institute for Cosmic Ray Research (ICRR), The University of Tokyo,
Kashiwa City, Chiba 277-8582, Japan} 

\author{Sanjit Mitra}
\email{sanjit@iucaa.in}
\affiliation{Inter-University Centre for Astronomy and Astrophysics (IUCAA), Pune 411007, India}

\author{Anirban Ain}
\email{anirban.ain@pi.infn.it}
\affiliation{Istituto Nazionale di Fisica Nucleare (INFN) sezione Pisa, 56126 Pisa, Italy }

\begin{abstract}
The stochastic gravitational-wave background (SGWB) created by astrophysical sources in the nearby Universe is likely to be anisotropic. Upper limits on SGWB anisotropy have been produced for all major data-taking runs by the ground-based laser interferometric detectors. However, due to the challenges involved in numerically inverting the pixel-to-pixel noise covariance matrix, which is necessary for setting upper limits, the searches accounted for angular correlations in the map by using the spherical harmonic basis, where regularization was relatively easier. This approach is better suited though for extended sources. Moreover, the upper-limit maps produced in the two different bases are seemingly different. While the upper limits may be consistent within statistical errors, it was important to check whether the results would remain consistent if the full noise covariance matrix was used in the pixel basis. Here, we use the full pixel- to-pixel Fisher information matrix to create upper-limit maps of SGWB anisotropy. We first perform an unmodeled search for persistent, directional gravitational-wave sources using folded data from the first (O1) and second (O2) observing runs of Advanced LIGO and show that the results are consistent with the upper limits published by the LIGO-Virgo Collaboration (LVC). We then explore various ways to account for the pixel-to-pixel Fisher information matrix using singular-value decomposition and Bayesian regularization schemes. We do not find evidence for any SGWB signal in the data and the upper limits are consistent with the LVC results within statistical errors. Through an injection study, we show that they are all valid 95\% upper limits, that is, the upper limit in a pixel is less than the injected signal strength in less than 5\% of the pixels. Remarkably, we find that, due to nuances involved in the regularization schemes, the simplest method of using the convolved (dirty) map with a normalized variance, which was used in the LVC analysis, provides as good upper limits as the elaborate schemes with the full noise covariance matrix. Hence, we recommend continuing to use this simple method, though more regularization schemes may be explored to obtain stronger upper limits.
\end{abstract}

\maketitle

\section{Introduction}
Stochastic gravitational-wave backgrounds (SGWBs) are generated by the incoherent superposition of gravitational waves (GWs) from unmodeled or unresolved sources~\cite{MAGGIORE2000283,satya}. A number of different mechanisms may generate SGWBs, which include cosmological backgrounds composed of, e.g., inflationary gravitational waves~\cite{PhysRevD.50.1157,PhysRevD.85.023534,CROWDER201366}, or phase transitions in the early Universe~\cite{vonHarling2020}, and astrophysical backgrounds composed of a large number of sources or events, e.g., mergers of compact binaries~\cite{Regimbau_2008,Regimbau_2011} or isolated pulsars~\cite{PhysRevD.84.083007}. These backgrounds are expected to have different characteristic signatures in the frequency spectrum~\cite{Regimbau_2011} and angular distribution~\cite{CONTALDI20179,AnisoSGWB_Giulia,AnisoSGWB_Jenkins}.

The searches for isotropic and anisotropic SGWBs are fundamentally based on cross correlation of data from pairs of detectors, though the algorithms vary depending on the basis in which the search is being conducted~\cite{Allen_Romano,Ballmer_2006,Radiometer,Eric,Romano2017}. Upper limits have been set on isotropic~\cite{iso_O2} and anisotropic backgrounds using Advanced LIGO's first (O1) and second (O2) observation run data~\cite{O2paper,renzini1,renzini2} and past major data-taking runs.
The sky maps were made using the GW radiometer algorithm~\cite{Ballmer_2006,Radiometer}, primarily to probe localized point sources (e.g., a ``hot spot'' created by a large number of millisecond pulsars in a galaxy cluster \cite{PhysRevD.84.083007}), which is analogous to Earth rotation synthesis imaging used in radio astronomy, where data from pairs of detectors are cross correlated with a time-varying phase delay that accounts for the light-travel time delay between the detector sites for a given direction on the sky.
A method was proposed to fold cross-spectral data to one sidereal day~\cite{Folding} and a pipeline {\tt PyStoch}~\cite{PyStoch} has been developed to take full advantage of the folded data to map the anisotropies in SGWBs in \hpx~\cite{HEALPix,Zonca}. Folding and {\tt PyStoch} together provide more than a few hundred times computational speed up in the broad-band radiometer analysis in comparison to conventional pipeline. In this paper, we present the upper-imit maps produced using {\tt PyStoch} applied to folded O1-O2 data and show that the results agree, constituting an important validation step for the data and the pipeline.

In the GW radiometer analysis with the two LIGO detectors, as was the case for the O1-O2 analysis, noise is correlated across very different directions (pixels) on the sky. Despite this, only the variance [diagonal components of the noise covariance matrix (NCVM)] has been used for the pixel-based analysis of data~\cite{S5_Dir}. This is partly because the numerical computation of the full NCVM in the pixel basis remained computationally challenging at the present pixel resolution used for the analysis of LIGO-Virgo data~\cite{O3paper} until the advent of data folding. Moreover, it would be difficult to incorporate the NCVM in the analysis due to its ill-conditioned nature. %
{\em However, no study was performed to show that the noise covariance does not have a significant effect. This was an untested assumption.} %
In this paper, we use for the first time the combination of folding and {\tt PyStoch} to produce the full pixel space NCVM for O1 and O2 data and show that at the present sensitivity the covariances can be ignored.

Synthesis imaging requires multiple lengths of the baselines to faithfully reproduce the image of a source, which is why arrays of telescopes are used in radio astronomy along with Earth's rotation to observe the sky with different effective baseline lengths. Due to the same reason, the sky-map information captured by the GW radiometer formed with only two detectors is highly limited. The situation is likely to improve with the addition of multiple detectors to the network. As was shown in Ref. ~\citet{Radiometer}, an angular correlation pattern for a single baseline, described by the pixel-to-pixel noise covariance matrix which is proportional to the beam matrix for the chosen normalization, spans an extended pattern on the sky with a strong maximum at the ``pointing'' direction. Different baselines scan the sky with different orientations~\cite{Multi_baseline}. The resultant correlation patterns have the maximum at the same point, but the extended structures are very different. Thus, the correlation pattern of the combined map has a much stronger peak near the pointing direction compared to the case for a single baseline, making the NCVM significantly more diagonal, and hence, better conditioned. This can be shown numerically by comparing the singular value spectrum obtained using singular-value decomposition (SVD) of the matrices for separate and combined baselines~\cite{Eric}.

The most obvious approach of incorporating the full NCVM in the Likelihood for the dirty map, to create upper limit maps in the absence of a GW signal and to compute the significance ($p$-value) of pixel values, leads to huge inaccuracies, because the inverse of the (ill-conditioned) NCVM is difficult to estimate. We explore several potential regularization schemes based on singular value decomposition and Bayesian regularization to create a deconvolved clean map and to obtain an effective inverse of the NCVM, which will be necessary to incorporate the full NCVM in the analysis. We estimate upper limits and $p$-value from both dirty and clean maps using these schemes, for which the detailed expressions are listed in the paper. However, only a selected set of results, which are precise enough, are presented here.

Part of this approach is similar to what was used in the analysis published by the LVC~\cite{S5_Dir} in the spherical harmonic basis. There, a singular-value decomposition techniques was used to regularize and invert the NCVM to create a deconvolved clean map. The significance was estimated via simulations using a transformed NCVM that can closely represent the NCVM of the regularized clean map. However, only the diagonal components of the above transformed NCVM (but computed with slightly different regularization scheme to account for the loss of modes in obtaining the clean map) was used in the likelihood to set a conservative upper limit. We apply a similar method here, but with the full pixel space NCVM for the significance and upper-limit calculations. Even though the regularization scheme introduces a bias (while in the spherical harmonic searches this bias could be assumed to be small~\cite{S5_Dir,Eric}), we account for it in the Likelihood to obtain more accurate upper limits.
 
To test the reliability of the regularization schemes, we perform an injection study by adding a weak (essentially not detectable) signals to the noise and then comparing the upper limits with the injected values. A conservative $95\%$ upper limit must ensure that the upper limit is more than the injected value in more than $95\%$ of the pixels. The schemes we report here all satisfy this criterion. We finally compare our results with the results published by the LVC.

 The paper is organized as follows. In Sec.~\ref{sec:method} we briefly review GW radiometer algebra, folding, and {\tt PyStoch}. We discuss the details about the full-covariance matrix calculation and the deconvolution procedures in the same section. In Sec.~\ref{sec:covaraince} we discuss assigning the significance of an event or setting the upper limits with the full covariance matrix. Section~\ref{sec:data} summarizes the details about the data set used, and details about the injection study are given in Sec.~\ref{sec:clean_inj}. We summarize our results in Sec.~\ref{sec:results} and make concluding remarks in Sec.~\ref{sec:conclusion}.
%

\section{Methods}
\label{sec:method}
In this section, we review the map-making techniques in the SGWB searches and present recipes to regularize the covariance matrix which is a necessary condition to remove the effect of the point spread function~\cite{Ballmer_2006,Radiometer,Romano2017} of the detector from the estimator of SGWB power.

\subsection{GW radiometry}

SGWBs are typically characterized by the GW energy density parameter. GWs arriving from the direction $\mathbf{\hat{\Omega}}$, having an energy density of $\rho_\mathrm{gw}$, with observed frequency ranging from $f$ to $f+\dd{f}$, measured in units of critical energy density $\rho_c=3H_0^2c^2/8\pi G$ for a flat universe, can be written as
\begin{equation}
\OGW (f,\mathbf{\hat{\Omega}})\equiv\frac{1}{\rho_\mathrm{c}}\frac{\dd[3]{\rho_\mathrm{gw}}}{\dd{(\ln f)}\dd{\mathbf{\hat{\Omega}}}}=\frac{8\pi Gf}{3H_0^2c^2}\frac{\dd[3]{\rho_\mathrm{gw}}}{\dd{f}\dd{\mathbf{\hat{\Omega}}}}\, .
\end{equation}
Here $H_0$ is the Hubble constant at the current epoch, $c$ is the speed of light and $G$ is the universal constant of gravitation. We further assume that $\OGW (f,\mathbf{\hat{\Omega}})$ can be decomposed into an angular power spectrum, $\propto \mathcal{P}(\mathbf{\hat{\Omega}})$, and a spectral shape, $H(f)$. Then the above GW energy density can be expressed as
\begin{equation}
\OGW (f,\mathbf{\hat{\Omega}})=\frac{2\,\pi^2}{3H_0^2}\,f^3\,H(f)\,\mathcal{P}(\mathbf{\hat{\Omega}})\,.
\label{eq:Omega_H_P}
\end{equation}
From the analysis perspective we can also define the energy flux in units of erg cm$^{-2}$ s$^{-1}$ Hz$^{-1}$ sr$^{-1}$ as
\begin{equation}
\mathcal{F} (f,\mathbf{\hat{\Omega}})=\frac{c^3 \pi}{4G}\,f^2\,H(f)\,\mathcal{P}(\mathbf{\hat{\Omega}})\, .
\label{eq:Flux}
\end{equation}

In the radiometer analysis~\cite{Allen_Romano,Radiometer}, one can expand the GW power spectrum in a given basis as
\begin{equation}
\mathcal{P}(\mathbf{\hat{\Omega}})=\sum_{p} P_p\, e_{p}(\mathbf{\hat{\Omega}})\, ,
\label{power_expansion}
\end{equation}
where $e_p (\mathbf{\hat{\Omega}})$ is the $p$th basis function, which can be the spherical harmonics $Y_{lm}(\mathbf{\hat{\Omega}})$ for extended sources and $\delta^{2}(\mathbf{\hat{\Omega}}-\mathbf{\hat{\Omega}_0})$ to search for a point source in an arbitrary direction $\mathbf{\hat{\Omega}_0}$. In both the cases, we take the spectral shape $H(f)$ to be characterized by a power law, given as
 \begin{equation}
 H(f) \equiv \left ( \frac{f}{f_{\rm{ref}}} \right)^{\alpha - 3} \, .
 \label{spectral_shape}
 \end{equation}
Here $f_{\rm{ref}}$ is a reference frequency, which is often set to be 25Hz~\cite{O1paper,O2paper}, and $\alpha$ is the spectral index that characterizes different source models. We consider three values for the index, $\alpha = 0,2/3,3$, corresponding to SGWBs from cosmological sources, the population of compact binary coalescence, and spinning neutron stars (pulsars, magnetars) respectively. 

The primary output of the radiometer search represents the GW sky seen through the response matrices of the detectors: the dirty map~\cite{Radiometer}. For a baseline $\mathcal{I}$, formed with a pair of detectors $\mathcal{I}_1$ and $\mathcal{I}_2$, we can write this quantity as
\begin{equation}
X_p =  \sum_{\mathcal{I}ft} \gamma^{\mathcal{I}*}_{ft,p} \frac{ H(f)} {P_{\mathcal{I}_1}(t;f) P_{\mathcal{I}_2}(t;f)}  C^{\mathcal{I}} (t;f)\,. 
\label{eq:Dirty_map}
\end{equation}
Here $C^{\mathcal{I}} (t;f)$ is the cross-spectral density (CSD), which is the product of the Fourier transform of time series strain data from one detector at time $t$ and the complex conjugate of the same from the other detector~\cite{Folding}, while $P_{\mathcal{I}_1}(t;f)$ and $P_{\mathcal{I}_2}(t;f)$ denote the one-sided noise power spectral density (PSD) of the detectors $\mathcal{I}_1$ and $\mathcal{I}_2$ respectively. The uncertainty in this estimation can be quantified by the Fisher information (or NC) matrix~\cite{Eric},
\begin{equation}
\Gamma_{pp'} = \sum_{\mathcal{I}ft} \frac{H^2(f)}{P_{\mathcal{I}_1}(t;f) \, P_{\mathcal{I}_2}(t;f)} \,\gamma^{\mathcal{I}*}_{ft,p} \, \gamma^{\mathcal{I}}_{ft,p'} \, .
\label{eq:Fisher}
\end{equation}
In both Eqs.~(\ref{eq:Dirty_map}) \& (\ref{eq:Fisher}) the variable $\gamma$ is a detector geometry dependent function called the overlap reduction function ~\cite{Radiometer}, expressed as 
\begin{equation}
\gamma_{ft,p} ^{\mathcal{I}} := \sum_{A} \int_{S^2} d \mathbf{\hat \Omega} F^{A}_{\mathcal{I}_1}(\mathbf{\hat \Omega},t) 
F^{A}_{\mathcal{I}_2}(\mathbf{\hat\Omega},t) e^{2\pi i f \frac{\mathbf{\hat \Omega}\cdot {\mathbf{\Delta x}_{\mathcal{I}} (t)}}{c}} e_p(\mathbf{\hat \Omega}) \, .
\label{eq:ORF_general}
\end{equation}
$F^{A}_{\mathcal{I}_{1,2}}(\mathbf{\hat \Omega},t)$ represents the antenna pattern function, which records the response of the detector pair, as the baseline separation $\mathbf{\Delta x}_{\mathcal{I}} (t)$ varies with time.

The observed dirty map $\mathbf{X}$ is a convolution of the true sky $\bm{\mathcal{P}}$ with $\mathbf{\Gamma}$ and contains additive Gaussian noise $\mathbf{n}$, whose covariance is also given by $\mathbf{\Gamma}$~\cite{Eric},
\begin{equation}\label{eq:convolution}
    \mathbf{X} = \mathbf{\Gamma} \cdot \bm{\mathcal{P}} + \mathbf{n} \,.
\end{equation}
Here $\mathbf{X}$ and $\bm{\mathcal{P}}$ are vectors in the chosen basis and $\mathbf{\Gamma}$ is a square matrix of order equaling to the total number of components in the chosen basis.

Then the Maximum Likelihood (ML) estimate of the intensity of the SGWB sky, $\mathcal{P}(\mathbf{\hat{\Omega}})$, obtained from the above equation is given by~\cite{Radiometer,Eric}
\begin{equation}
\bm{\hat{\mathcal{P}}}=\mathbf{\Gamma}^{-1} \, \mathbf{X}\,, 
\label{eq:ML_power}
\end{equation}
which requires the inversion of the Fisher information matrix $\mathbf{\Gamma}$. This procedure is non-trivial due to the ill-conditioned nature of the matrix. For the pixel-based radiometer analysis~\cite{S5_Dir,O1paper,O2paper,O3paper}, the correlation between the neighboring pixels is ignored to obtain the signal-to-noise ratio (SNR) of the estimator, defined as
\begin{equation}
\rho_{\mathbf{\hat\Omega}}= \left[(\Gamma_{\mathbf{\hat\Omega}, \mathbf{\hat\Omega}}) ^{-1} X_{\mathbf{\hat\Omega}}\right] /\, (\Gamma_{\mathbf{\hat\Omega}, \mathbf{\hat\Omega}})^{-1/2} \,.
\label{eq:pixel_SNR}
\end{equation}

The time translation symmetry by a sidereal day in the radiometer search~\cite{Radiometer} can be utilized to fold the entire observation data to one sidereal day~\cite{Folding}. One can easily rewrite the summation over all time segments as $\sum_t \equiv \sum_{i_{day}} \sum_{t_s}$, where the index $i_{day}$ takes the values up to the total number of sidereal days for which the data is processed, while $t_s$ runs over all the time segments in one sidereal day [see Eqs.~(\ref{eq:Dirty_map}) $\&$ (\ref{eq:Fisher})]. We can use this folded data to compute the dirty map and Fisher matrix in an efficient way using much less computational resources~\cite{Folding}.

In this paper, we took advantage of the {\tt PyStoch}~\cite{PyStoch} pipeline, which can efficiently analyze the folded data set (and produce $\mathbf{X}$ and $\mathbf{\Gamma}$) taking advantage of its compactness and can perform all types of analyses for persistent stochastic sources~\cite{pystoch_SpH} (both modeled and unmodeled search) with much less computational time compared to the conventional pipeline running on unfolded data. We also validate the folded data set by comparing the obtained results with the previous LVC results~\cite{O2paper}. 

\subsection{Clean Map : Regularization Recipes}\label{sec:reg_recipe}
``Clean map,'' the estimator of the true source distribution on the sky, is obtained by eliminating the effects of the response function of the detectors, through a deconvolution procedure~\cite{Radiometer,Eric,regDeconv}. Any deconvolution process requires the calculation of the Fisher information matrix ($\mathbf{\Gamma}$) at all the pixels on the sky. The Fisher matrix for a single baseline has poorly observed modes. This adversely affects the deconvolution process. Additional noise introduced in the clean map due to these insensitive modes of $\mathbf{\Gamma}$ makes the deconvolution process nontrivial and impractical. The ML estimation of the true SGWB sky given by Eq.~(\ref{eq:ML_power}) exists only when the Fisher matrix is well conditioned and hence invertible. This problem leaves us with two solutions: either by linearly solving the convolution equation or by applying appropriate regularization to $\mathbf{\Gamma}$ before the inversion. In this paper, we will limit our discussions to two types of regularization recipes, SVD and Norm Regularization~\cite{regDeconv}. The SVD regularization has been used in the earlier studies~\cite{Eric,S5_Dir,O3paper} to characterize and condition the Fisher matrix. On the other hand, since we are focusing on the SGWB in the pixel-basis, which is well suited to search for point-like sources, norm regularization is an apt choice. 

\subsubsection{SVD Regularization}
The Fisher matrix as defined in Eq.~(\ref{eq:Fisher}), is Hermitian. Hence its SVD~\cite{Numerical_recipes,hansen_book} takes the form
\begin{equation}
\mathbf{\Gamma}= \mathbf{U \Sigma V}^{\dagger} \, ,
\end{equation}
where $\mathbf{U}$ and $\mathbf{V}$ are unitary matrices, and $\Sigma_{ij}=S_i\,\delta_{ij}$ is a diagonal matrix consisting of singular values $S_i$, whose nonzero elements are the real and positive eigenvalues of the Fisher matrix, arranged in descending order. Using the above decomposition of the Fisher matrix, we can rewrite the estimator of the true sky [see Eq.~(\ref{eq:ML_power})] as
\begin{equation}
\hat{\mathcal{P}_{k}}=\sum_j\,V_{kj}\frac{(\mathbf{U}^{\dagger}\mathbf{X})_j}{S_{j}} \, .
\end{equation}
The above equation shows that the solution converges if $|\mathbf{U_j}^{\dagger}\mathbf{X}|<S_j$ is satisfied. However, due to the noise contamination in the observed dirty map, the quantity $|\mathbf{U_j}^{\dagger}\mathbf{X}|$ does not decrease monotonically to zero and instead settles at a threshold depending on the noise level. The singular values which are less than this threshold contribute to the further enhancement of the noise.

Now, one can replace the eigenvalues of these problematic components with $\infty$ or alternatively their amplitude can be increased depending on the regularization scheme to obtain the SVD regularized covariance matrix as 
\begin{equation}
\mathbf{\Gamma}_S = \mathbf{U} \mathbf{\Sigma'} \mathbf{V}^{\dagger} \,,
\label{eq:reg_Cov_svd}
\end{equation}
where $\Sigma'_{jk}=S'_{j}\,\delta_{jk}$ and $S'_j$ are the regularized singular values defined by
\begin{align}
    S_i'=
    \begin{cases}
      S_i ~, &  S_i>S_{\mbox{cut}} \\
      \infty ~, & \text{otherwise}
    \end{cases}\, .
    \label{eq:reg_s_svd}
\end{align}
Here $S_{\mbox{cut}}$ is the singular value threshold, below which value we will be modifying the singular values for regularization. Since the choice of $S_{\mbox{cut}}$ plays a crucial role in the deconvolution process, a detailed discussion is laid out in the coming sections. Using the above regularized covariance matrix, the clean map can be written as
\begin{equation}
\label{eq:clean_SVD}
 \bm{\hat{\mathcal{P}}}_S = (\mathbf{\Gamma'}_S)^{-1}\mathbf{X} \, .
\end{equation}

\subsubsection{Norm Regularization} 
\label{subsec_norm}

In a Bayesian framework, assuming that the noise follows the Gaussian distribution, the posterior of the true map parameter, given dirty map $\mathbf{X}$ and covariance matrix $\mathbf{\Gamma}$, can be written as
\begin{equation}
    P(\bm{\mathcal{P}}|\mathbf{X},\mathbf{\Gamma},\lambda)=\frac{P(\mathbf{X}|\bm{\mathcal{P}},\mathbf{\Gamma})P(\bm{\mathcal{P}}|\lambda)}{\mathcal{Z}}\,, 
\end{equation}
where $\lambda$ is regularization parameter (or strength) and $\mathcal{Z}$ is normalization constant. The ML estimator in Eq.~(\ref{eq:ML_power}) maximizes the likelihood function $P(\mathbf{X}|\bm{\mathcal{P}},\mathbf{\Gamma})$ as well as the posterior but with a uniform prior.
Two broadly different Bayesian regularization schemes, namely norm and gradient regularization, were studied in Ref. \cite{regDeconv}. It was shown that the norm regularization which minimizes the total power in the map (thus suppressing noise), is more suitable to search for localized point-like sources. While, the gradient regularization, that uses a prior to prefer a smoother sky distribution, is better suited to look for a diffuse background. Here, since we limit the analysis to the pixel basis, which is more appropriate for point sources, we use norm regularization scheme, with the corresponding prior given by
%
\begin{equation}
    P(\bm{\mathcal{P}}|\lambda)=\frac{1}{2}\,e^{-\frac{1}{2}\lambda||\bm{\mathcal{P}}||^2}\,,
\end{equation}
which suppresses the noise efficiently for an optimal choice of the regularization parameter $\lambda$. The estimator of $\bm{\mathcal{P}}$ that maximizes the above posterior, the norm regularized clean map, is given as,
\begin{equation}
    \bm{\hat{\mathcal{P}}_N}=(\mathbf{\Gamma'_N})^{-1}\mathbf{X}=(\mathbf{\Gamma}+\lambda \mathbf{I})^{-1}\mathbf{X}\,,
    \label{eq:clean_Norm}
    \end{equation}
where $\mathbf{I}$ is the identity matrix.
As we discussed for SVD regularization, the norm regularization suppresses the enhancement of the noise by introducing some modification to the eigenvalues. This is achieved by adding the regularization constant $\lambda$ to each eigenvalue of $\mathbf{\Gamma}$.

The covariance matrices of the clean map estimators for both the regularization schemes can be written in a compact form as
\begin{equation}
 \mathbf{C}=(\mathbf{\Gamma'})^{-1}\mathbf{\Gamma}(\mathbf{\Gamma'})^{-1}\,.
\end{equation}
We can also write the SNR of the clean-map estimator as the ratio of the estimator and the square root of its variance. 

Both SVD and norm regularization affect the elements of the covariance matrix in different ways. Norm regularization increases the diagonal elements by equal amounts while preserving the non-diagonal elements. On the other hand, the SVD scheme changes the diagonal elements by unequal amounts. Unlike the norm regularization scheme, SVD changes the nondiagonal elements also. The SVD regularization scheme discards the problematic modes during the inversion, but the norm regularization modifies the contribution from such modes.

\subsection{Towards an Optimal `Reconditioning' of  NCVM}\label{sec:NMSE}
Even though the regularization schemes discussed in this paper rely on the statistical properties of both the noise and the source, they introduce nonzero bias. This implies that the regularized solution given in Eqs.~(\ref{eq:clean_SVD}) $\&$ (\ref{eq:clean_Norm}) are biased estimators of the true map. Since we opted to ignore the poorly observed modes of the covariance matrix, the SVD scheme introduces a bias in the estimator and the expected bias is given by
\begin{equation}
\langle\bm{\mathcal{P}}-\bm{\hat{\mathcal{P}}_S}\rangle=(\mathbf{I}-(\mathbf{\Gamma'_S})^{-1}\mathbf{\Gamma})\bm{\mathcal{P}}\,.
\label{eq:bias_SVD}
\end{equation}
Similarly for the norm regularization, the expected bias can be written as
\begin{equation}
 \langle \bm{\mathcal{P}}-\bm{\hat{\mathcal{P}}_N}\rangle = (\mathbf{I}-(\mathbf{\Gamma'_N})^{-1}\mathbf{\Gamma})\bm{\mathcal{P}}\,.\\
\label{eq:bias_Norm}
\end{equation}
Thus the bias is dependent on the difference between the unregularized and regularized covariance matrices. To understand its behavior, we consider two scenarios that differ in the regularization strength. First, the bias [Eq.~(\ref{eq:bias_SVD})] tends to increase with respect to an increase in $S_{\mbox{cut}}$ or $\lambda$. However, with larger $S_{\mbox{cut}}$ or $\lambda$ values, the solution [Eq.~(\ref{eq:clean_SVD})] becomes stable against the rounding-off errors or the number of iteration in a linear equation solver. Second, for small values of $S_{\mbox{cut}}$ or $\lambda$, the clean map is found to be dominated by noise (solution is under-smoothed, i.e., the contribution of high frequency components is dominated). Here, opposite to the first case, the bias is small and the solution (clean map) does not converge with increasing iterations. This type of behavior demands a trade-off between the bias and a better deconvolution. 

The condition of the covariance matrix to achieve a stable solution can be redefined in terms of the \textit{condition number}, which is the ratio of its largest eigenvalue to the smallest. A large condition number ($>10^6$) for a given observation is an indication that too many modes are getting reconstructed and hence making the map estimation an ill-conditioned problem. 

For the covariance matrix computed for broadband radiometer search from recent Advanced LIGO's observing run data~\cite{O1paper,O2paper,renzini1,renzini2}, it is found that the condition number $\kappa$ for $3072$ equal area pixels is greater than $\mathcal{O} (10^{17})$, hence the matrix is highly ill-conditioned. We are redefining the problem of setting the threshold on $S_{\mbox{cut}}$ or $\lambda$ as the problem of setting a threshold on the target condition number which optimally ``recondition" the NCVM.

Since the choice of regularization (and the corresponding target condition number) plays an important role in the quality of reconstruction we need to construct an estimator for this. However, there is no unique way of determining this quality of the recovered map. For this work, since we are interested in point-like sources, we consider the normalized mean squared error (NMSE) as the figure of merit. The NMSE is defined as 
\begin{equation}
    \mbox{NMSE}=\frac{||\bm{\mathcal{A}}-\bm{{\mathcal{B}}}||}{||\bm{\mathcal{A}}||}\,,
    \label{eq:def_NMSE}
\end{equation}
where $\bm{\mathcal{A}}$ and $\bm{\mathcal{B}}$ are the source map and reconstructed map, respectively. The values of NMSE give us some insight into the normalized bias of the estimator. In this study, we will rely on injection studies to find an optimal target condition number for both SVD ($\kappa(\mathbf{\Gamma'}_S)=\kappa_S$) and norm ($\kappa(\mathbf{\Gamma'}_N)=\kappa_N$) regularization methods, such that it minimizes NMSE and ensures that the clean map is not dominated by the noise.  

\section{Noise covariance matrix and significance}
\label{sec:covaraince}
The main challenge in a signal detection problem is usually associated with the methods being used to quantify the significance of the true signal (in our case, astrophysical) from noisy data. A detection statistic is formed to ascertain the presence of a signal (alternative hypothesis) against noise (null hypothesis). The goal of this section is to use different methods for identifying the signals and determining the associated significance. In the past SGWB searches~\cite{S5_Dir,O1paper,O2paper}, this was achieved by considering the highest SNR pixel in the sky map and calculating the expected probability distribution of the maximum SNR ($\rho_{\rm{max}}$) for $N$ independent background realizations using the recipe from Ref.~\citet{S5_Dir}. In all of these calculations, the covariance between the pixels plays a crucial role. Before the introduction of {\tt PyStoch}, calculating the expected NCVM was severely limited by the computational cost. As a workaround, NCVM from a Spherical Harmonic basis was utilized to perform the significance calculations in the pixel basis. In this paper, we incorporate the full NCVM in pixel basis. We compute the $p$-value to quantify the significance (in the absence of any significant events, we quote Bayesian upper limits at $95\%$ confidence). We discuss different methods to calculate $p$-value and the upper limits below. 

\subsection{$p$-value}\label{sec:p_value}

\subsubsection{Noise simulation}\label{sec:p_value_simulation}
The subtle nature of the distribution of the maximum SNR ($\rho_{\rm{max}}$), due to the nonzero covariance between the pixels on the sky plays an important role in the significance calculations. On the other hand, the folded data for an observation run is expected to be nearly distributed in a Gaussian manner due to the central limit theorem, as each folded data segment has been averaged over several sidereal days. Using these properties of the analysis, the probability distribution of $\rho_{\mbox{max}}$ can be obtained by simulating many realizations of the dirty maps, using the full NCVM. We use the standard Python library NUMPY's built-in multivariate pseudorandom number generator to perform this simulation. We carry out $N_{\mbox{sim}} = 10^3$ background realizations. Then the $p$-value is given by,
\begin{equation}\label{eq:p_realisation}
    p\text{-value}=\frac{N \left(\rho>\rho_{\mbox{max}}\right)}{N_{\mbox{sim}}}\,,
\end{equation}
where the numerator describes the number of simulated backgrounds that exceed $\rho_{\mbox{max}}$. 

\subsubsection{Conditional Multivariate Gaussian Probability}\label{sec:p_value_MVN}

The additive noise [see Eq.~(\ref{eq:convolution})] is supposed to follow a multivariate Gaussian (MVG) distribution with mean $\vec{0}$ and NCVM $\mathbf{\Gamma}$. In our analysis, the null hypothesis is that the observed sky map consists of pure noise, in which case the probability distribution of dirty map ($\mathbf{X}$) or clean map ($\bm{\hat{\mathcal{P}}}$) also follows a MVG with mean zero. The calculation of the significance using the analytical likelihood is impossible due to the ill-conditioned behavior of the NCVM
without regularization. To understand this problem in detail, we will explore $p$-value calculations using both dirty-map and clean-map covariance matrices.

First, we define the log-likelihoods for observing a noise-only dirty map (null hypothesis) using the SVD and Norm regularized covariance matrices respectively as
\begin{eqnarray}
    \mathcal{L}_1&=&A_1-\frac{1}{2} \mathbf{X}^T{\mathbf{\Gamma'}_S}^{-1}\mathbf{X}\,,\\
    \mathcal{L}_3&=&A_3-\frac{1}{2} \mathbf{X}^T{\mathbf{\Gamma'}_N}^{-1}\mathbf{X}\,,
\end{eqnarray}
where $A_1$ and $A_3$ are normalization constants. 

The clean map $\bm{\hat{\mathcal{P}}}$ and the corresponding covariance matrix can also be used to calculate the $p$-value. Given the covariance matrix $\mathbf{C}$ of both the SVD-regularized and norm-regularized clean map are noninvertible, we assume that $\mathbf{C}^{-1}=\mathbf{\Gamma'}$ (with this assumption, standard deviation of clean map is overestimated and hence the estimated $p$-value is expected to be biased). Now the corresponding log-likelihood functions can be written as
\begin{eqnarray}\label{eq:p_L4_lkhd}
     \mathcal{L}_2 &=& A_2-\frac{1}{2} \bm{\hat{\mathcal{P}}}_S^T\mathbf{\Gamma'}_S\bm{\hat{\mathcal{P}}}_S\,, \\
     \mathcal{L}_4 &=& A_4-\frac{1}{2} \bm{\hat{\mathcal{P}}}_N^T\mathbf{\Gamma'}_N\bm{\hat{\mathcal{P}}}_N\,,
\end{eqnarray}
where $A_2$ and $A_4$ are normalization constants.

Here we are interested in defining the significance of the maximum SNR pixel, because if the maximum SNR pixel is not significant enough to host a potential signal, any smaller SNR pixel will obviously not have enough significance. The correlation between the pixels affects the observed statistic (and the significance) of individual pixels. Hence, defining the significance demands the calculation of the conditional probability of the observed value of $X_{M}$ or $\hat{\mathcal{P}}_M$, given other elements of $\mathbf{X}$ or $\bm{\hat{\mathcal{P}}}$. Now the $p$-value can be written as
\begin{eqnarray}
     p\text{-value}\, (X^{\mbox{obs}}_{M})&=&P(X_M>X^{\mbox{obs}}_{M}\,|\,\mathbf{X}_{\neq M},\mathbf{\Gamma'})\nonumber\\
    &=&1-P(\mathbf{X}\leq X^{\mbox{obs}}_{M}\,|\,\mathbf{X}_{\neq M},\mathbf{\Gamma'})\,.
    \label{eq:p_dirty}
\end{eqnarray}
Following the above equation, we can also rewrite the $p$-value calculation using the clean-map estimator for its maximum SNR pixel.

\begin{figure*}
\centering
\includegraphics[width = 0.32\textwidth]{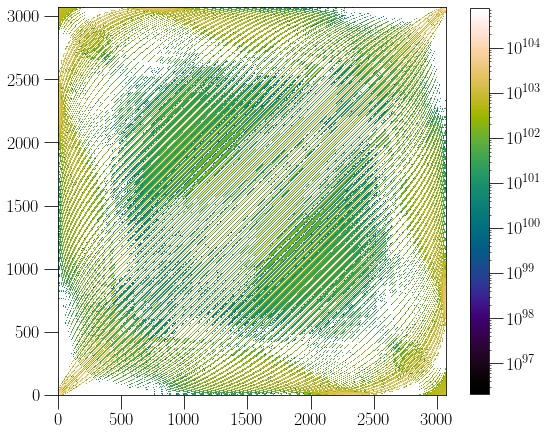}
\includegraphics[width = 0.32\textwidth]{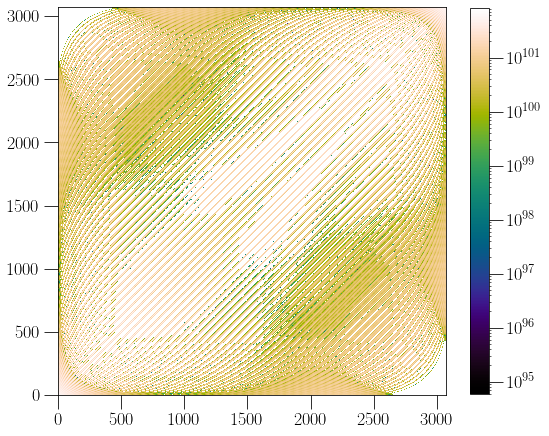} 
\includegraphics[width = 0.32\textwidth]{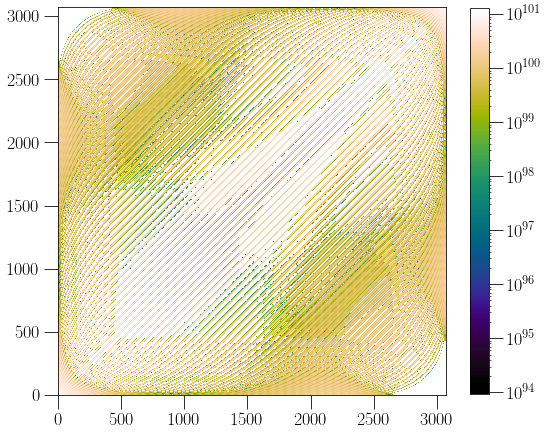} 
\caption{Covariance matrix (real value) calculated from O1 and O2 folded data sets using {\tt PyStoch} for the spectral indices $\alpha=3, 2/3, 0$ are shown from left to right, respectively.}
\label{fig:fisher_matrix}
\end{figure*}

\subsection{Upper Limit}\label{sec:UL}
In the absence of a detection, we can set a Bayesian upper limit on the strength of the source characterized by GW flux $\mathcal{F} (f,\mathbf{\hat{\Omega}})$, which is related to $\bm{\mathcal{P}}$ through Eq.~(\ref{eq:Flux}).

Using Bayes' theorem, the posterior of $\bm{\mathcal{P}}$ given the estimator $\bm{\hat\mathcal{P}}$, its covariance matrix $\mathbf{C}$, the likelihood function $L(\bm{\hat\mathcal{P}}|\bm{\mathcal{P}},\mathbf{C})$, and prior $P(\bm{\mathcal{P}})$ is given by
\begin{equation}\label{eq:comb_posterior}
    P(\bm{\mathcal{P}}|\bm{\hat\mathcal{P}})=\frac{L(\bm{\hat\mathcal{P}}|\bm{\mathcal{P}},\mathbf{C})\,P(\bm{\mathcal{P}})}{\int\,L(\bm{\hat\mathcal{P}}|\bm{\mathcal{P}},\mathbf{C})\,P(\bm{\mathcal{P}}) d\bm{\mathcal{P}}}\,.
\end{equation}
(A similar equation can be derived for dirty map as well.) Following the discussions of Ref.~\citet{Whelan_2014}, due to the calibration uncertainty of the detectors in a baseline, the $\bm{\hat\mathcal{P}}$ became the estimator of $\eta\bm{\mathcal{P}}$, where $\eta$ is an unknown calibration factor from a baseline with uncertainty $\epsilon$. Hence, the likelihood $L(\bm{\hat\mathcal{P}}|\bm{\mathcal{P}},\mathbf{\Sigma})$ is obtained by marginalizing over calibration uncertainty $\eta$. Using Eq.~(\ref{eq:comb_posterior}), the combined Bayesian upper limit $\bm{\mathcal{P}}_{UL}$ with confidence level (C.L.) is given as
\begin{equation}
     \text{CL}=\int_0^{\bm{\mathcal{P}}_{UL}}\,P(\bm{\mathcal{P}}|\bm{\hat\mathcal{P}})\, d\bm{\mathcal{P}}\,.
\end{equation}

To obtain an upper limit on $\bm{\mathcal{P}}$ for all sky directions, we can either marginalize over the directions or use the confidence contours in the $N_{\mbox{pix}}$ dimensional space. The results published by LVC~\cite{S5_Dir,O1paper,O2paper} used an estimator derived from the diagonal terms of the Fisher matrix which ignores pixel correlation, i.e., $\mathcal{\hat{P}}_{\mathbf{\hat\Omega}}=(\Gamma_{\mathbf{\hat\Omega}, \mathbf{\hat\Omega}})^{-1}X_{\mathbf{\hat\Omega}}$ with standard deviation $(\Gamma_{\mathbf{\hat\Omega}, \mathbf{\hat\Omega}})^{-1/2}$ . Then, the likelihood for $\mathcal{{P}}_{\mathbf{\hat\Omega}}$ is given by a Gaussian distribution with mean $\mathcal{\hat{P}}_{\mathbf{\hat\Omega}}$ and standard deviation $(\Gamma_{\mathbf{\hat\Omega}, \mathbf{\hat\Omega}})^{-1/2}$. In this work, we make a comparison of the upper limits derived from this likelihood, which will be called the \textit{conventional} likelihood, with new likelihoods formed with a clean map. First, the likelihood is formed with the clean map $(\bm{\hat{\mathcal{P}}}_S$ or $\bm{\hat{\mathcal{P}}}_N)$ and the diagonal of the clean map covariance matrix $\mathbf{C}$, i.e., each pixel is treated independently,
\begin{equation}\label{eq:norm_gaussian_lkhd}
    L(\mathcal{\hat{P}}_{\mathbf{\hat\Omega}}|\mathcal{P}_{\mathbf{\hat\Omega}},\bm{\mathbf{C}})=\frac{1}{\sqrt{2\pi C_{\mathbf{\hat\Omega}, \mathbf{\hat\Omega}}}}\,\mbox{exp}\left(\frac{-(\mathcal{\hat{P}}_{\mathbf{\hat\Omega}}-(\mathbf{\Gamma'}^{-1}\mathbf{\Gamma}\bm{\mathcal{P}})_{\mathbf{\hat\Omega}})^2}{2\,C_{\mathbf{\hat\Omega},\mathbf{\hat\Omega}}}\right)\,. 
\end{equation}
The clean map ($\bm{\hat{\mathcal{P}}}$) is a biased estimator of $\bm{\mathcal{P}}$. We incorporate this information into the likelihood by replacing $\bm{\mathcal{P}}$ by $\langle\bm{\hat{\mathcal{P}}}\rangle=\mathbf{\Gamma'}^{-1}\mathbf{\Gamma}\bm{\mathcal{P}}$ [see Eqs.~(\ref{eq:bias_SVD}) $\&$ (\ref{eq:bias_Norm})], where the angular brackets represent averaging over noise realizations.

The marginalization over the other directions suppresses the effect of correlation between pixels on the upper limit. Hence to take the pixel correlation into account, we will use the conditional MVG likelihood to set an upper limit for each direction across the sky. As we described in the significance calculation section, we have both dirty-map ($\mathbf{X}$) and clean-map ($\bm{\hat{\mathcal{P}}}$) estimators of the SGWB sky, along with their corresponding covariance matrices. Then, the log-likelihoods for observing a specific dirty map or clean map in presence of a source distribution $\bm{\mathcal{P}}$ can be formed,
\begin{eqnarray}
  \mathcal{L}_1&=&A_1-\frac{1}{2}\,(\mathbf{X}-\eta\mathbf{\Gamma}\bm{\mathcal{P}})^T  \mathbf{\Gamma'}_S^{-1} (\mathbf{X}-\eta\mathbf{\Gamma}\bm{\mathcal{P}})\,,\\
 \mathcal{L}_3&=&A_3-\frac{1}{2}(\mathbf{X}-\eta\mathbf{\Gamma}\bm{\mathcal{P}})^T \mathbf{\Gamma'}_N^{-1}(\mathbf{X}-\eta\mathbf{\Gamma}\bm{\mathcal{P}})\,, \\
\mathcal{L}_2&=&A_2-\frac{1}{2} (\bm{\hat{\mathcal{P}}}_S-\eta\mathbf{\Gamma'}_S^{-1}\mathbf{\Gamma}\bm{\mathcal{P}})^T \mathbf{\Gamma'}_S (\bm{\hat{\mathcal{P}}}_S-\eta\mathbf{\Gamma'}_S^{-1}\mathbf{\Gamma}\bm{\mathcal{P}})\,, \\
 \mathcal{L}_4&=&A_4-\frac{1}{2} (\bm{\hat{\mathcal{P}}}_N-\eta\mathbf{\Gamma'}_N^{-1}\mathbf{\Gamma}\bm{\mathcal{P}})^T \mathbf{\Gamma'}_N (\bm{\hat{\mathcal{P}}}_N-\eta\mathbf{\Gamma'}_N^{-1}\mathbf{\Gamma}\bm{\mathcal{P}}) \label{eq:UL_lkhd_L4}\,. 
\end{eqnarray}

To reduce the computational cost, we assume that $\bm{\mathcal{P}}$ for all pixels, excluding the one we are interested in, is zero. With this assumption, the upper limit can be computed using
\begin{equation}
     \text{CL}=\int_0^{\mathcal{P}_{i,UL}}\,P(\mathcal{P}_i|\bm{\hat\mathcal{P}},\mathcal{P}_{j\ne i}=0)\, d\mathcal{P}_i\,,
\end{equation}
where $i$ is the index corresponding to the pixel of interest in our SGWB search. 

In the case of multiple datasets (e.g., networks consisting of multiple detectors such as the LIGO observatories and Virgo or multiple observing runs such as O1 and O2), the statistic cannot be combined before marginalizing over $\eta$, due to different calibration uncertainties. Hence one has to consider the combined likelihood, which is given by~\cite{Whelan_2014},
\begin{equation}
    L(\bm{\hat\mathcal{P}}|\bm{\mathcal{P}},\bm{\eta})=\prod_{\beta}\,L(\bm{\hat\mathcal{P}}^{\beta}|\bm{\mathcal{P}},\bm{\mathbf{C}}^{\beta}) ,
\end{equation}
where $\beta$ is the index for the dataset. Further, we assume a uniform prior for the SGWB estimator $\bm{\mathcal{P}}$. We test these likelihoods with injection studies and rank them based on their ability to discriminate between signal and noise by assigning significance. The ``best'' likelihood can then be selected for setting the upper limit with the O1-O2 data.

\begin{figure*}
\centering
\includegraphics[width = 0.3\textwidth]{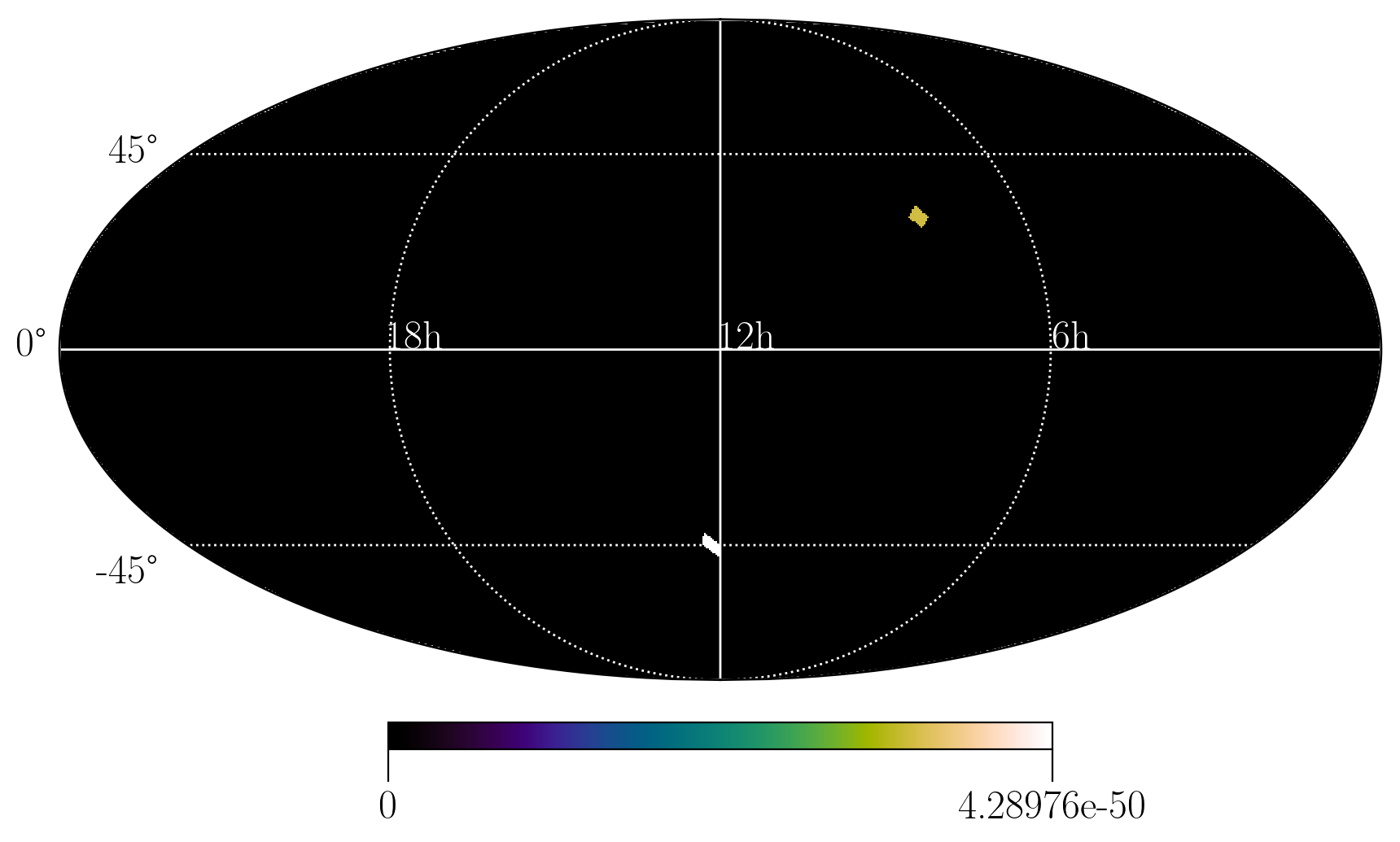} 
\includegraphics[width = 0.3\textwidth]{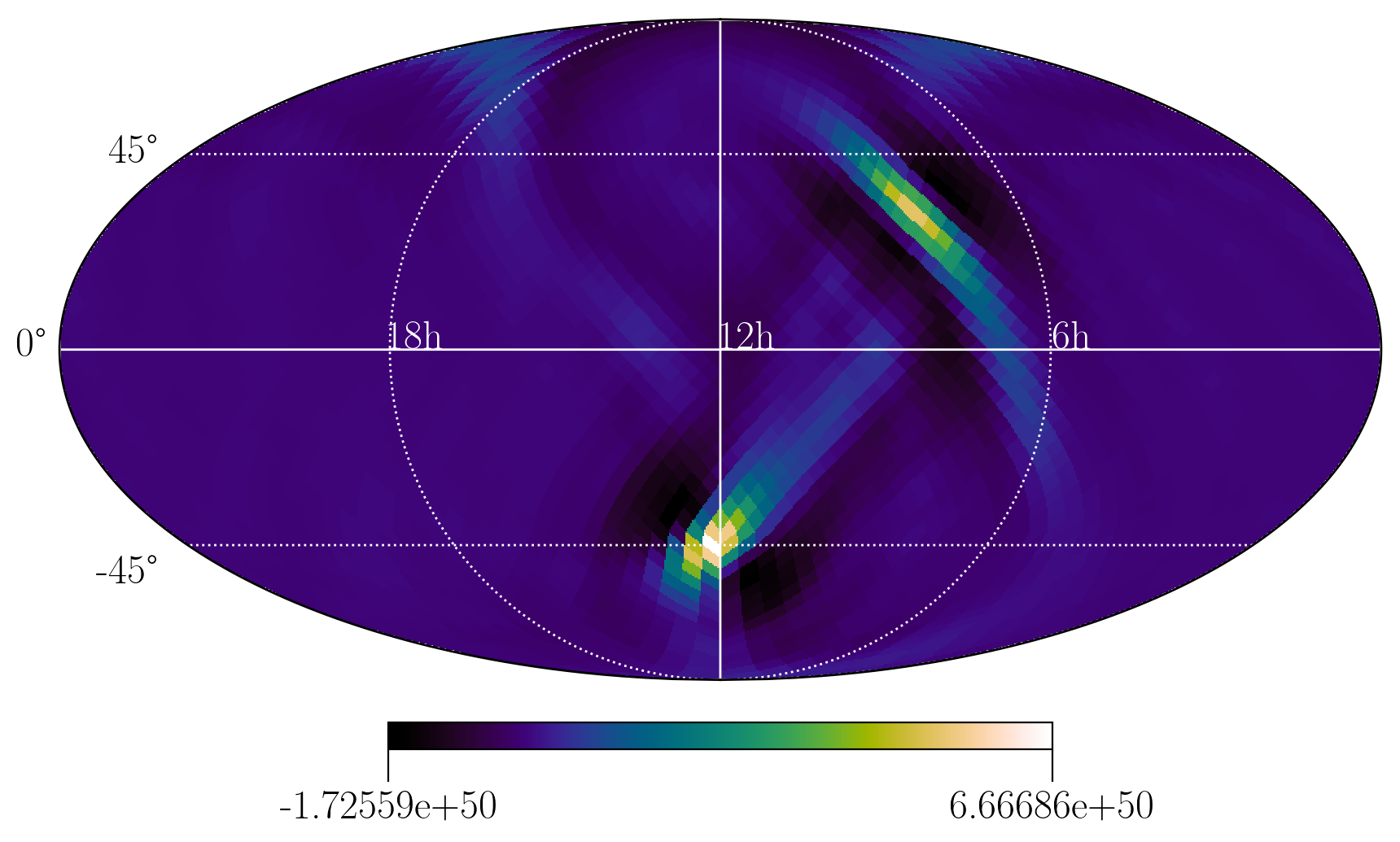} 
\includegraphics[width = 0.3\textwidth]{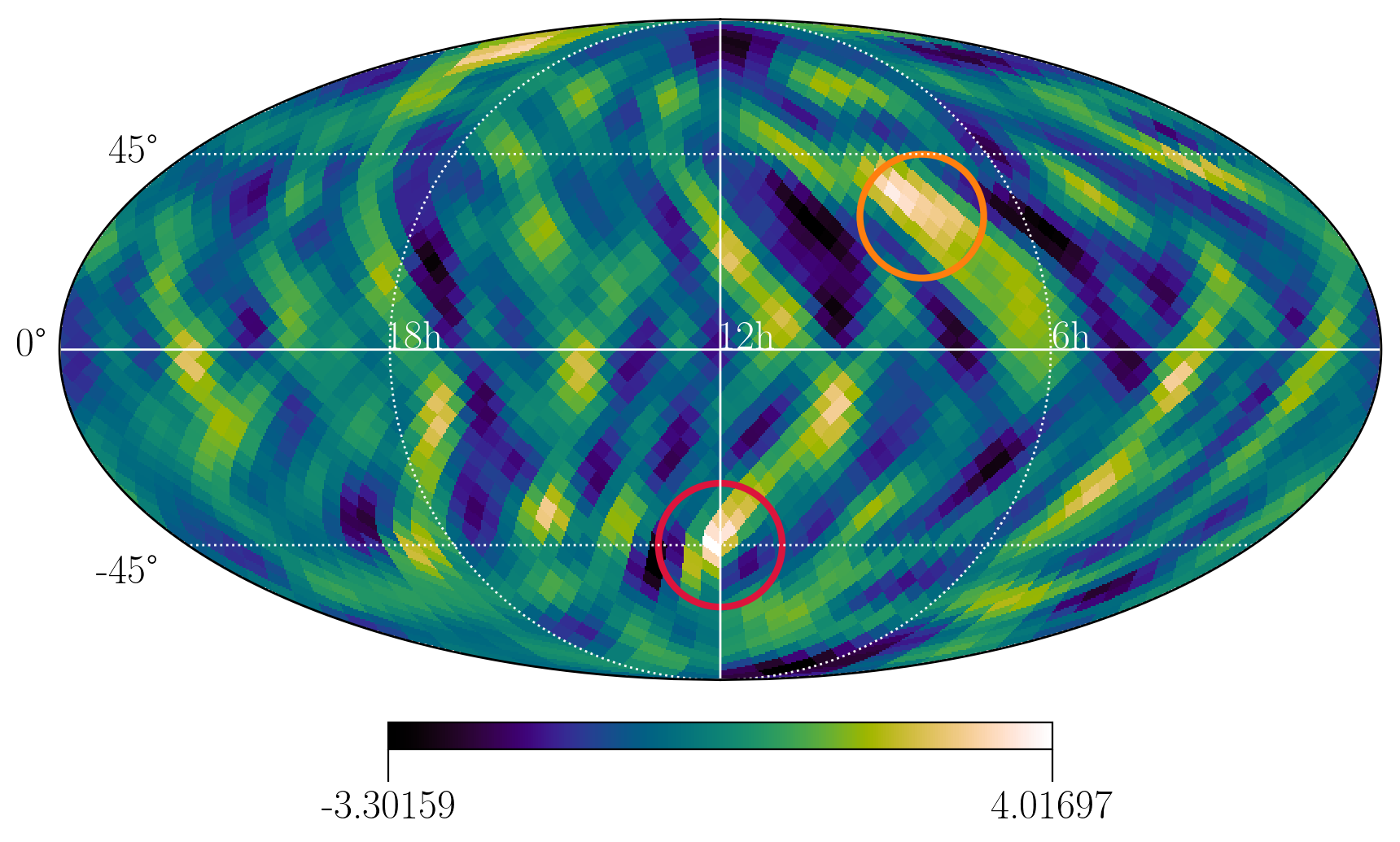} \\
\vspace{0.3cm}
\includegraphics[width = 0.3\textwidth]{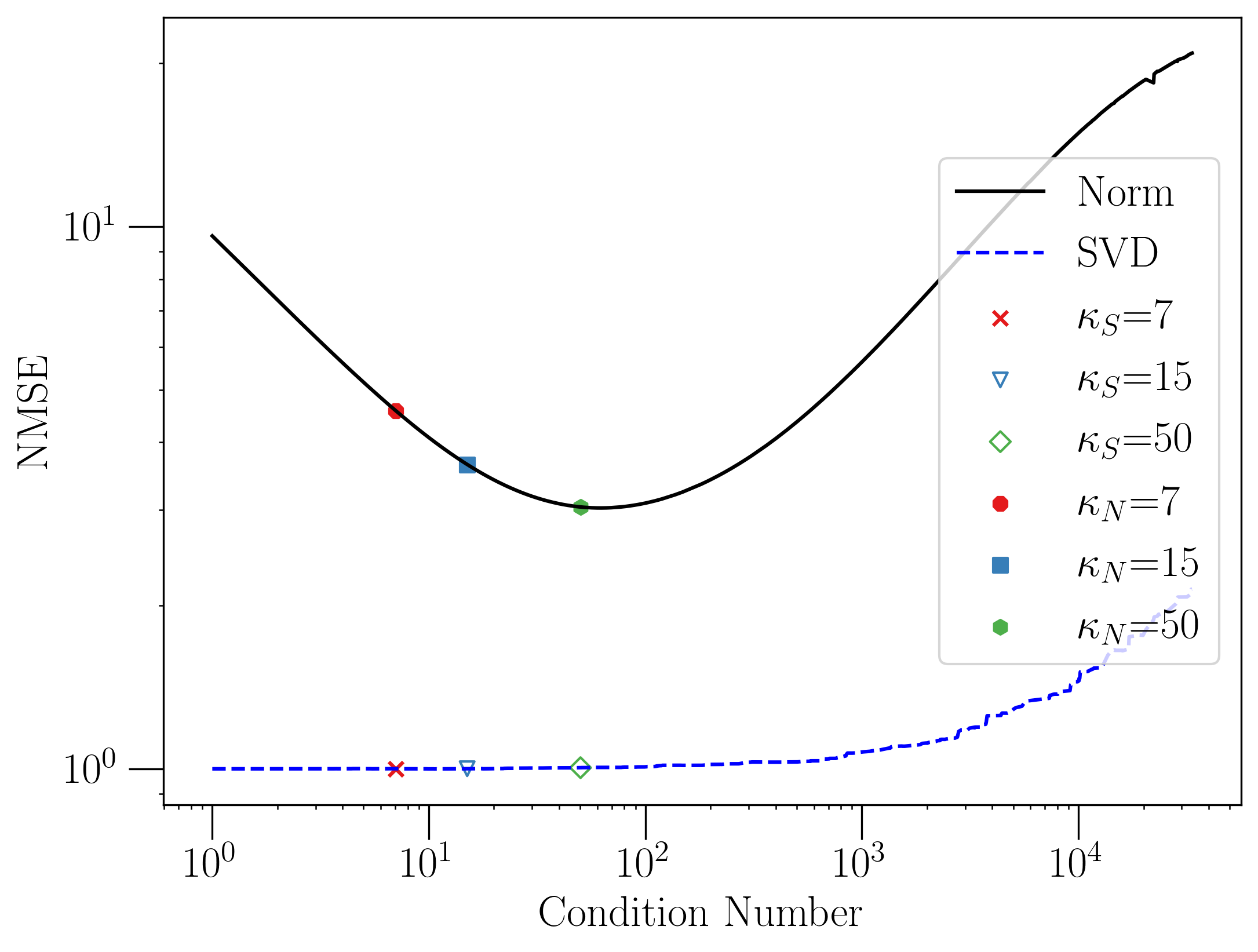}
\includegraphics[width = 0.3\textwidth]{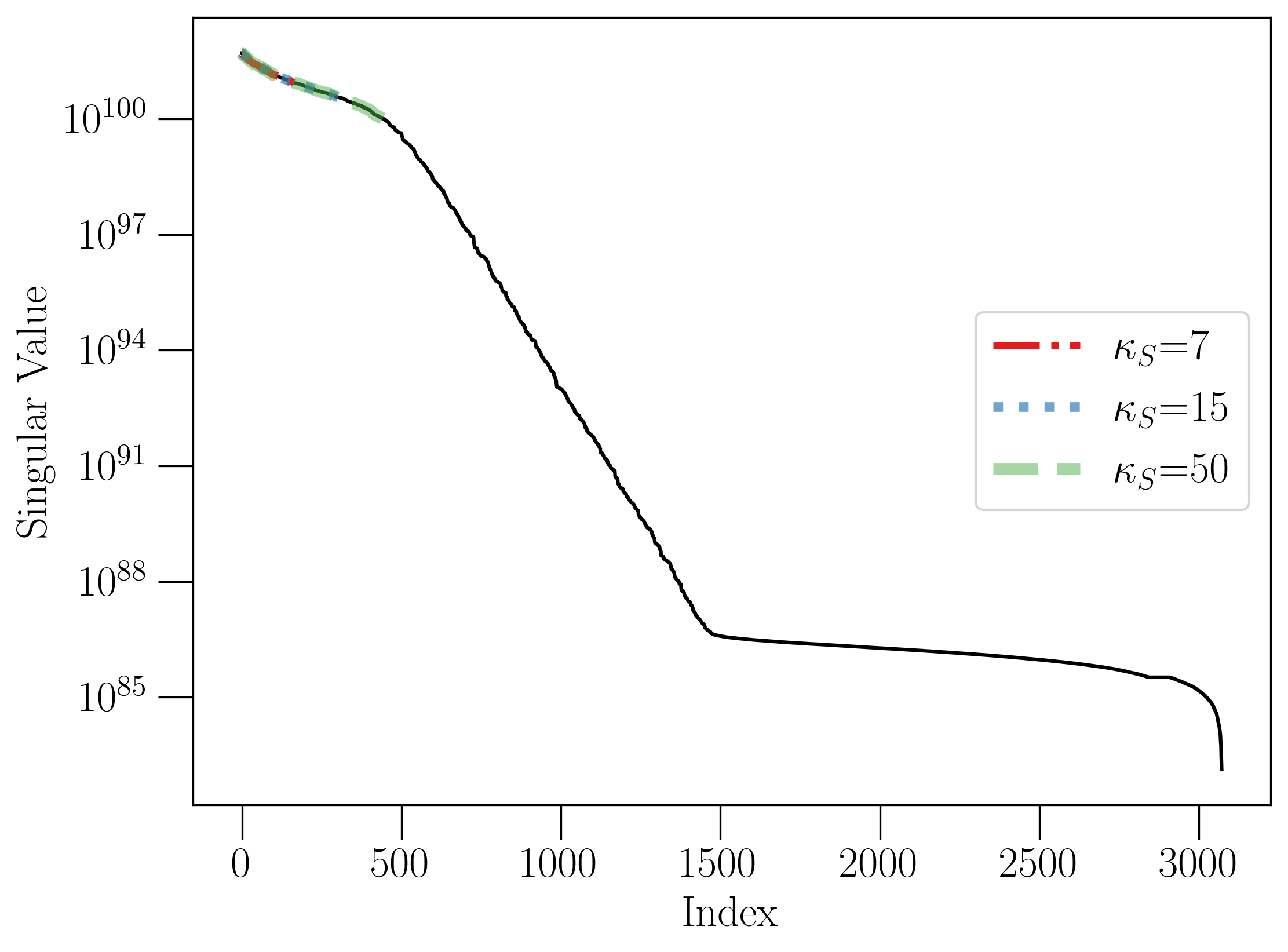}
\includegraphics[width = 0.3\textwidth]{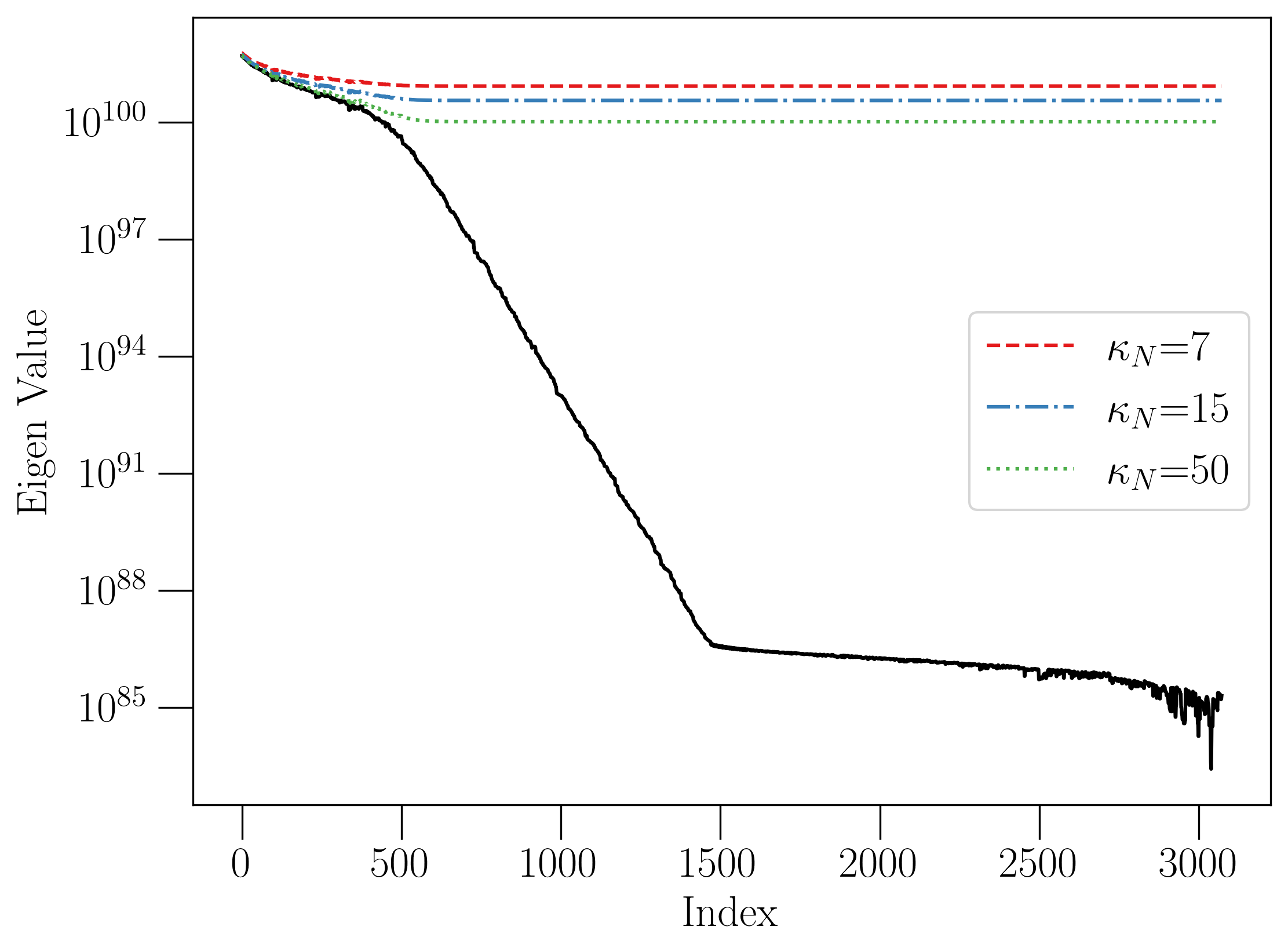} \\
\vspace{0.3cm}
\includegraphics[width = 0.3\textwidth]{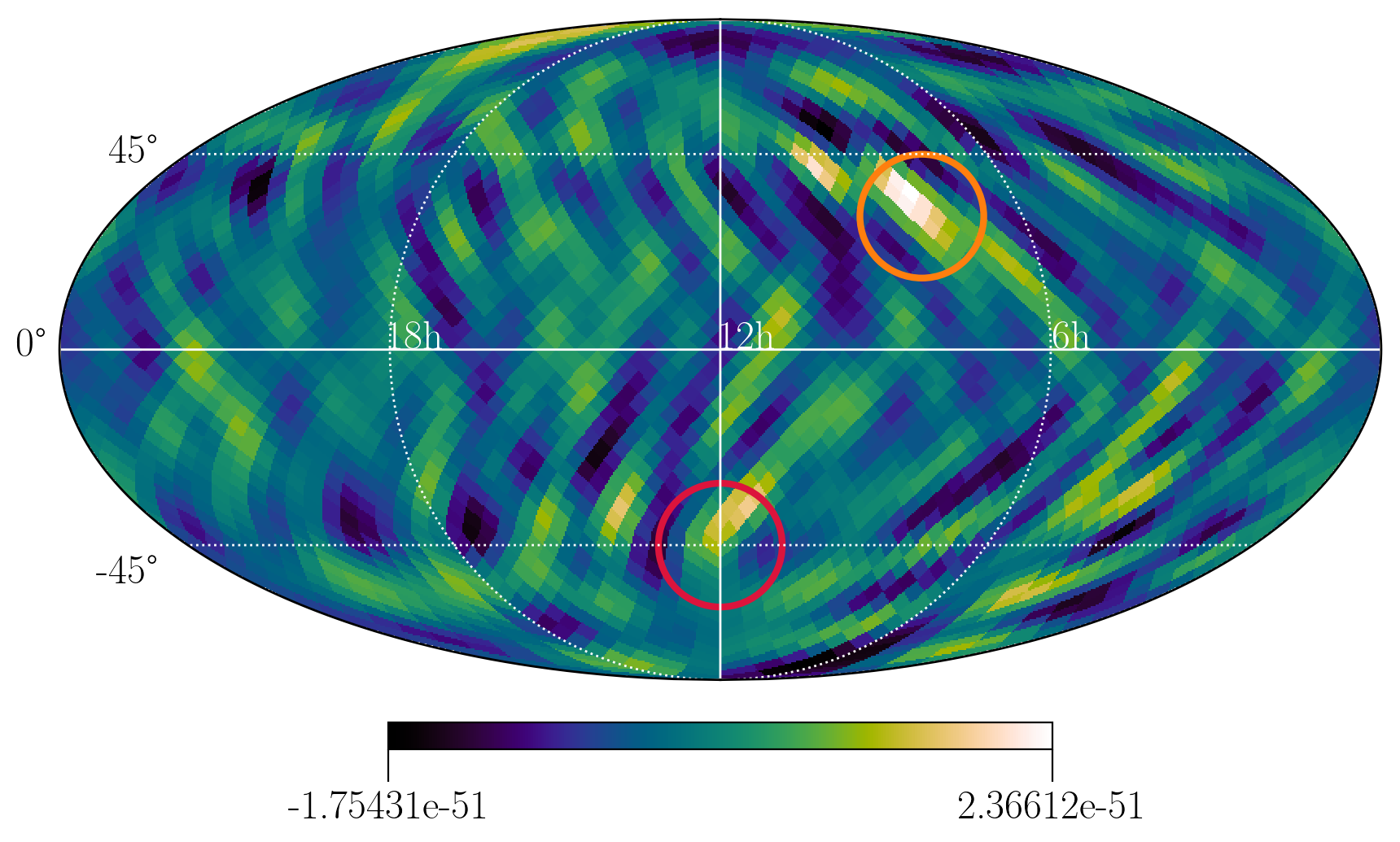} 
\includegraphics[width = 0.3\textwidth]{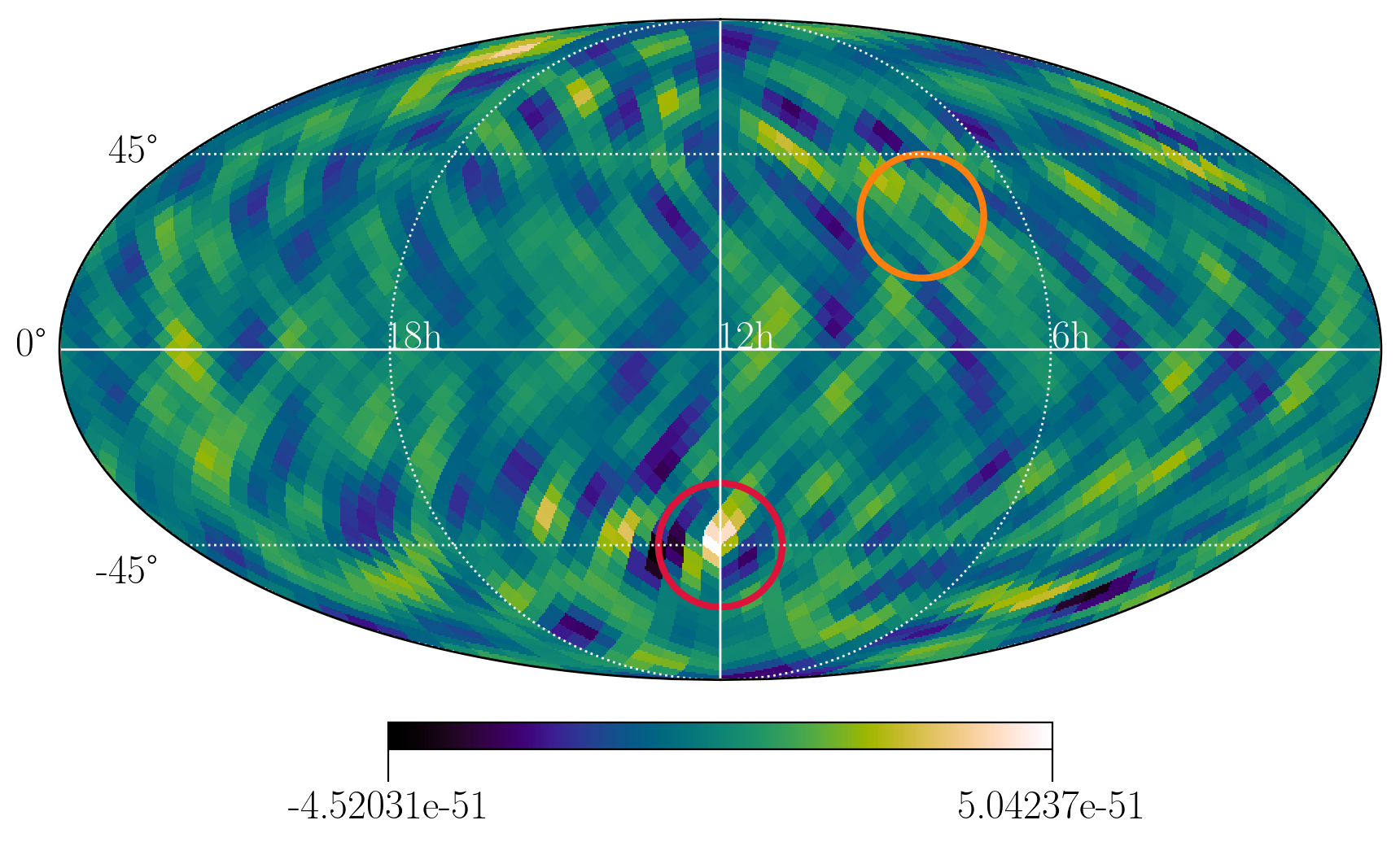} 
\includegraphics[width = 0.3\textwidth]{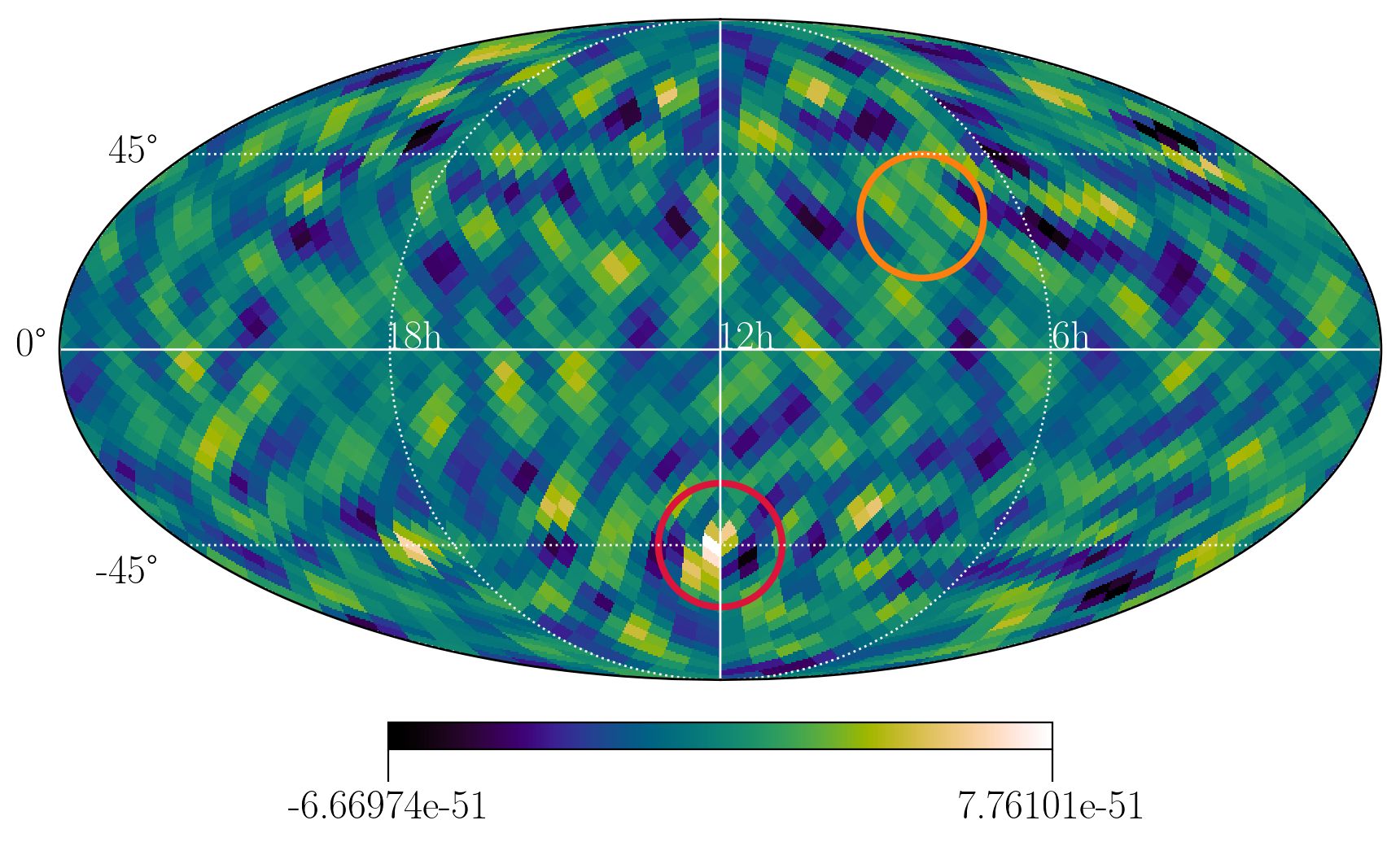} \\
\vspace{0.3cm}
\includegraphics[width = 0.3\textwidth]{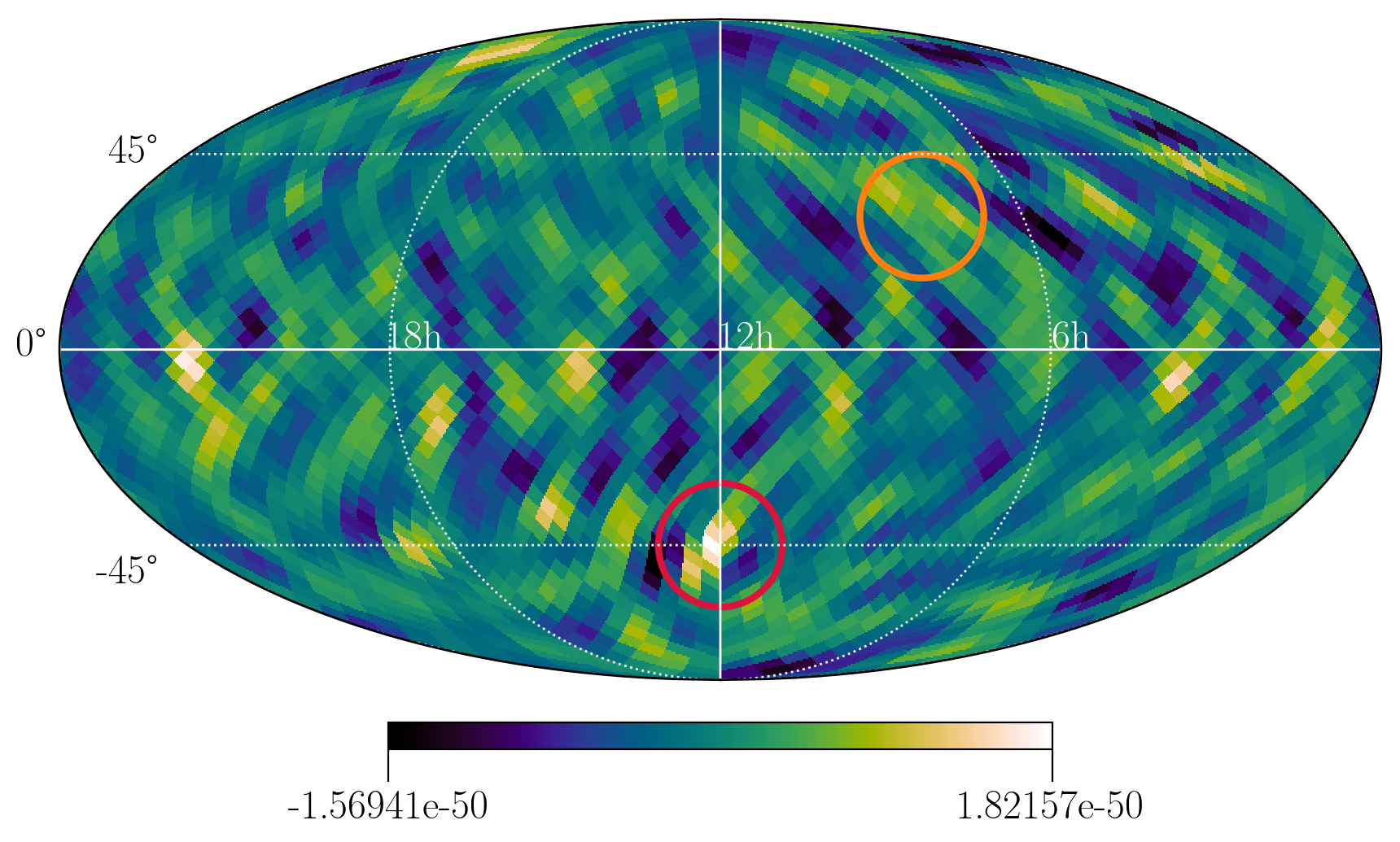} 
\includegraphics[width = 0.3\textwidth]{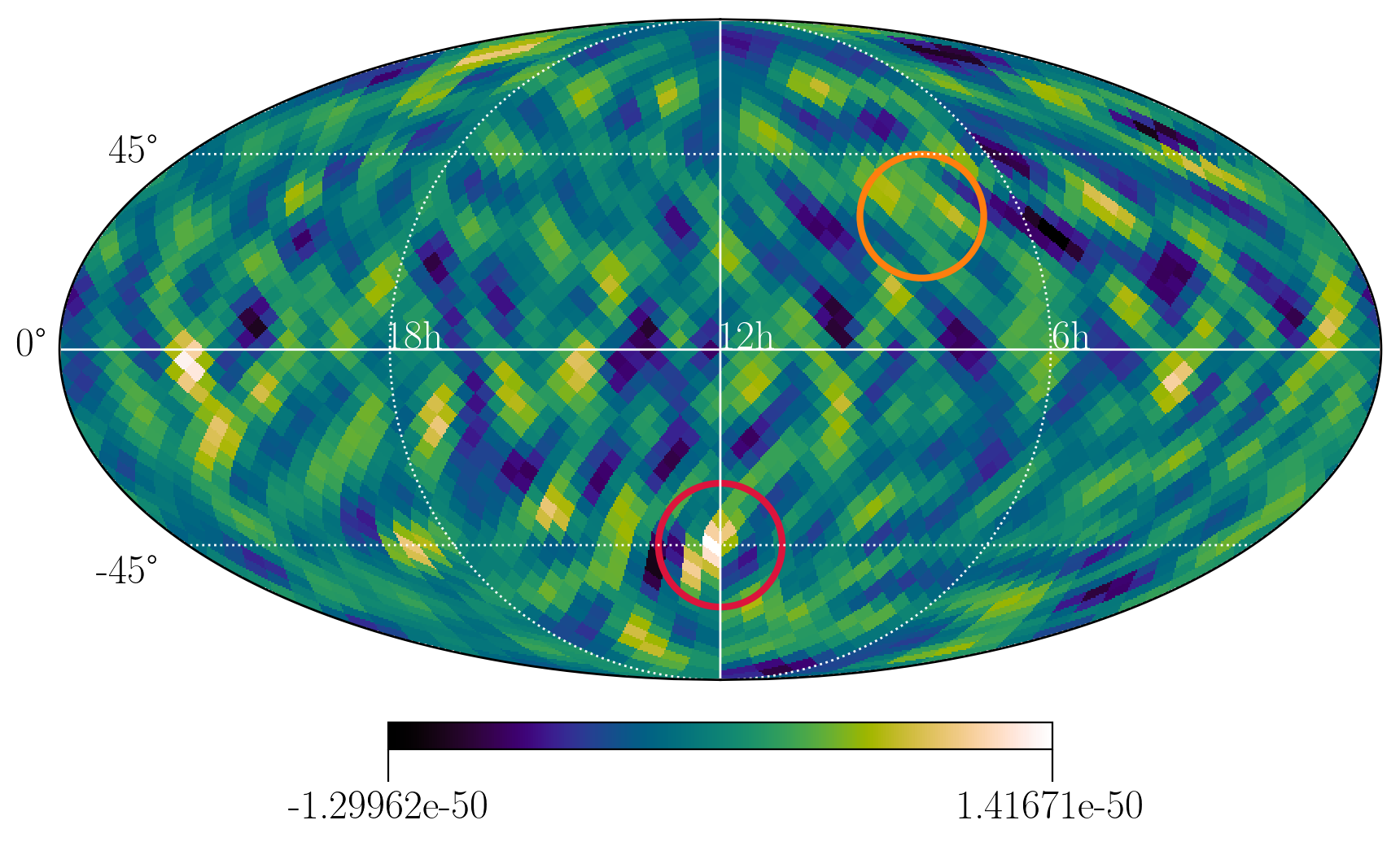} 
\includegraphics[width = 0.3\textwidth]{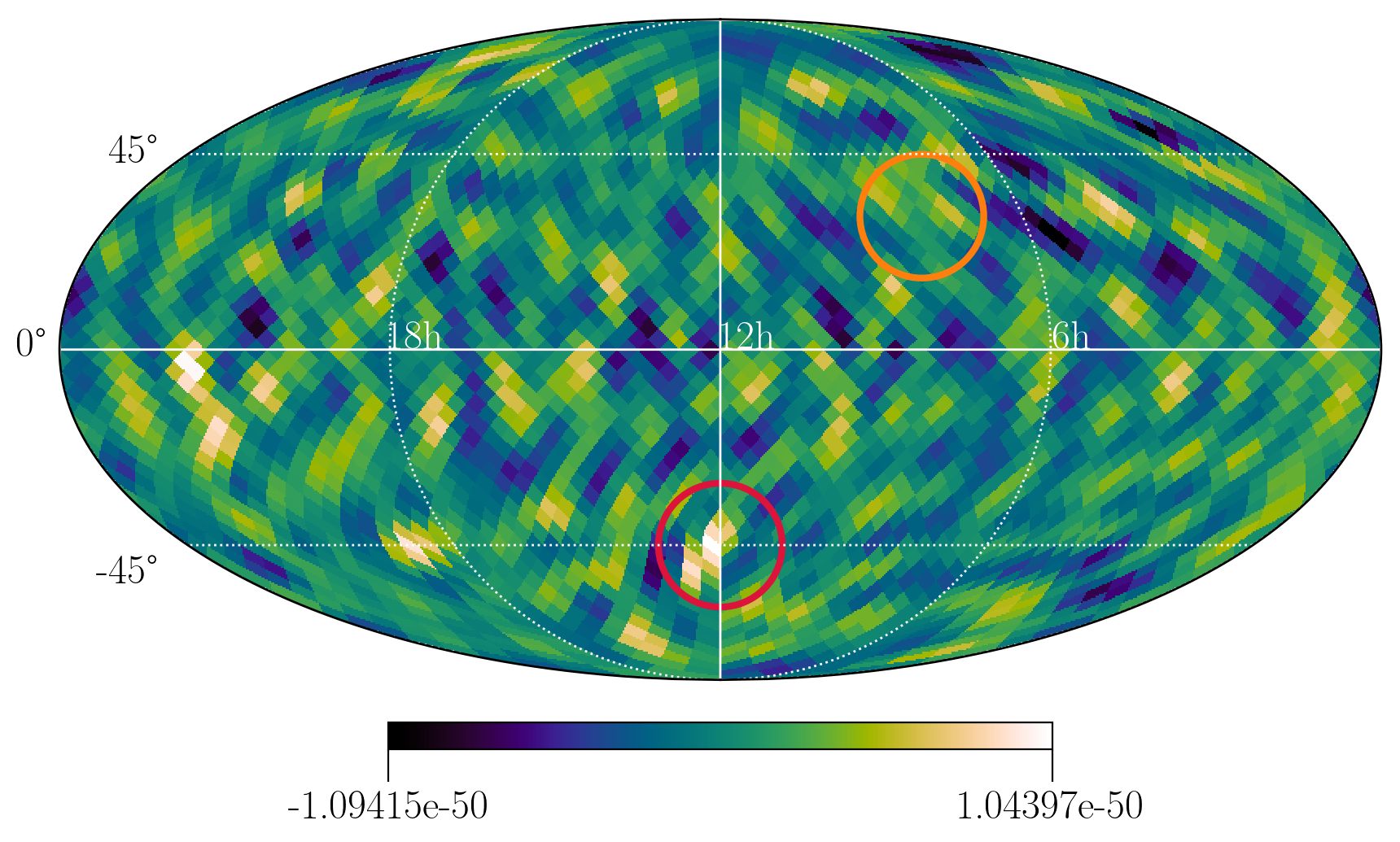} \\

\caption{Results of the injection study carried out to understand the effect of regularization recipes on deconvolution for $\alpha=3$ point source injections. The first row, respectively, shows the injected source power map, the clean map convolved with $\mathbf{\Gamma}$ without noise, and SNR of the dirty map with noise. In the second row, the leftmost plot shows the variation of the NMSE with target condition number ($\kappa_S$ or $\kappa_N$). We have chosen three target condition number (marked in the plot) to show their effect on the recovery of injections. The middle plot shows the singular value spectrum of $\mathbf{\Gamma}$ along with reconditioned covariance matrix to obtain the chosen target condition number. The rightmost plot shows the eigenvalue spectrum of $\mathbf{\Gamma}$ along with $\mathbf{\Gamma'}_N$ regularized with the chosen target condition number. The third row shows the ``scaled" clean-map power with SVD regularization with $\kappa_S=[7.0,15.0,50.0]$ from left to right, where the regularization keeps $10\%$, $14\%$, and $16\%$ of the singular values, respectively. The fourth row shows the ``scaled" clean-map power with Norm regularization with $\kappa_N=[7.0,15.0,50.0]$ from left to right. The quantitative results are summarized in Table~\ref{table:Injection study}. All the maps are represented as a color bar plot on a Mollweide projection of the sky in ecliptic coordinates.}
\label{fig:inj_4_a3}
\end{figure*}

\begin{table*}
\begin{tabular}{|c|cc|cccc|cccc|}
\hline
\multicolumn{1}{|c|}{$\alpha$}&\multicolumn{2}{c|}{Injections}& \multicolumn{4}{c|}{SVD Regularization}& \multicolumn{4}{c|}{Norm Regularization}\\
\multicolumn{1}{|c|}{}&\multicolumn{1}{c}{Power($\times10^{-50}$)}&\multicolumn{1}{c|}{SNR}& \multicolumn{1}{c}{$\kappa_S$}&\multicolumn{1}{c}{$s_{cut}(\times10^{97})$}& \multicolumn{1}{c}{Power($\times10^{-50}$)}&\multicolumn{1}{c|}{SNR}&\multicolumn{1}{c}{$\kappa_N$}&\multicolumn{1}{c}{$\lambda(\times10^{97})$}& \multicolumn{1}{c}{Power($\times10^{-50}$)}&\multicolumn{1}{c|}{SNR}\\
 \hline
3 & 3.9,4.3     & 3,4       & 7.0  & 7380 & 0.2 , 0.13 & 0.24 , 4.25 & 7.0  & 3630   & 0.54 , 1.82 & 0.84 , 4.19  \\
 &       &        & 15.0  & 3400 & 0.02 , 0.5 & 0.24 , 4.25 & 15.0  & 3630   & 0.26 , 1.42 & 0.84 , 4.19  \\
  &             &           & 50.0 & 1020  & 0.05 , 0.78 & 0.28 , 3.94 & 50.0  & 1040 & 0.10 , 1.04 & 0.40 , 3.73    \\
\hline
2/3 & 50,147.  & 3,4       & 32.6 & 4.5 & 2.09 , 3.47 & 2.03 , 2.96  & 32.6 & 4.6 & 19.0 , 14.9 & 2.19 , 1.51  \\
    &          &           & 45.4 & 3.2  & 1.48 , 2.36 & 1.29, 1.63    & 45.4 & 3.27    & 15.9 , 11.5  & 1.97 , 1.25 \\
    &          &           & 100.9  & 1.44 & 7.22 , 1.43 & 3.51 , 0.58  & 100.9 & 1.45 & 10.0 , 6.36 & 1.37 , 0.77 \\
\hline
0   & 157,295   & 3,4     & 26.4 & 1.85 & 2.8 , 3.2 & 1.84 , 2.78  & 26.4 & 26 & 20.6 , 48.4 & 1.38 , 2.83 \\
    &           &         & 45.4   & 1.2  & 2.3 , 6.6 & 1.33 , 4.25    & 45.4 & 1.29  & 12.6 , 35.2  & 1.01 , 2.39 \\
    &           &         & 102.0   & 0.56  & 1.6 , 7.2  & 0.73 , 2.42&  100.9 & 0.58  & 5.79 , 21.8 & 0.54 , 1.73 \\
\hline
\end{tabular}
\caption{Results from the injection study performed to learn the effect of the regularization recipes on deconvolution. A comparison of injected and recovered ``scaled" power (or SNR) with SVD and norm regularization with three target condition numbers for different power laws ($\alpha={3,2/3,0}$) with point sources is shown. The visualization of results is shown in Figs.~\ref{fig:inj_4_a3},~\ref{fig:inj_4_a23} $\&$~\ref{fig:inj_4_a0} for $\alpha=3,2/3,$ and 0, respectively.}
\label{table:Injection study}
\end{table*}

\section{Data}
\label{sec:data}

To perform the SGWB searches, we fetch the strain data from the first (O1) and second (O2) observing runs of Advanced LIGO detectors located in Hanford (H1) and Livingston (L1). The O1 data used here is collected from 120 days of observation starting from September 18, 2015 15:00 UTC to January 12, 2016 16:00 UTC. The collected O2 data consists of 265 days of observation from November 30, 2016 16:00 UTC to August 25, 2017 22.00 UTC. We followed the same data-processing methods used in Refs.~\cite{O1paper,O2paper,Folding}. Initially, the fetched time-series data is down sampled to 4096 Hz from 16 kHz. These data are divided into 192 s duration, $50\%$ overlapping, Hann-windowed segments. These are then high-pass filtered through a 16th-order Butterworth digital filter with a knee frequency of 11Hz. We then generate the CSD data from the two detectors (H1L1 baseline) and coarse-grained to a 1/32 Hz frequency resolution. These correlated data sets and their estimated variances are called stochastic intermediate data (SID). Following the steps described in Ref.~\citet{Folding}, SID are folded to one sidereal to form folded stochastic intermediate data. To account for the non-Gaussian features in the data, we identify segments containing known GW signals, segments associated with instrumental artifacts, and hardware injections. We also identify the segments which exhibit non-Gaussian behavior. We remove these nonstationary data and other ``bad segments'', during the folding process~\footnote{These cuts removed $35\%$ and $16\%$ of the data from O1 and O2, respectively~\cite{O1paper,O2paper}.}. In addition to this, we also identify the frequency bins associated with known artifacts. These frequency-domain cuts are applied to {\tt PyStoch} while performing the individual analyses~\footnote{These frequency cuts removed $21\%$ of the observing band in O1 data whereas in O2 it was $15\%$.}. 

The analysis is performed ``blindly'' on the folded data set to obtain the dirty map and the Fisher matrix corresponding to each run. We can form a combined Fisher matrix and dirty map by adding these results from individual runs~\cite{Romano2017}, O1 and O2,
\begin{eqnarray}
 \mathbf{\Gamma} &=& \mathbf{\Gamma}^{O1}+\mathbf{\Gamma}^{O2}\,, \\
 \mathbf{X} &=& \mathbf{X}^{O1} + \mathbf{X}^{O2} \,.
\label{eq:combined_dirty_fisher}
\end{eqnarray}
Following the procedure in Ref.~\citet{PyStoch}, both $\mathbf{\Gamma}$ (see Fig.~\ref{fig:fisher_matrix}) and $\mathbf{X}$ are computed using the O1 and O2 folded data and {\tt PyStoch} with a \hpx~\cite{HEALPix} resolution of $N_{\mbox{side}}=16$. Using the above equation along with Eq.~(\ref{eq:pixel_SNR}), one can easily construct the estimators of the GW power on the sky. 

\section{Injection Studies}
\label{sec:clean_inj}

In this section, we compare and investigate the regularization recipes described in Sec.~\ref{sec:reg_recipe} by injecting weak point sources, which will be the type of source we look for in a SGWB directed search. The injection study is expected to help in finding the optimal target condition number as well. In the next step, we perform the injection study to investigate the performance of the methods discussed in Sec.~\ref{sec:p_value}. The results from this study will give us an insight about the ``optimal" regularization recipe and the corresponding target condition number, and how to select the likelihood function among the ones listed above for significance and upper limit calculations.

The details of the injection study for point sources with the spectral index $\alpha=3$ are shown in Fig.~\ref{fig:inj_4_a3}. The full covariance matrix $\mathbf{\Gamma}$ is computed using the O1 folded data with a \hpx~\cite{HEALPix} resolution of $N_{\mbox{side}}=16$ using {\tt PyStoch}. We perform the point source injections by assigning a broadband intensity value to the corresponding pixels (see the first row of Fig.~\ref{fig:inj_4_a3}). These intensity values can be easily translated to the source strength and hence the SNR ($\sim 3-4$ in our case) of the maps. Then the injected map is convolved with the covariance matrix and combined with simulated noise to obtain the dirty map. As is evident from the first row of Fig.~\ref{fig:inj_4_a3}, the obtained dirty map displays a significant leakage of power in directions other than the injected source's directions (large point spread function). 

\begin{table*}
\begin{tabular}{|c|c|c|c|c|c|c|c|}
\hline
\multicolumn{1}{|c|}{$\alpha$}&\multicolumn{1}{c|}{Dirty map max snr}&\multicolumn{6}{c|}{$p$-value(\%)}\\
\multicolumn{1}{|c|}{}&\multicolumn{1}{c|}{}&\multicolumn{1}{c}{Simulation}&\multicolumn{1}{c}{Cond no.}&\multicolumn{1}{c}{$\mathcal{L}_1$}&\multicolumn{1}{c}{$\mathcal{L}_2$}&\multicolumn{1}{c}{$\mathcal{L}_3$}& \multicolumn{1}{c|}{$\mathcal{L}_4$}\\
 \hline
 3 & 2.68 & 98 & 7.03  & 12.93 & 0.73 & 28.32 & 31.64 \\
   &       &      & 15.00  & 0.85 & 3.47  & 28.40 & 24.59 \\
\hline
2/3 & 2.97 & 29.80 & 32.60  & 0.07 & 0.23  & 39.43 & 20.84 \\
    &       &      & 45.40  & 0.22 & 0.50  & 39.46 & 32.06 \\
   \hline
0 & 1.60 & 95.80 & 26.40  & 35.27 & 26.87  & 45.73 & 45.96 \\
  &       &      & 45.40  & 33.63 & 23.90  & 45.19 & 44.85 \\
  \hline
\end{tabular}
\caption{$p$-values calculated for a noise-only dirty map. Maximum SNR of the dirty map for each spectral index are tabulated along with the $p$-value obtained by the different approaches.}
\label{table:p_values_noise_inj}
\end{table*}

\begin{table*}
\begin{tabular}{|c|c|c|c|c|c|c|c|}
\hline
\multicolumn{1}{|c|}{$\alpha$}&\multicolumn{1}{c|}{Dirty map max snr}&\multicolumn{6}{c|}{$p$-value(\%)}\\
\multicolumn{1}{|c|}{}&\multicolumn{1}{c|}{}&\multicolumn{1}{c}{Simulation}&\multicolumn{1}{c}{Cond no.}&\multicolumn{1}{c}{$\mathcal{L}_1$}&\multicolumn{1}{c}{$\mathcal{L}_2$}&\multicolumn{1}{c}{$\mathcal{L}_3$}& \multicolumn{1}{c|}{$\mathcal{L}_4$}\\
 \hline
 3 & 3.67 & 16.60 & 7.03  & 5.44 & 4.05  & 24.32 & 23.74 \\
   &       &      & 15.00  & 0.01 & 0.01 & 25.12 & 10.56 \\
   \hline
2/3 & 3.80 & 1.80 & 32.60  & 0.06 & 0.13  & 36.00 & 15.19 \\
    &       &      & 45.37  & 0.05 & 0.06  &  35.86 & 10.91\\
   \hline
0 & 3.72 & 2.30 & 26.43  & 2.46 & 0.014  & 43.42 & 19.54 \\
  &       &      & 45.37  & 2.05 & 0.013 & 43.60 & 41.18 \\
  \hline
\end{tabular}
\caption{Results from the injection study to test the performance of different methods to calculate $p$-values in the presence of a source with a realistic signal strength.}
\label{table:p_values_signal_inj}
\end{table*}
 
In the next step, we perform regularized deconvolution, employing both SVD and norm-regularization schemes to reconstruct the true sky map from the observed dirty map. We first compute the singular values of the covariance matrix using the SVD scheme. We use these singular values to regularize the matrix and obtain the clean map [Eq.~(\ref{eq:clean_SVD})]. We then study the quality of source reconstruction using NMSE as the metric, by varying the target condition number and repeating the above steps to obtain the proper condition number-NMSE trade-off. On the other hand, the clean map with a norm-regularization scheme is obtained by solving Eq.~(\ref{eq:clean_Norm}) with an built-in conjugate gradient solver (CGS) in the {\tt PYTHON} {\tt{SciPy}} package for different target condition numbers. Using a maximum of $20$ iterations with the CGS module, we obtain the stable solution with a tolerance of $\sim 10^{-6}$. It is noticed that the strong regularization produces an estimator with multiplicative bias in the weak source case. We correct for this bias while calculating NMSE by scaling the recovered power map by this multiplicative bias, i.e., $\mbox{diag}(\mathbf{\Gamma'})/\mbox{diag}(\mathbf{\Gamma})$.

The second row of Fig.~\ref{fig:inj_4_a3} displays the condition number-NMSE plot along with singular values and eigenvalues. The NMSE is observed to first decrease and then increase with increasing $\kappa_N$, while NMSE always increases with increasing $\kappa_S$. To further demonstrate the effect of the choice of target condition number on deconvolution, we select three values of $\kappa_S$ and $\kappa_N$ from the region of the NMSE-condition number plot, where the NMSE is near to its minimum value. These condition numbers are marked in the plot and the corresponding effects on singular values (SVD regularization) and eigenvalues (norm regularization) are also shown (see the second row of Fig.~\ref{fig:inj_4_a3}). 

The third row of Fig.~\ref{fig:inj_4_a3} shows clean maps with a SVD regularization scheme for the three chosen target condition numbers. As we increase the condition number, the ``scaled" power of recovered sources increase (and the bias decreases). On the other hand, the amplitude of the noise also increases (increase in variance) which is apparent from the maps. This is indicated by the color bars of the recovered maps in the third row of Fig.~\ref{fig:inj_4_a3} and by the recovered SNR in the second row of Table~\ref{table:Injection study} as well. Both sources can be recovered with $\kappa_S=7$, but with $\kappa_S=15,50$, only the comparatively strong source is recovered and the recovery of the source with lower injection power is adversely affected by the noise boost. 
%
Note that, a recovered SNR can be significantly affected if we try to correct for the bias. However, since there is no way to guess the bias a priori, there is enormous uncertainty in bias-corrected SNR. Without bias correction, the clean-map SNR is less than dirty map SNR and hence not preferred, which is why, it is necessary to compare the injected value with the recovered value rather than the SNRs. 

The fourth row of Fig.~\ref{fig:inj_4_a3} shows clean maps with norm regularization. The injected sources having higher power (or SNR) is distinguishable (with recovered SNR$\sim$4) from the noise for all choices of the target condition number but it is not the case for injected sources with lower power. The recovered SNR of the source having higher power is not fluctuating with chosen target condition number, but for other source, it is observed to decrease with increase in target condition number (see second row of Table~\ref{table:Injection study}). This study suggests that the value of $\kappa_S$ and $\kappa_N$ in the range 7-15 are able to recover the injected weak sources with both regularization schemes for $\alpha=3$. 

We extend this study for $\alpha=2/3,0$. The results are summarized in the Appendix~\ref{sec:inj_a23_0} and quantitative results are presented in Table~\ref{table:Injection study}. This study suggests that we can recover the injected sources with both regularization schemes. We can use target condition numbers ($\kappa_S$ and $\kappa_N$) in the ranges [7-15, 32.6-45.4, 26.4-45.5] respectively for $\alpha=3,2/3,0$ in realistic scenarios where SNR$\sim4$.

Next, we investigate the ability of the methods to distinguish noise and the presence of a source by assigning significance to simulated realizations of noise and signal. An injection is performed following the method described earlier in such a way that the resulting dirty map will have a SNR$\sim3.5$ at the point of injection. Then, $p$-values are calculated following the methods described in Sec.~\ref{sec:covaraince}, and its maximum is normalized to $100\%$. Ideally, the $p$-value should be larger if the data set contains only noise, in comparison to the $p$-value obtained in the presence of a signal. 

\begin{figure}[ht]
\includegraphics[width=0.45\textwidth]{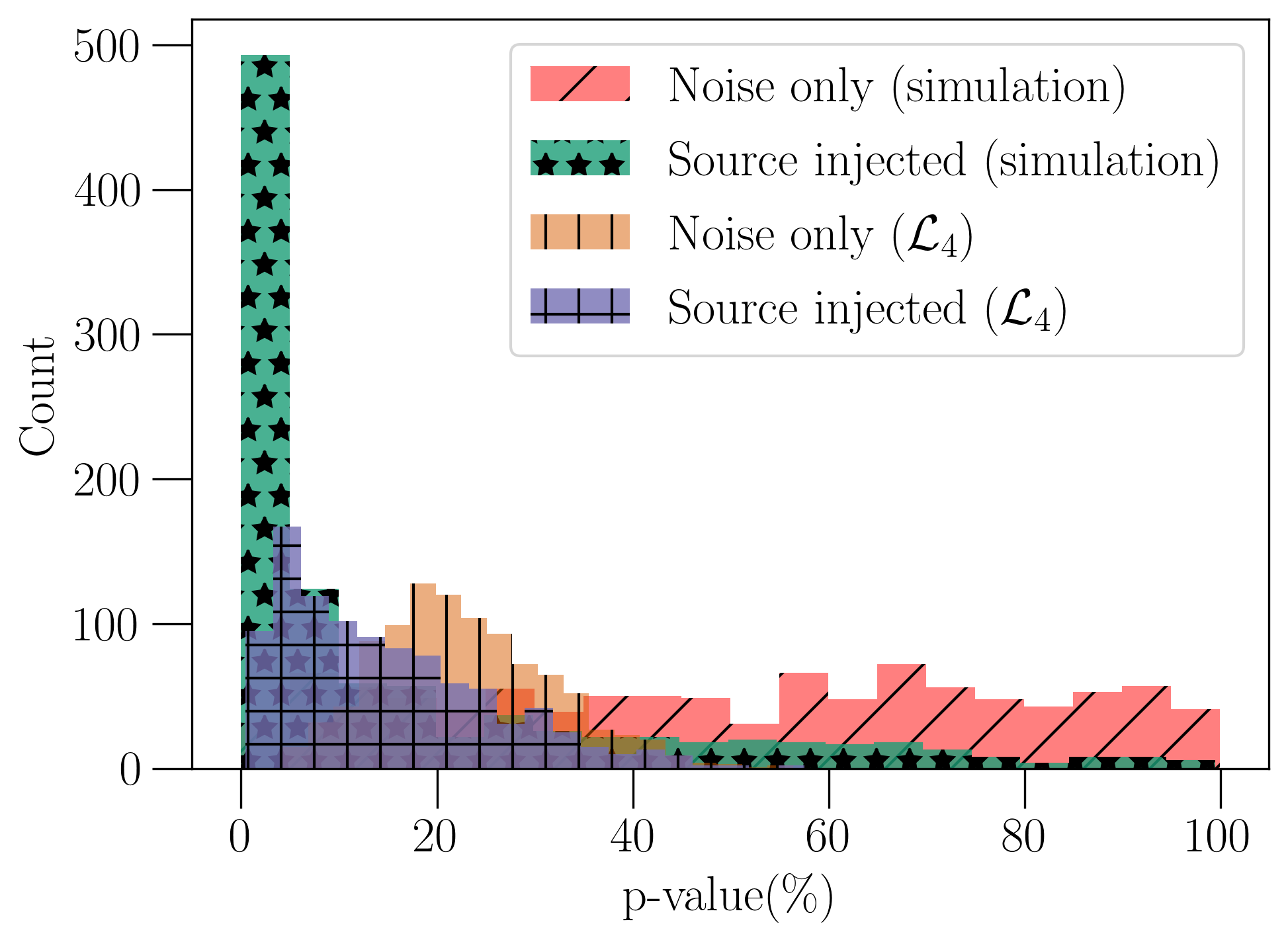}\\
\includegraphics[width=0.45\textwidth]{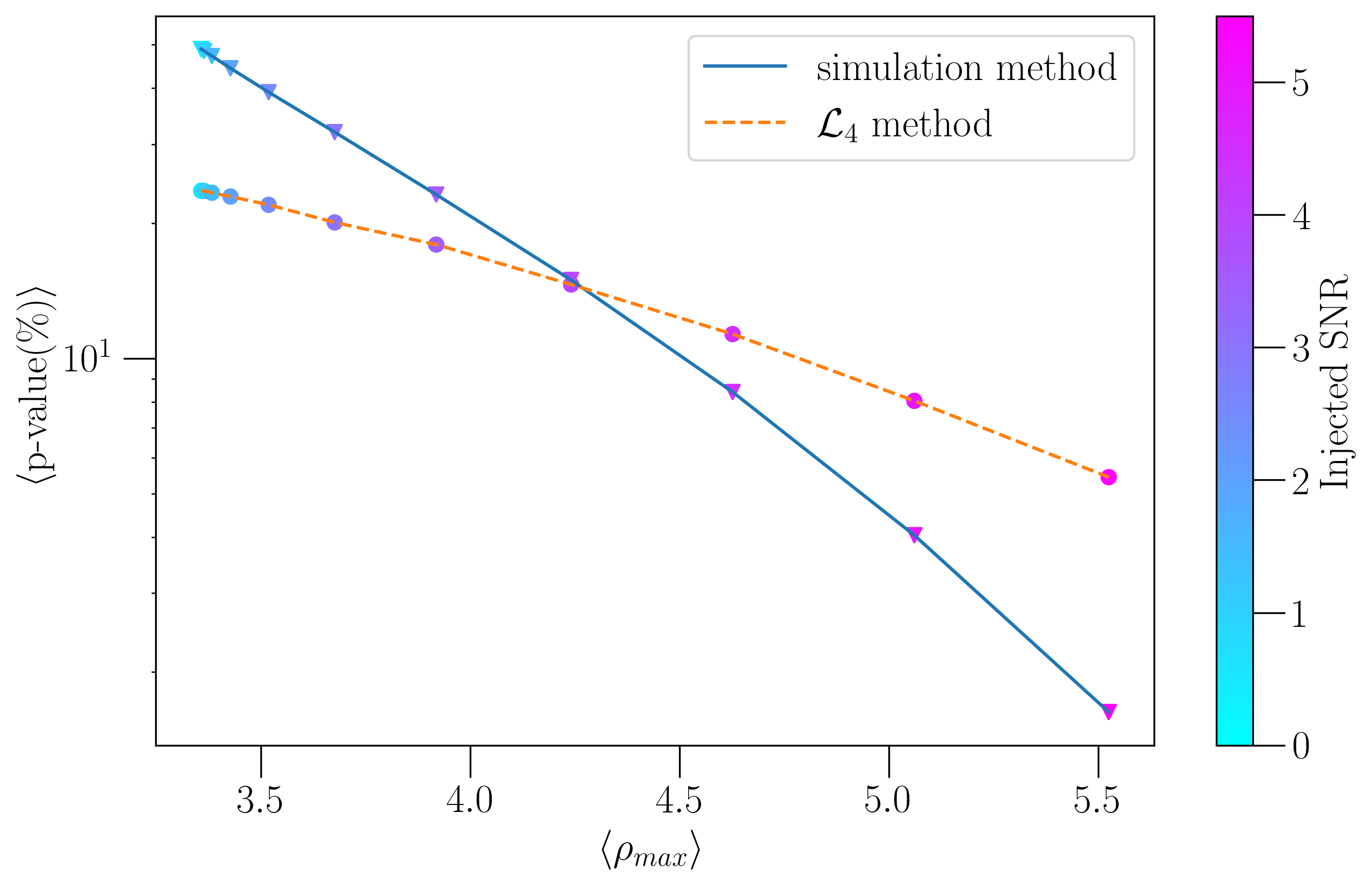}
\caption{Top: histogram of $p$-values calculated with 1000 realizations of noise with the covariance matrix $\mathbf{\Gamma}$ for a spectral index $\alpha=3$. It is compared with a histogram of $p$-values with the same noise realizations but in the presence of a constant source. The significance is calculated with simulations and the log-likelihood $\mathcal{L}_4$ method. Bottom: variation of $p$-values (\%) with maximum dirty-map SNR (averaged over 1000 noise realisations) for the case of a point source having a power spectrum with spectral index $\alpha=3$ and regularizing $\mathbf{\Gamma}$ with $\kappa_N=15.$ Note that mean dirty-map maximum SNR varies with the injected value of SNR (shown with the color bar). Considering the curves in the bottom panel, it is clear that the simulation method is better able to identify the noise (high $p$-value at low injected SNR) and also identify signals (lower $p$-value at higher injected SNR).}
    \label{fig:p_value_hist}
\end{figure}

For the $\alpha=3$ case, as shown in Table~\ref{table:p_values_noise_inj} and~\ref{table:p_values_signal_inj}, the $p$-value is significantly reduced in the presence of a source with all methods; however, with $\mathcal{L}_2$ and $\kappa_S=7.0$, the $p$-value is higher when a source is present than for the noise-only case. This is due to the incomplete recovery of the source. Instead, after performing the regularization with $\kappa_S=15,$ the source is recovered with higher SNR in comparison to the noise-only case. The $p$-values with norm regularization seem to be higher than those with the SVD regularization. The above study concludes that $\kappa_S=\kappa_N=15$ is an optimal choice to regularize NCVM for point sources with weak strength having spectral index $\alpha=3$. Similar studies have been carried out for $\alpha=2/3,0$. The calculated $p$-values are able to distinguish between the signal and noise.

The next step is to understand the threshold of $p$-values by considering many noise realizations. We create 1000 noise realizations with mean zero and covariance matrix $\mathbf{\Gamma}$ for $\alpha=3$ and calculate the $p$-value for each realization using methods described in Secs.~\ref{sec:p_value_simulation} and \ref{sec:p_value_MVN} with $\kappa_S=\kappa_N=15$. We inject a source of constant amplitude in these noise realizations and calculate the $p$-value with all methods. The $p$-values obtained by these methods cannot be compared directly, because the $p$-values obtained using the likelihood formula are biased. Their abilities to distinguish noise from signal injections can be judged by looking at the overlaps between the corresponding distributions. Among all of the methods, we observe that the noise simulations and $\mathcal{L}_4$ methods can distinguish noise against the presence of signals with the resolved $p$-value histograms. We present the comparison of histograms of the $p$-value for these two methods in the top panel of Fig.~\ref{fig:p_value_hist}. In the simulation method, the histogram of the $p$-value is significantly affected by the presence of a source and the peak is distinct. With the log-likelihood $\mathcal{L}_4$ the peaks of histograms are also distinct in the presence of the signal and noise. It is required to fix the $p$-value threshold such that if the $p$-value for the observed dataset is lower than this threshold. The chances of finding a true astrophysical source in the data is significant and, hence, may require further investigation. The statistical nature of the $p$-value is evident from the top panel of Fig.~\ref{fig:p_value_hist}. The threshold can be chosen such that the probability of getting a $p$-value less than that threshold is less than $1\%$ in a noise-only case. This can be decided based on the histograms. For $\alpha=3$, we get $0.25\%$ and $5\%$ thresholds for the simulation and $\mathcal{L}_4$ likelihood method.

To understand the effect of the presence of a source on significance, we test the variation of the $p$-values (averaged over 1000 noise simulations) with the mean of maximum dirty map SNR (bottom panel of Fig.~\ref{fig:p_value_hist}). Note that the variations in the SNR of injections are shown by the colorbar. The $p$-values are observed to decrease monotonically with the mean of the maximum dirty map SNR for the chosen methods. This study concludes that we can reliably assign significance to the observed estimator using both the noise simulation method and using the log-likelihood $\mathcal{L}_4$ with norm-regularized clean map and regularized covariance matrix using different thresholds on the $p$-value to claim a detection. Clearly, the conventional simulation method is superior in distinguishing signal from noise at the injected SNRs.
\begin{figure}[hb]
    \centering 
    \includegraphics[width=0.225\textwidth]{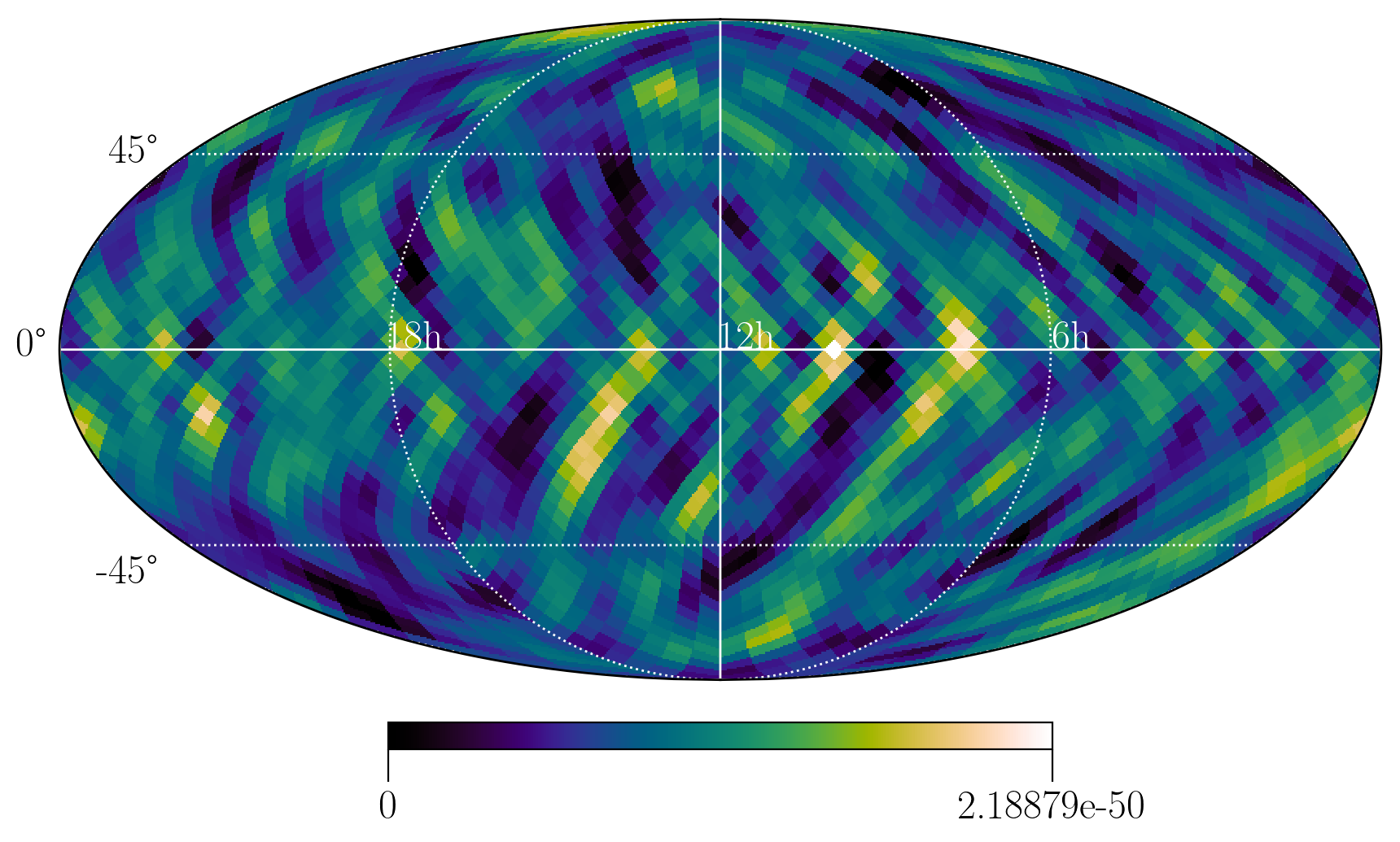}
    \includegraphics[width=0.225\textwidth]{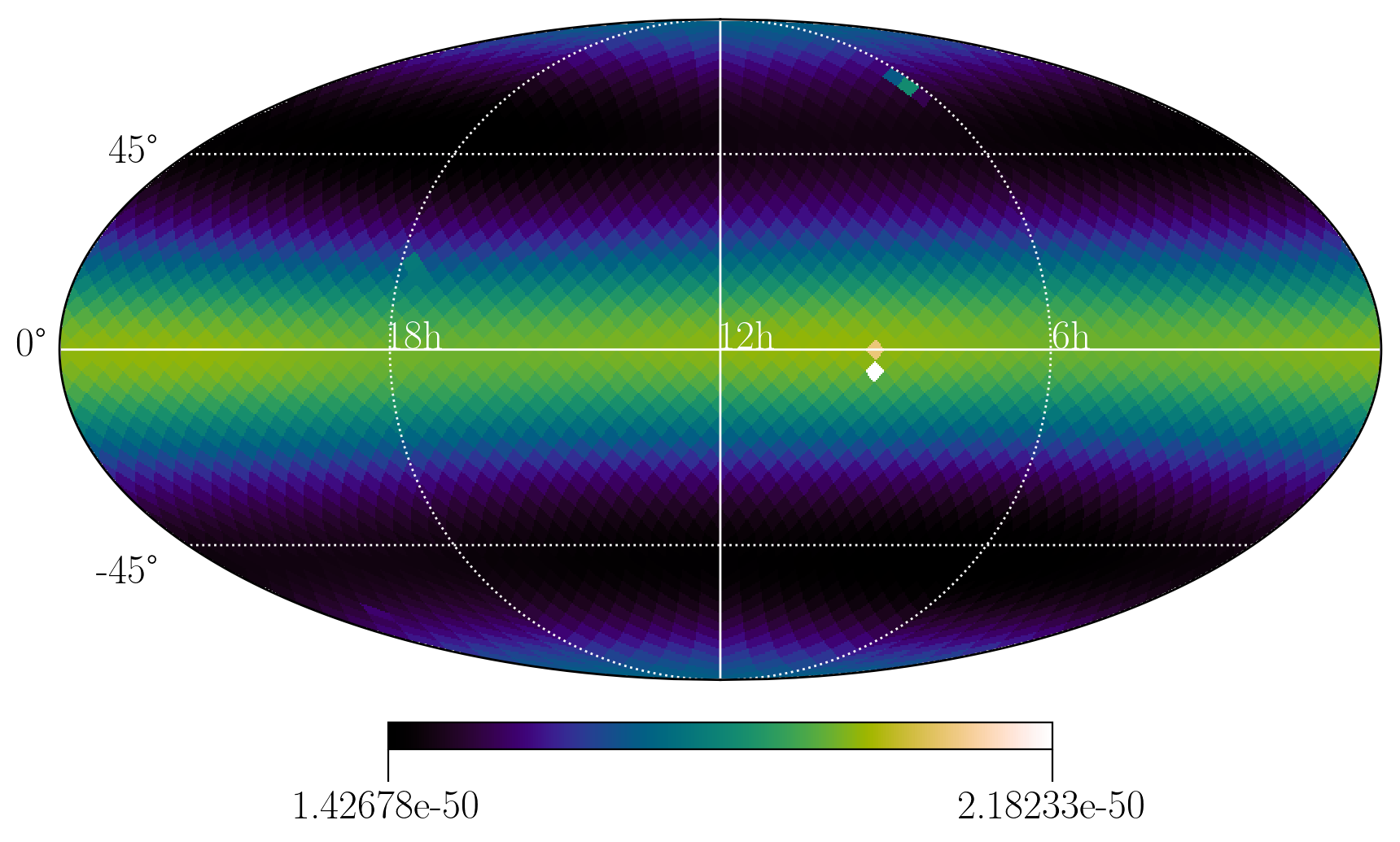}\\
    \includegraphics[width=0.225\textwidth]{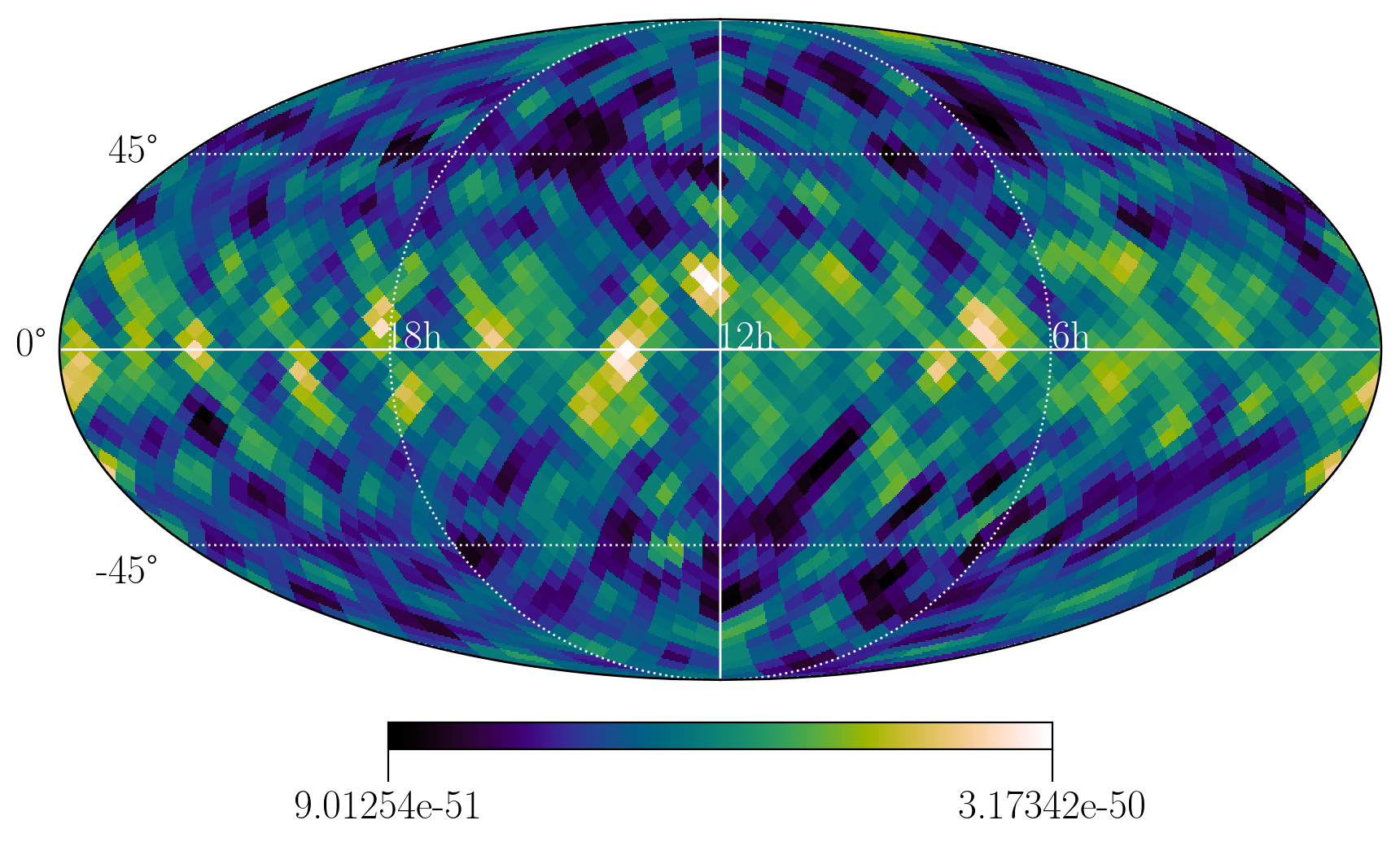}
    \includegraphics[width=0.225\textwidth]{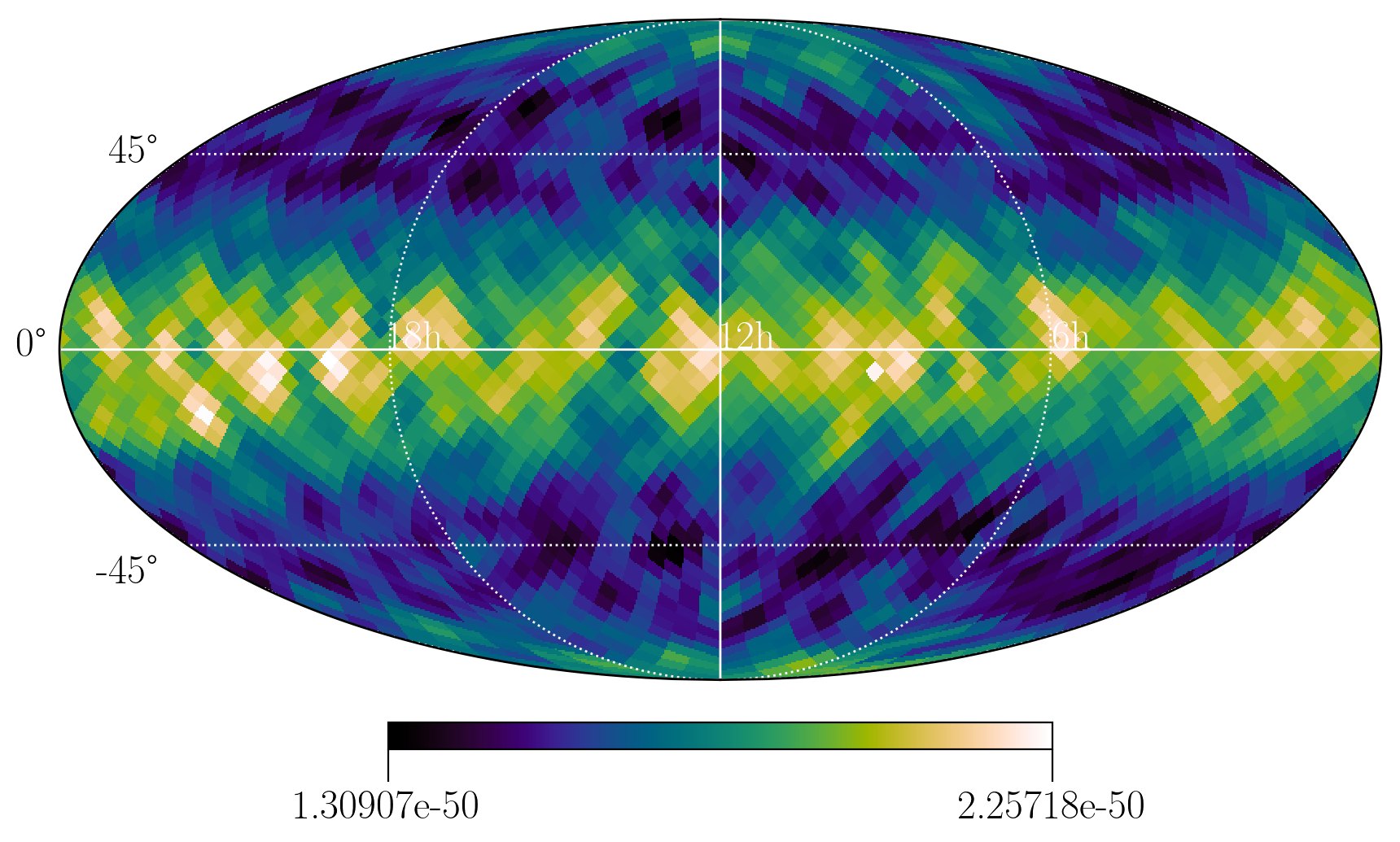}\\
    \includegraphics[width=0.4\textwidth]{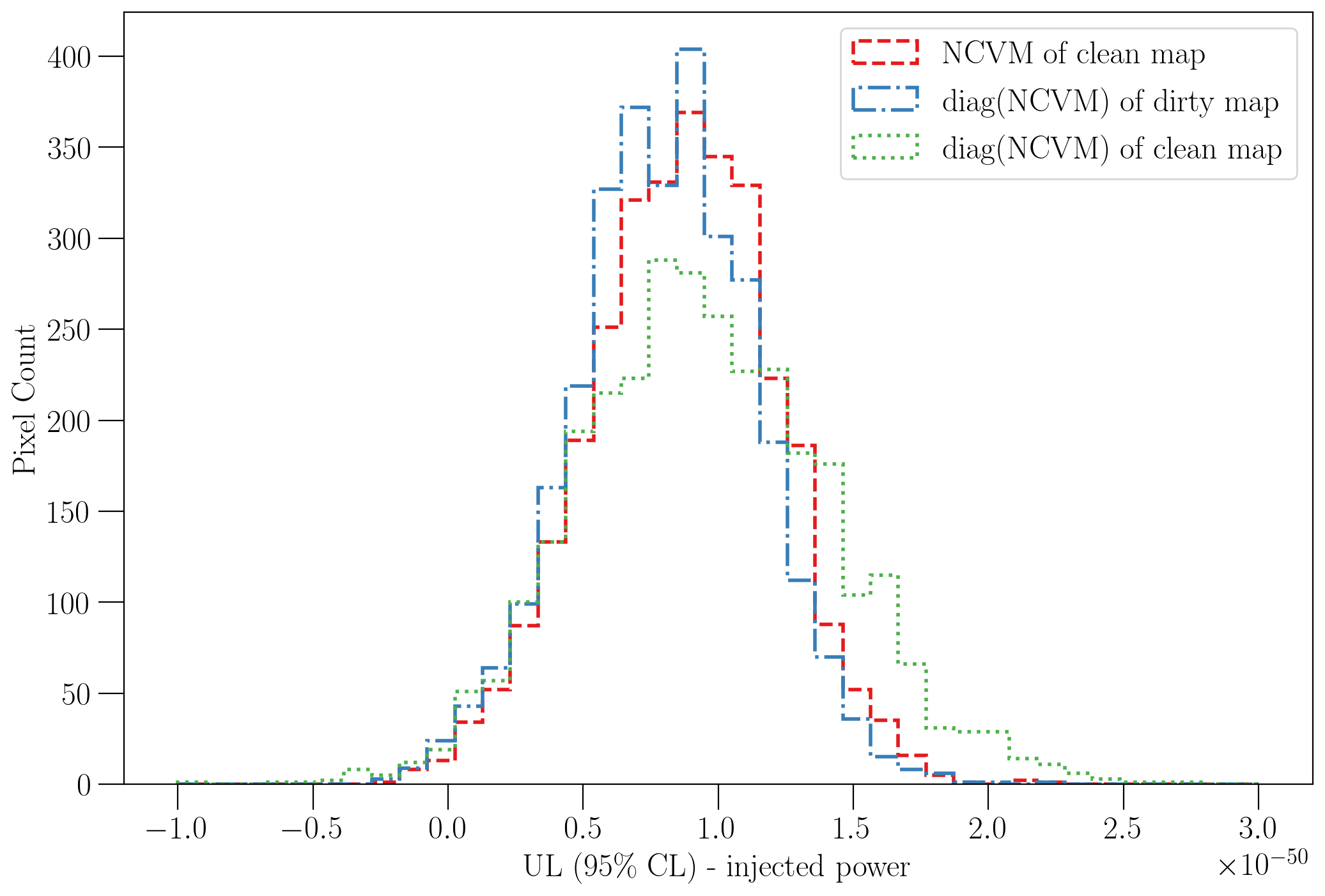}
\caption{Results of an upper limit (with $95\%$ confidence) injection study. First row: (left) sky map of injected power for visualization only (creating 3072 realizations by injecting source one by one into each pixel and picking a value from the injected pixel only) and (right) upper limit sky map with the conventional method, which looks smooth due to our choice of the injected value to create a reference upper limit map. Second row: (left) upper limit using likelihood with clean map and diagonals of its NCVM  [Eq.~(\ref{eq:norm_gaussian_lkhd})] and (right) upper limit using likelihood with the clean map and its (approximated) NCVM [Eq.~(\ref{eq:UL_lkhd_L4})]. All the maps are represented as a color bar plot on a Mollweide projection of the sky in ecliptic coordinates. Third row: histogram of the difference between the upper limit and the injected power. The injections are performed to achieve a dirty-map SNR $\sim2.5$. To achieve this, some pixels required unphysical negative power, which were set to zero instead. These pixels appear as artifacts in the upper-limit map with the conventional method (top-right panel). Though the upper-limit maps look very different, the histograms show that they are all consistent, that is, the number of pixels where the 95\% upper limit is less than the injected value (the area of the histograms for negative values) is less than 2\%.}
\label{fig:UL_injection}
\end{figure}

In the next step, we test the behavior of the likelihoods in determining upper limits with a norm-regularized clean map and its covariance matrix [Eqs.~(\ref{eq:norm_gaussian_lkhd}) and~\ref{eq:UL_lkhd_L4})], and the behavior of conventional likelihood. The results are shown in Fig.~\ref{fig:UL_injection}. We create 3072 dirty-map realizations injecting a point source with a $\alpha=3$ power law into different pixels of a noise-only dirty map one by one, resulting in dirty map SNR $\sim 2.5$ for each injection (top-left figure in Fig.~\ref{fig:UL_injection}). To obtain the dirty map SNR $\sim 2.5$, few pixels would require negative power injection, which would be unphysical, hence we set them to zero. These pixels appear as artifacts in the upper-limit sky map with the conventional method (top-right panel in Fig.~\ref{fig:UL_injection}). The obtained dirty maps are deconvolved using norm regularization with $\kappa_N=15$. The upper limits with $95\%$ confidence are calculated using all three likelihoods. The top-right panel of Fig.~\ref{fig:UL_injection} shows upper limits with the conventional method. The upper limits (middle-right panel of Fig.~\ref{fig:UL_injection}) are derived using the clean map and its approximated NCVM using Eq.~(\ref{eq:UL_lkhd_L4}), and are expected to be larger due to the broadening of the likelihood caused by overestimated clean-map variances. These upper limits are at a similar level as with the conventional method (see bottom histogram plot of Fig.~\ref{fig:UL_injection}) used in Refs.~\cite{S5_Dir,O1paper,O2paper}, though the upper limit map is over-smoothed in the latter case, due to the effect of the broad point spread functions of the detector pair. However, with the clean map, it is observed that using a full covariance matrix in the likelihood [see Eq.~(\ref{eq:UL_lkhd_L4})] does set stringent upper limits in comparison to a diagonal-only $\mathbf{\Gamma}_N$. The upper limit sky maps, one obtained using the clean map and the full NCVM (bottom-right panel of Fig.~\ref{fig:UL_injection}), show different patterns in the map, though, most importantly, the histograms plotted in Fig.~\ref{fig:UL_injection} are quite similar. In $\sim 9\%$ pixels, the former predicts stringent upper limits while the latter predicts loose upper limits, and in the other $\sim 9\%$ of pixels, the opposite behavior is noticed; $\sim 82\%$ of the histograms overlap. Though there are few pixels ($\sim 2\%$) for which conservative $95\%$ upper limits are lower than the injected value (Fig.~\ref{fig:UL_injection}), this is consistent with the criteria for $95\%$ upper limits. {\em Hence, we recommend continuing to use the conventional method which is simple and provides consistent upper limits taking only the diagonal elements of the pixel-to-pixel NCVM into account}. The $p$-value obtained by the conventional method through simulations can be used to claim a detection.

\begin{table*}
\begin{tabular}{|c|cc|cc|cc|ccc|}
\hline
\multicolumn{10}{|c|}{BBR O1+O2 Results}\\
\hline
\multicolumn{1}{|c|}{$\alpha$}&\multicolumn{1}{c}{H(f)}&\multicolumn{1}{c|}{$\Omega(f)$}&\multicolumn{2}{c|}{Dirty Map max SNR}&\multicolumn{2}{c|}{$p$-values($\%$)}&\multicolumn{3}{c|}{Upper Limit Ranges ($\times 10^{-8}$)}\\
\multicolumn{1}{|c|}{}&\multicolumn{2}{c|}{}&\multicolumn{1}{c}{{\tt PyStoch}}&\multicolumn{1}{c|}{Conventional}&\multicolumn{1}{c}{Simulation}&\multicolumn{1}{c|}{$\mathcal{L}_4$}&\multicolumn{1}{c}{Conventional}&\multicolumn{1}{c}{diag(NCVM) of $\bm{\hat{\mathcal{P}}}$}&\multicolumn{1}{c|}{NCVM of $\bm{\hat{\mathcal{P}}}$}\\
\hline
   0      &  $\propto f^{-3}$     &  constant            &  2.53 & 3.09  & 65.40  & 21.90 &  4.48-21.1    & 7.13-40.43 &  4.56-22.50\\
   2/3    &  $\propto f^{-7/3}$   &  $\propto f^{2/3}$   &  2.58 & 3.09 & 76.40   & 17.00 & 2.33-11.8    & 3.36-22.20 & 2.47-11.90\\
   3      &  constant             &  $\propto f^{3}$     &  3.12 & 3.27  & 91.50    & 5.00 & 0.05-0.32     & 0.06-0.42 & 0.05-0.33\\
  \hline
\end{tabular}
\caption{We present maximum dirty-map SNR across all sky positions obtained from O1 and O2 folded data using {\tt PyStoch} and compare it with LVC O1-O2 results~\cite{O2paper}. The values of $\rho_{\mbox{max}}$ differ from the LVC O1-O2 results because of the cumulative effect of two main differences, namely, we use an updated list of notched frequencies~\cite{O3paper} and a \hpx grid instead of a 1-degree Cartesian grid used in LVC analysis. The significance ($p$-value) of maximum SNR pixel is calculated using two different methods: the noise simulations and log-likelihood $\mathcal{L}_4$. The obtained $p$-values are consistent with Gaussian noise. Note that we cannot compare $p$-values obtained from different methods due to different biases involved in the methods. We present the ranges of upper limits with $95\%$ confidence on GW flux $\mathcal{F}(\mathbf{\hat{\Omega}})$ using likelihoods with a clean map ($\bm{\hat{\mathcal{P}}}$) as described in Eqs.~(\ref{eq:norm_gaussian_lkhd}) and (\ref{eq:UL_lkhd_L4}). Thus validating the folded data set and the analysis strategies we presented in this paper for setting significance and upperlimits.}
\label{table:BBR_results}
\end{table*}

\section{Results}
\label{sec:results}
\begin{figure*}
 \begin{tabular}{ccc}
 \includegraphics[width=0.3\textwidth]{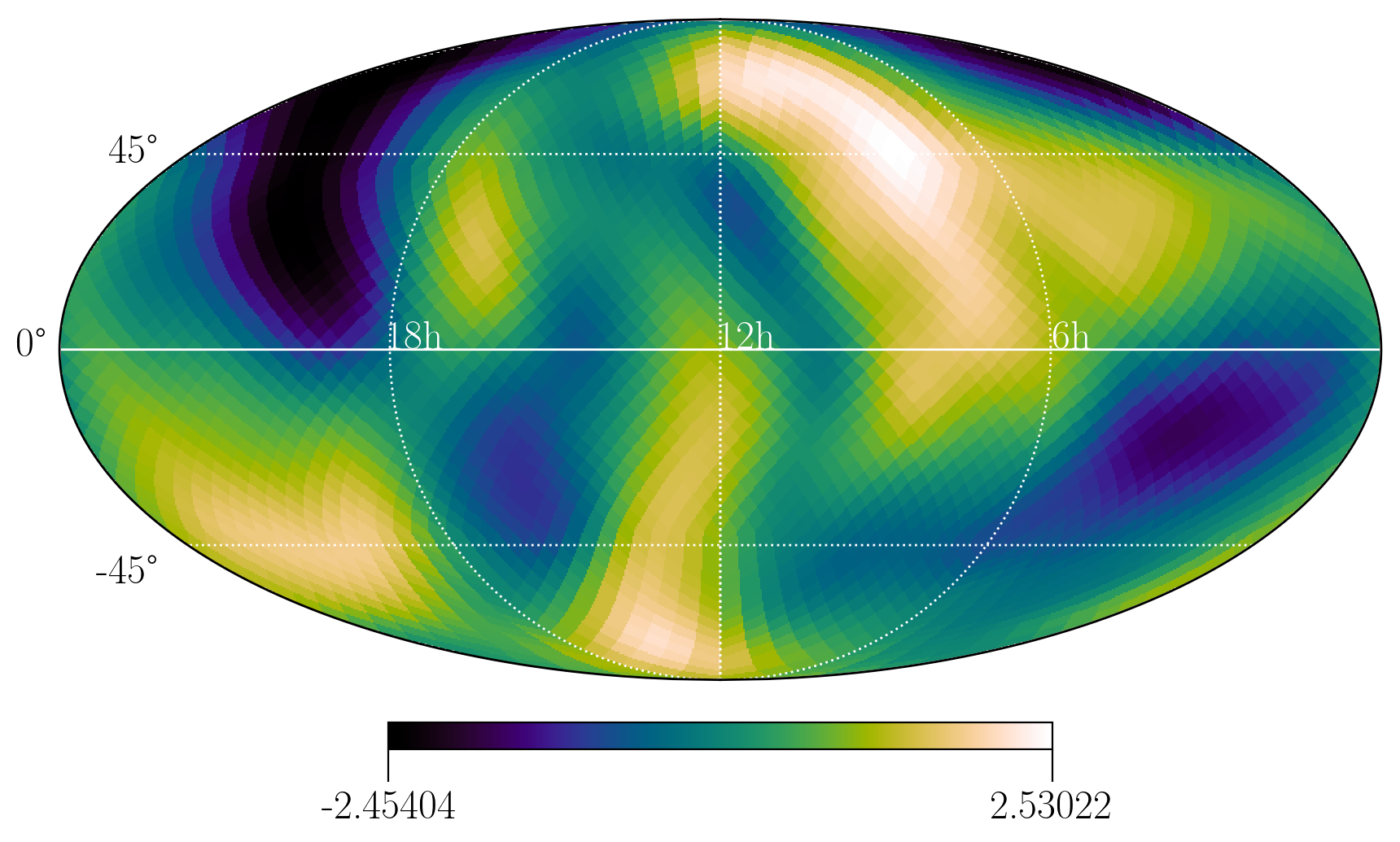} 
\includegraphics[width=0.3\textwidth]{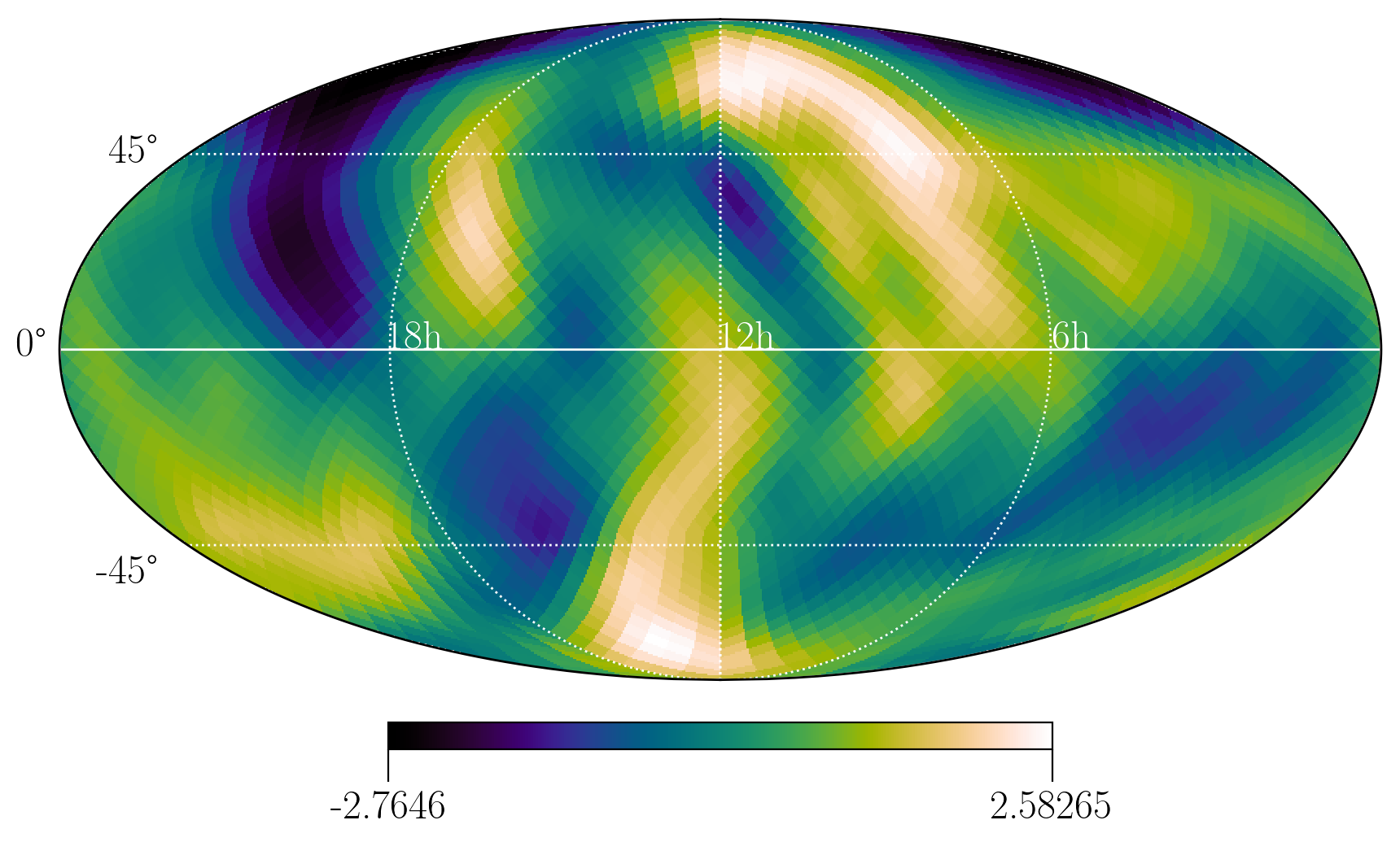} 
 \includegraphics[width=0.3\textwidth]{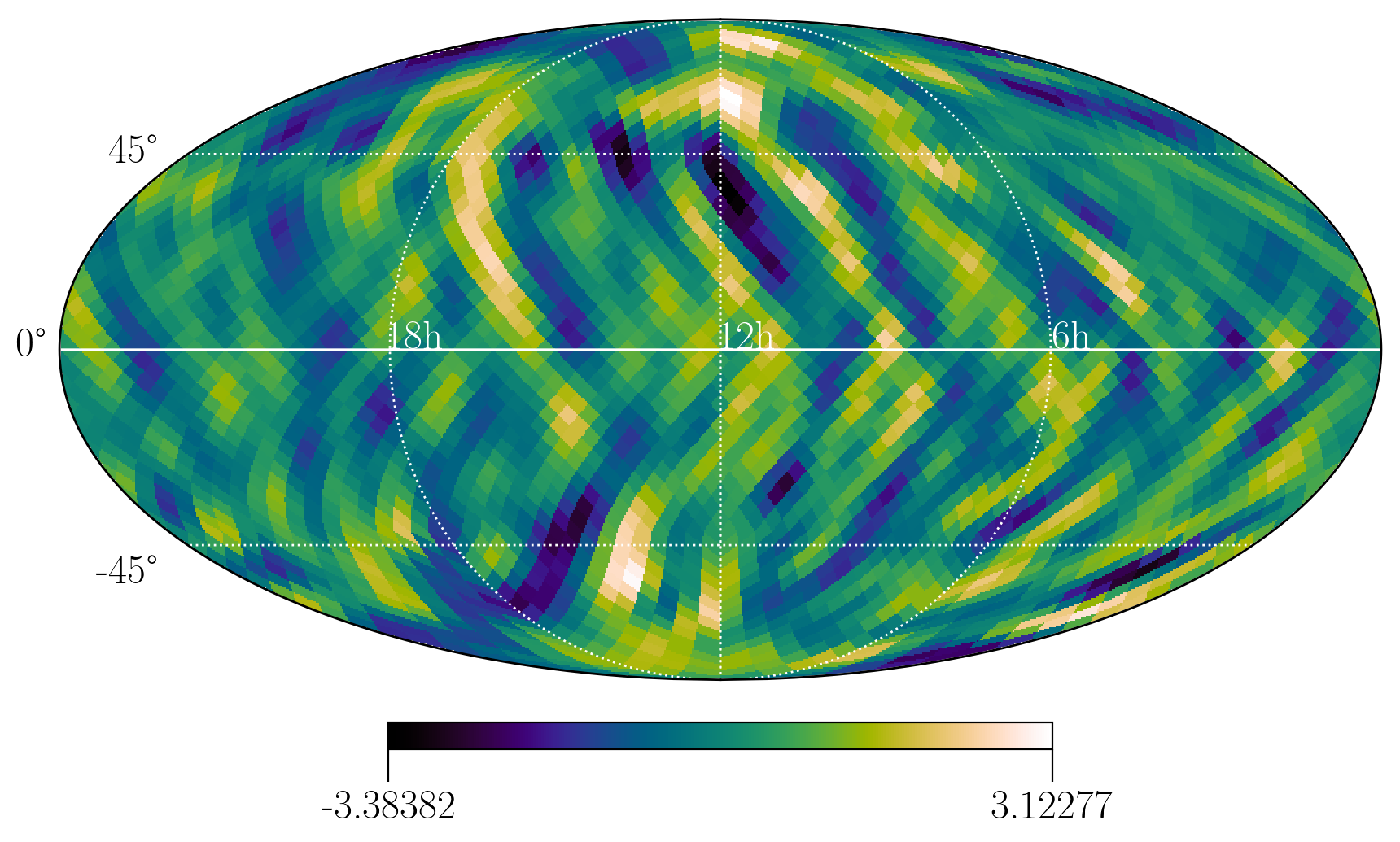}\\
\includegraphics[width=0.3\textwidth]{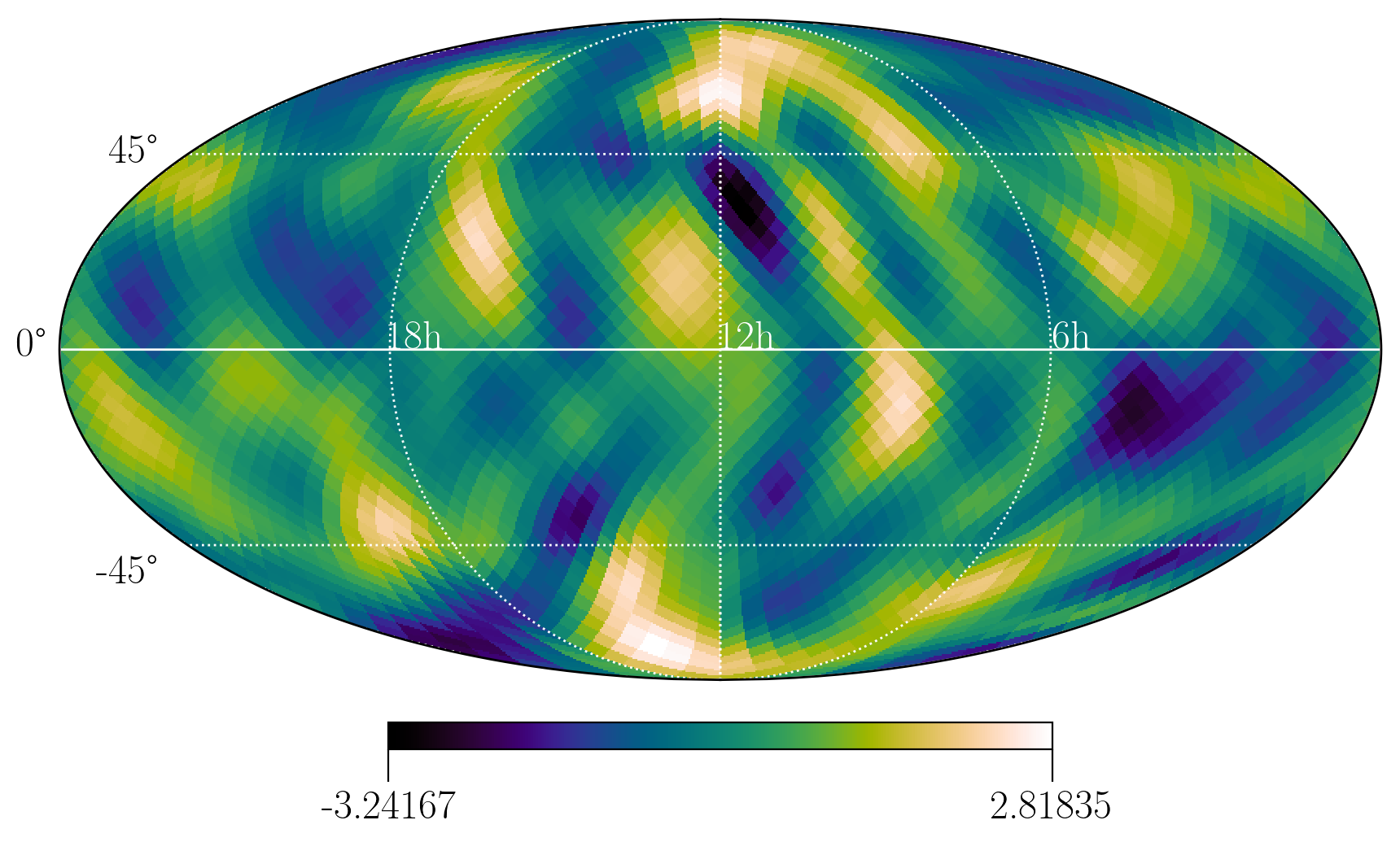} 
\includegraphics[width=0.3\textwidth]{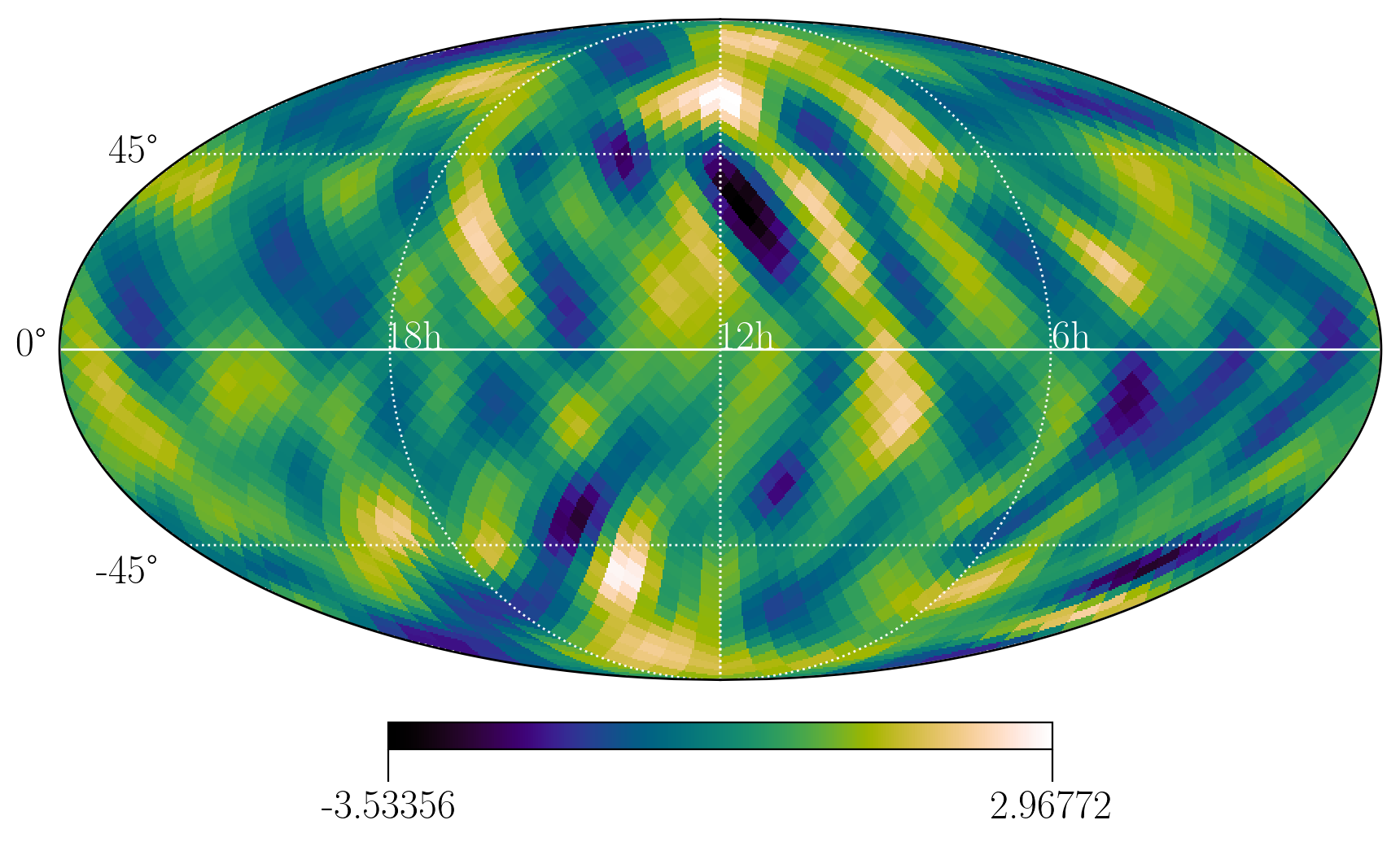} 
\includegraphics[width=0.3\textwidth]{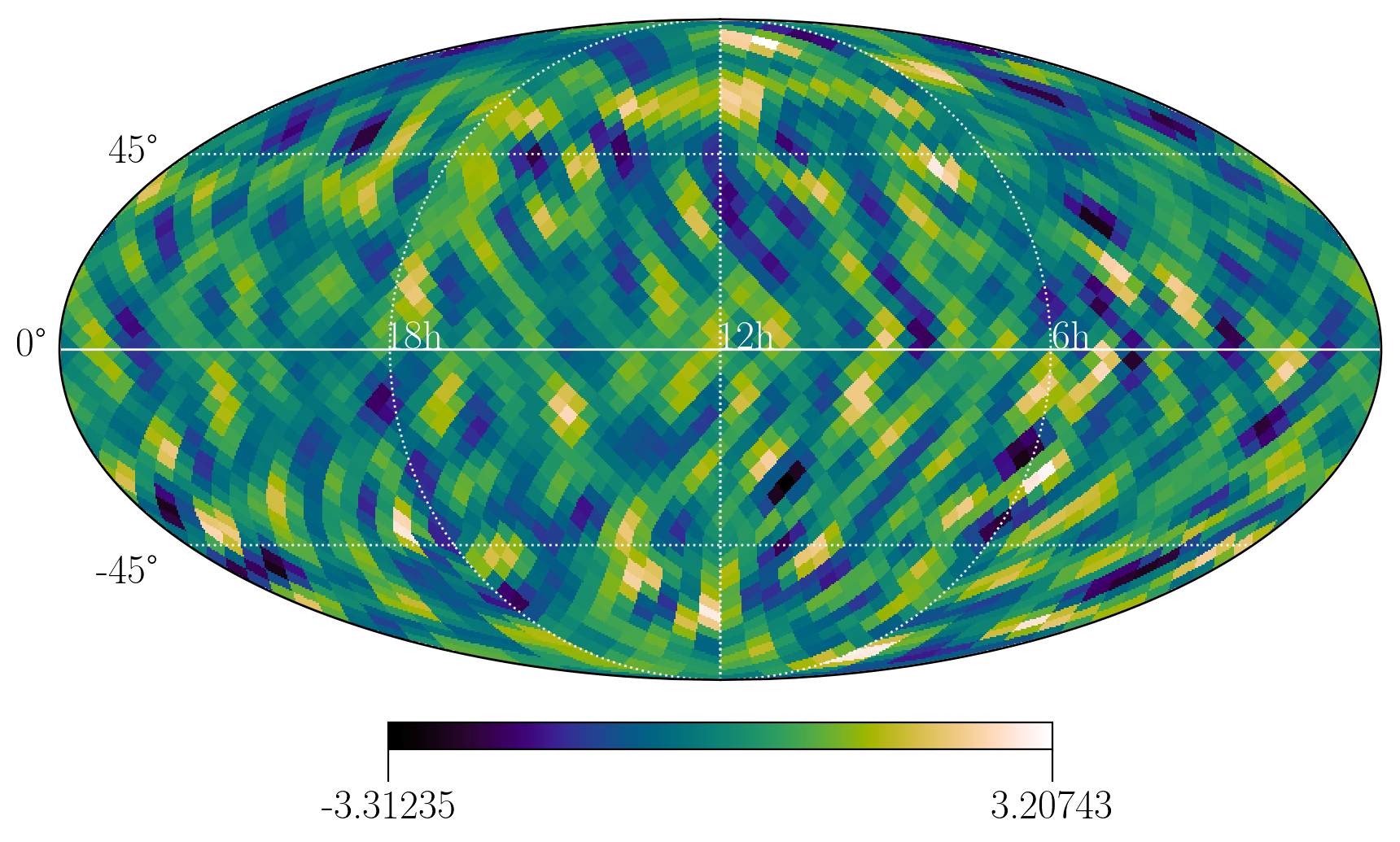}\\
\includegraphics[width=0.3\textwidth]{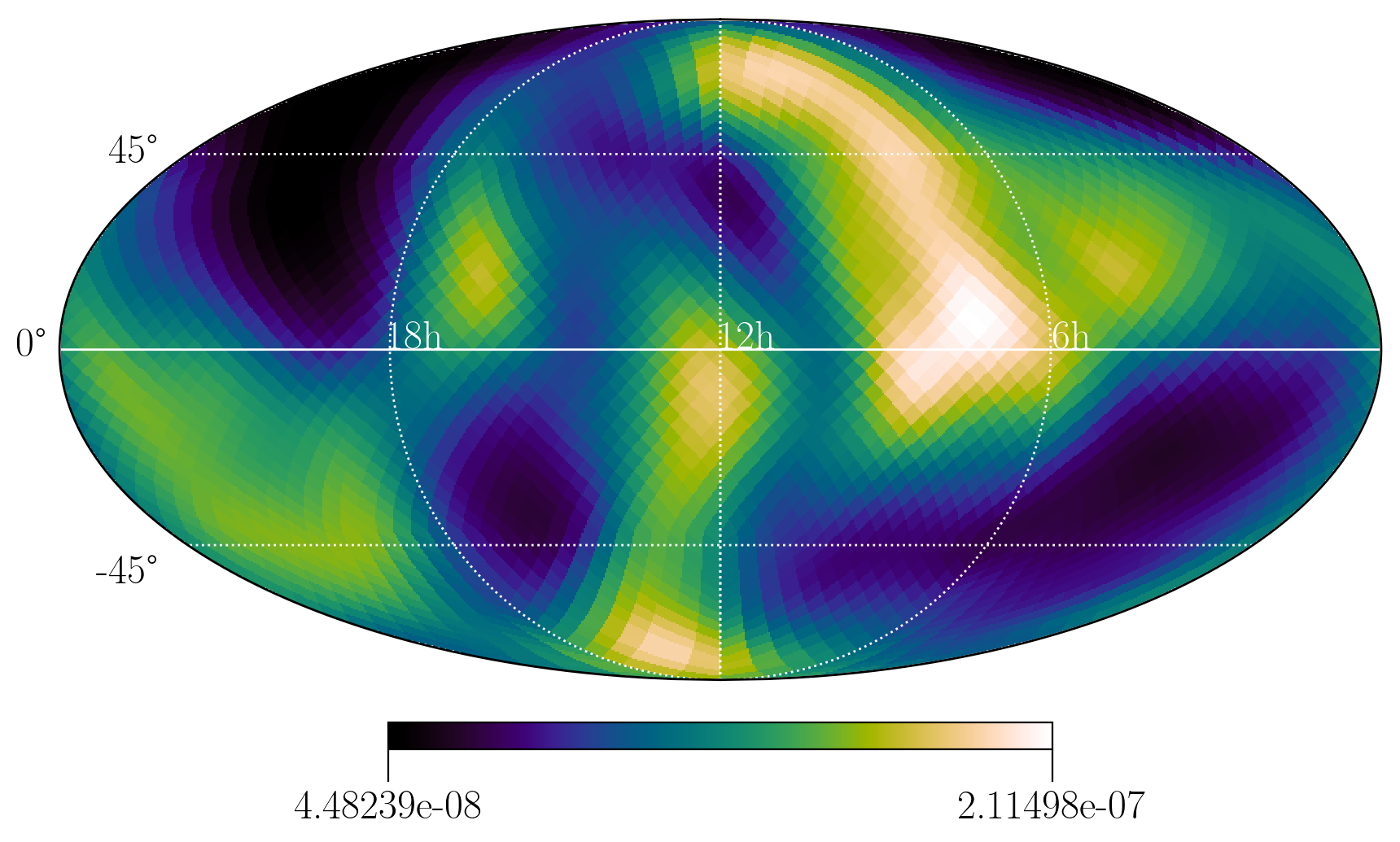}
\includegraphics[width=0.3\textwidth]{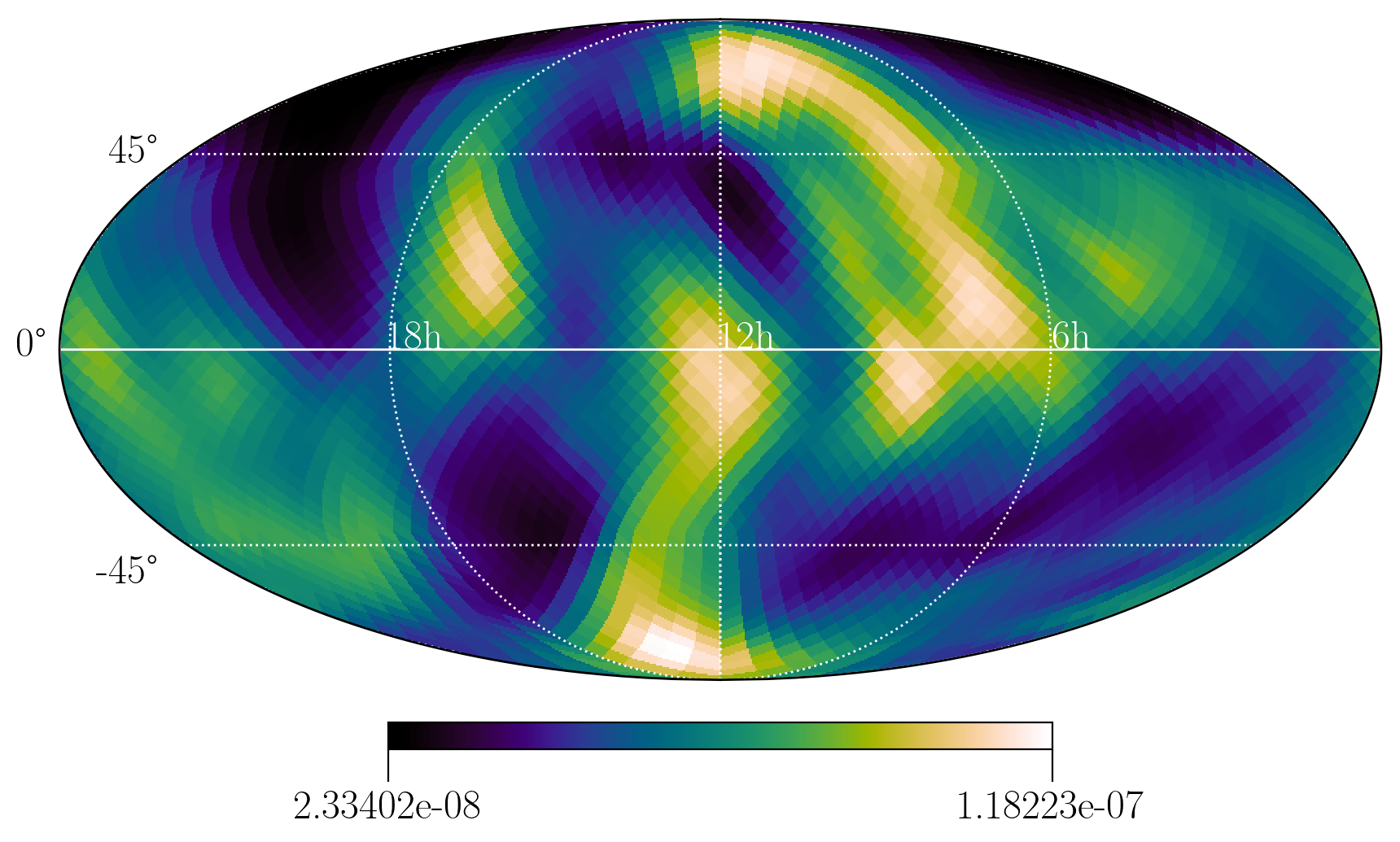}
\includegraphics[width=0.3\textwidth]{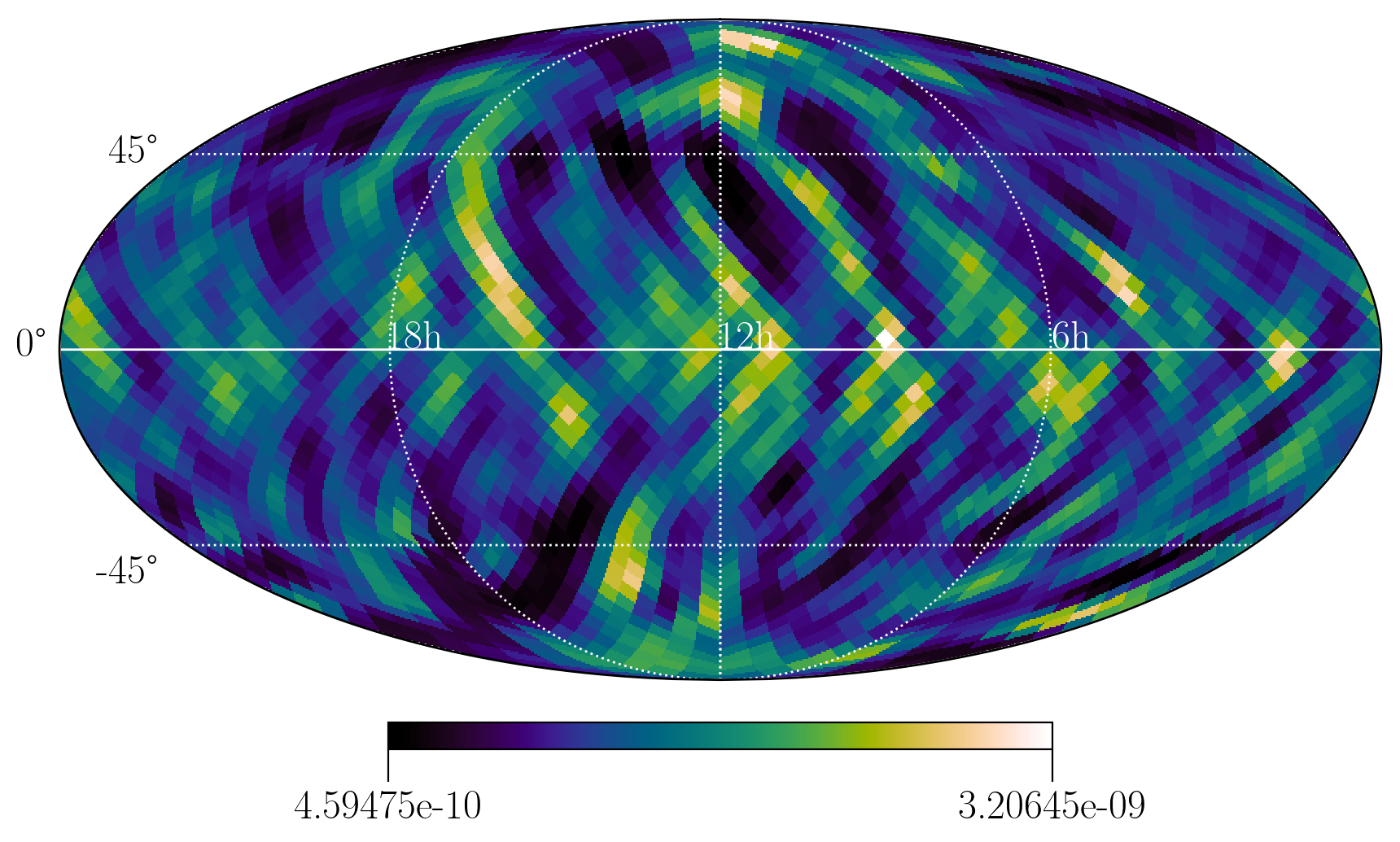}\\
\includegraphics[width=0.3\textwidth]{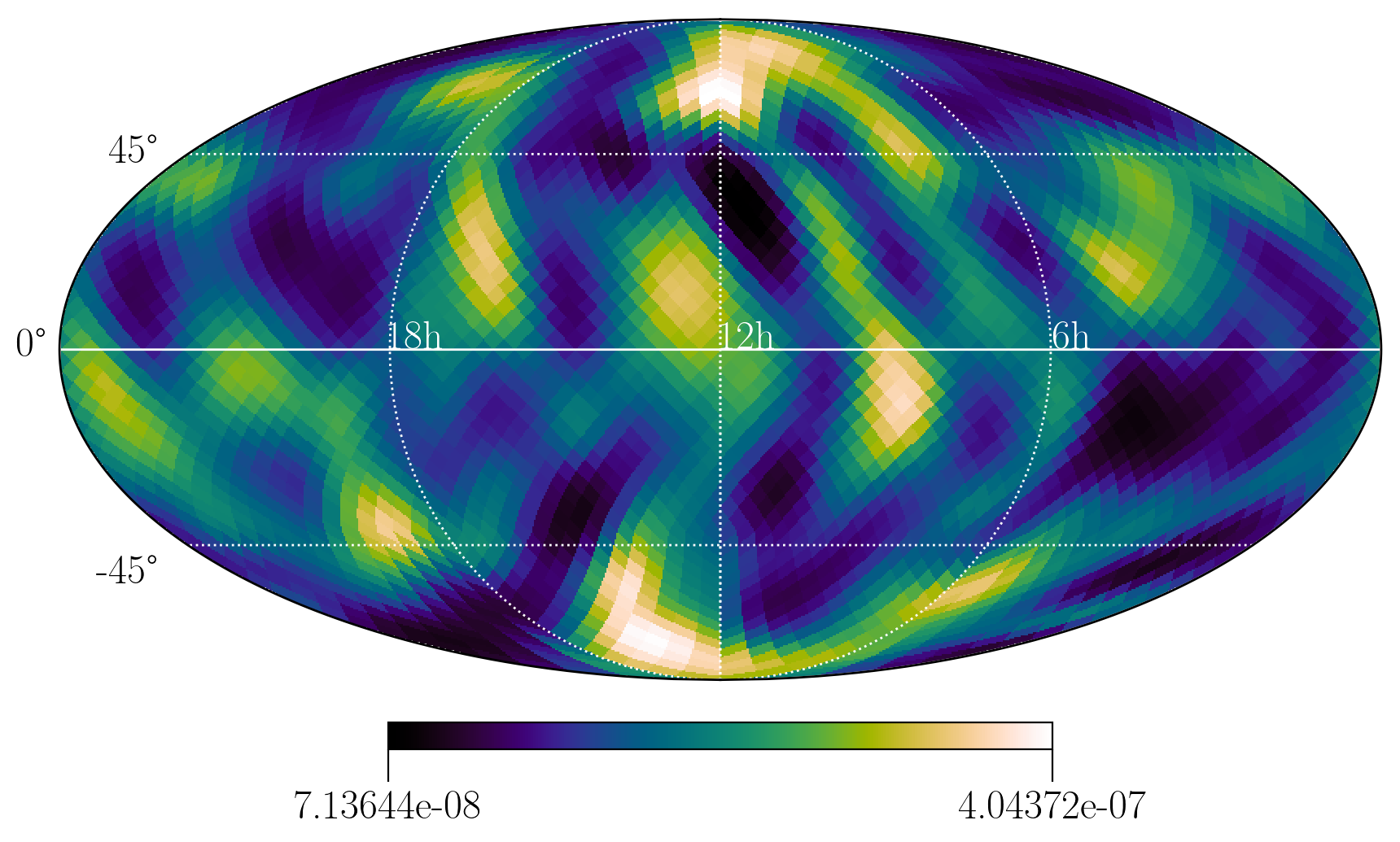}
\includegraphics[width=0.3\textwidth]{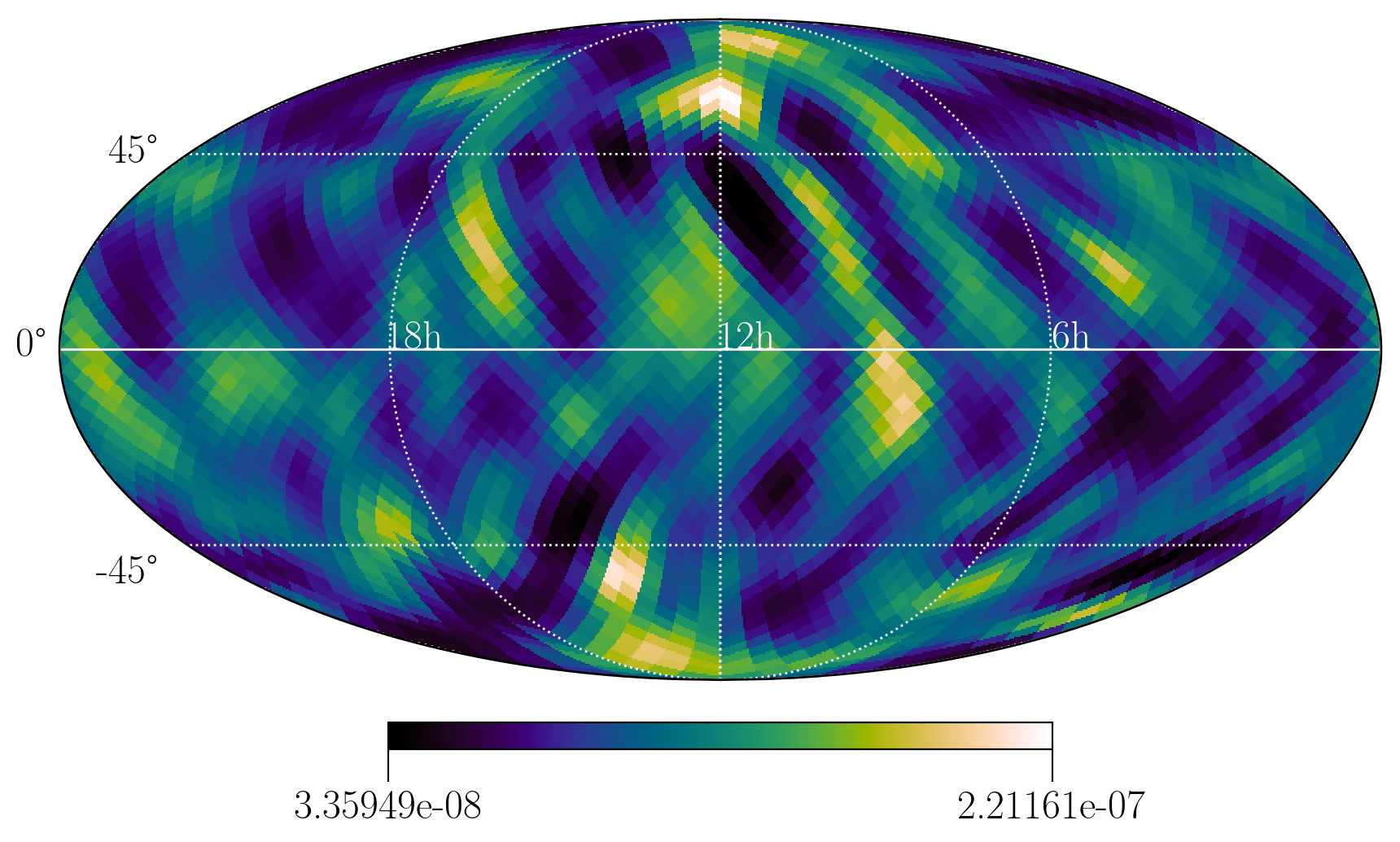}
\includegraphics[width=0.3\textwidth]{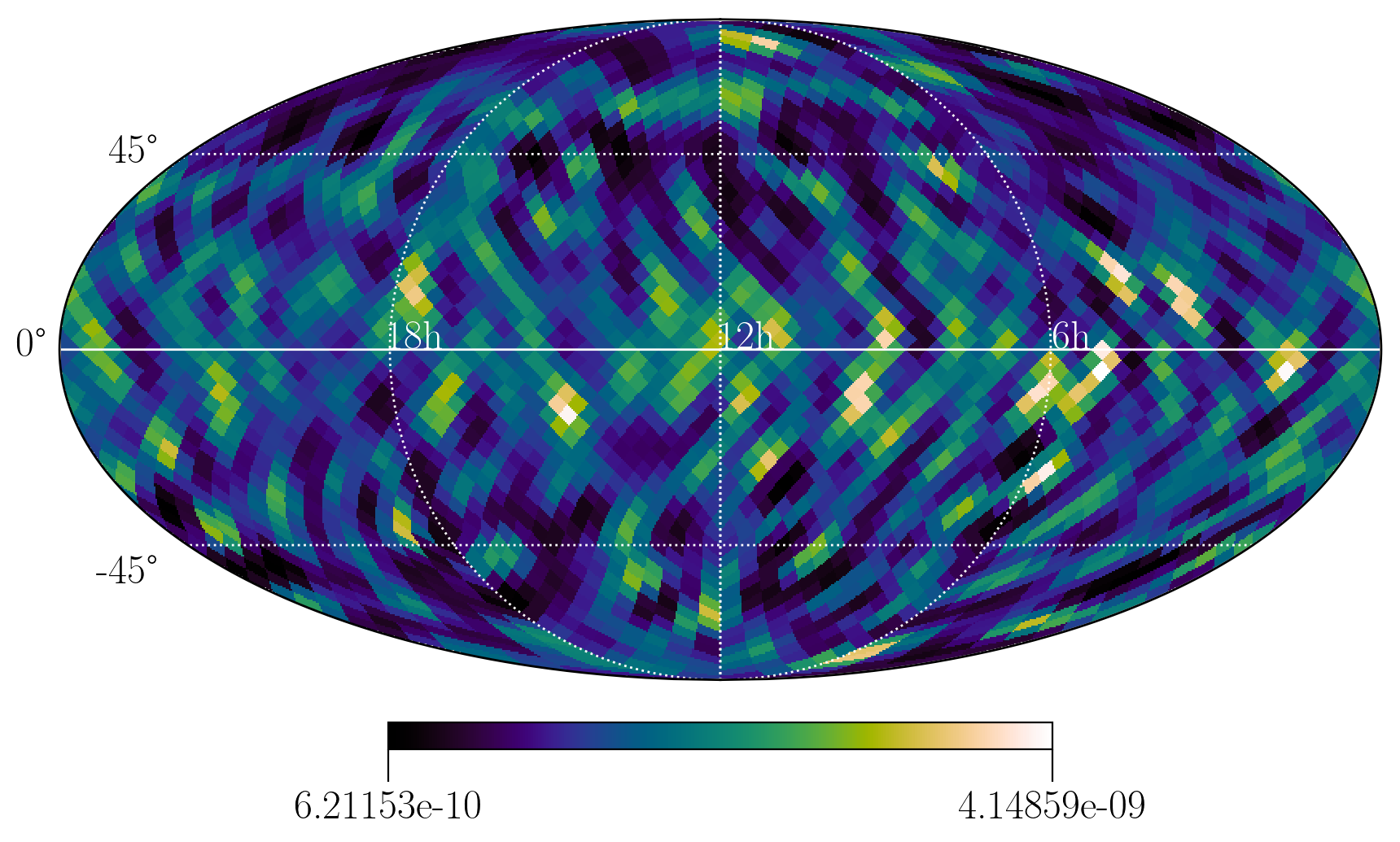}\\
 \includegraphics[width=0.3\textwidth]{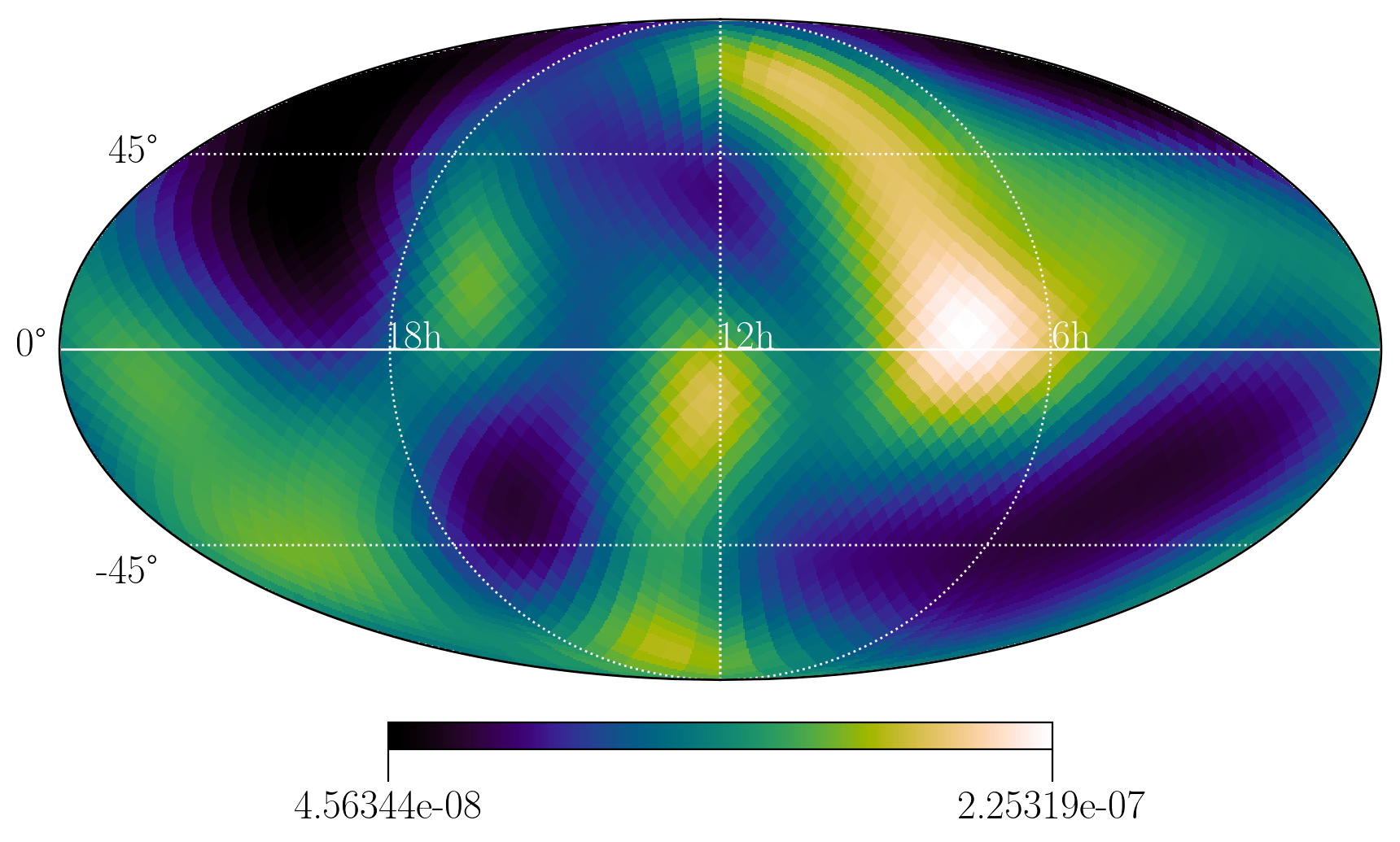}
\includegraphics[width=0.3\textwidth]{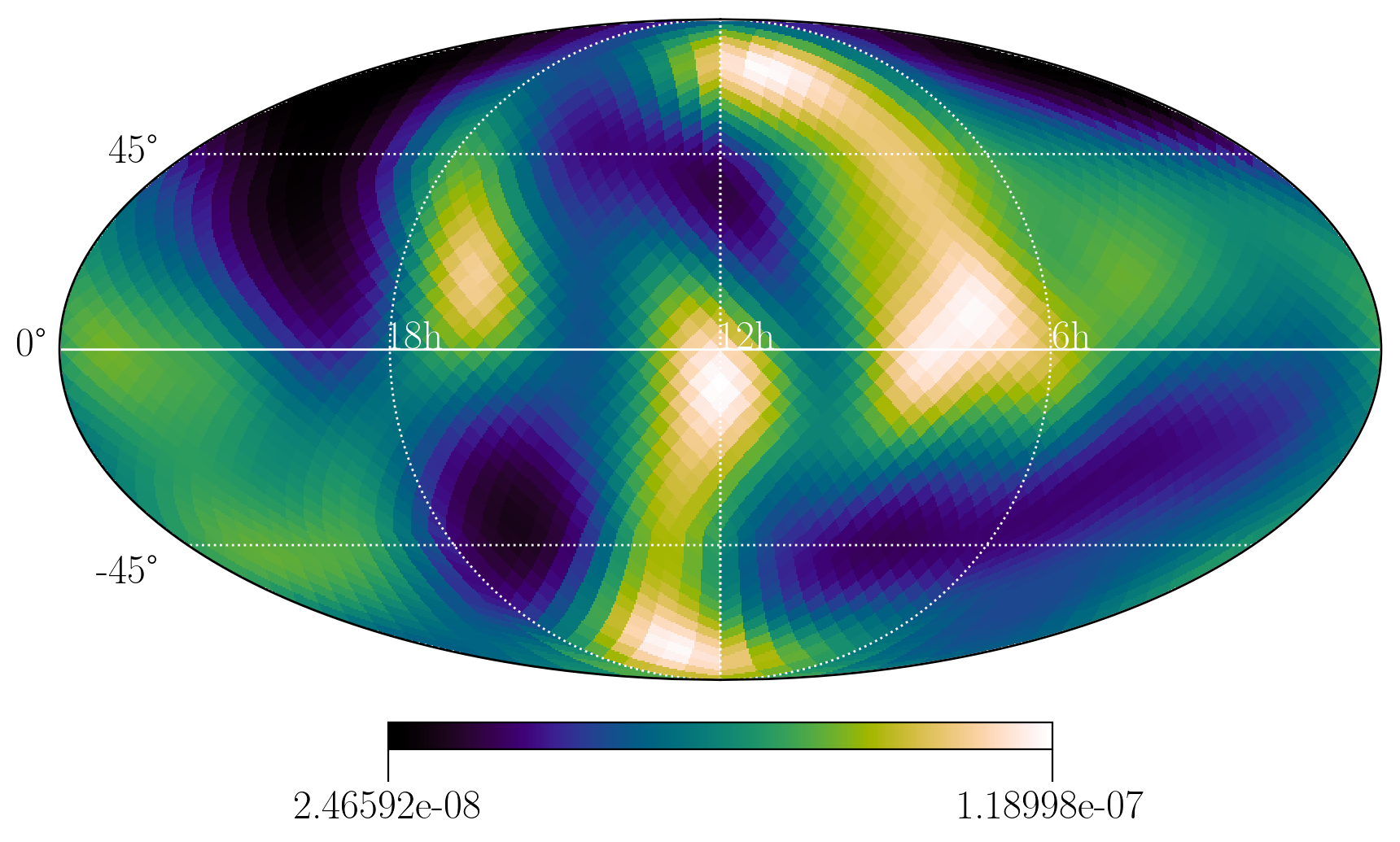}
\includegraphics[width=0.3\textwidth]{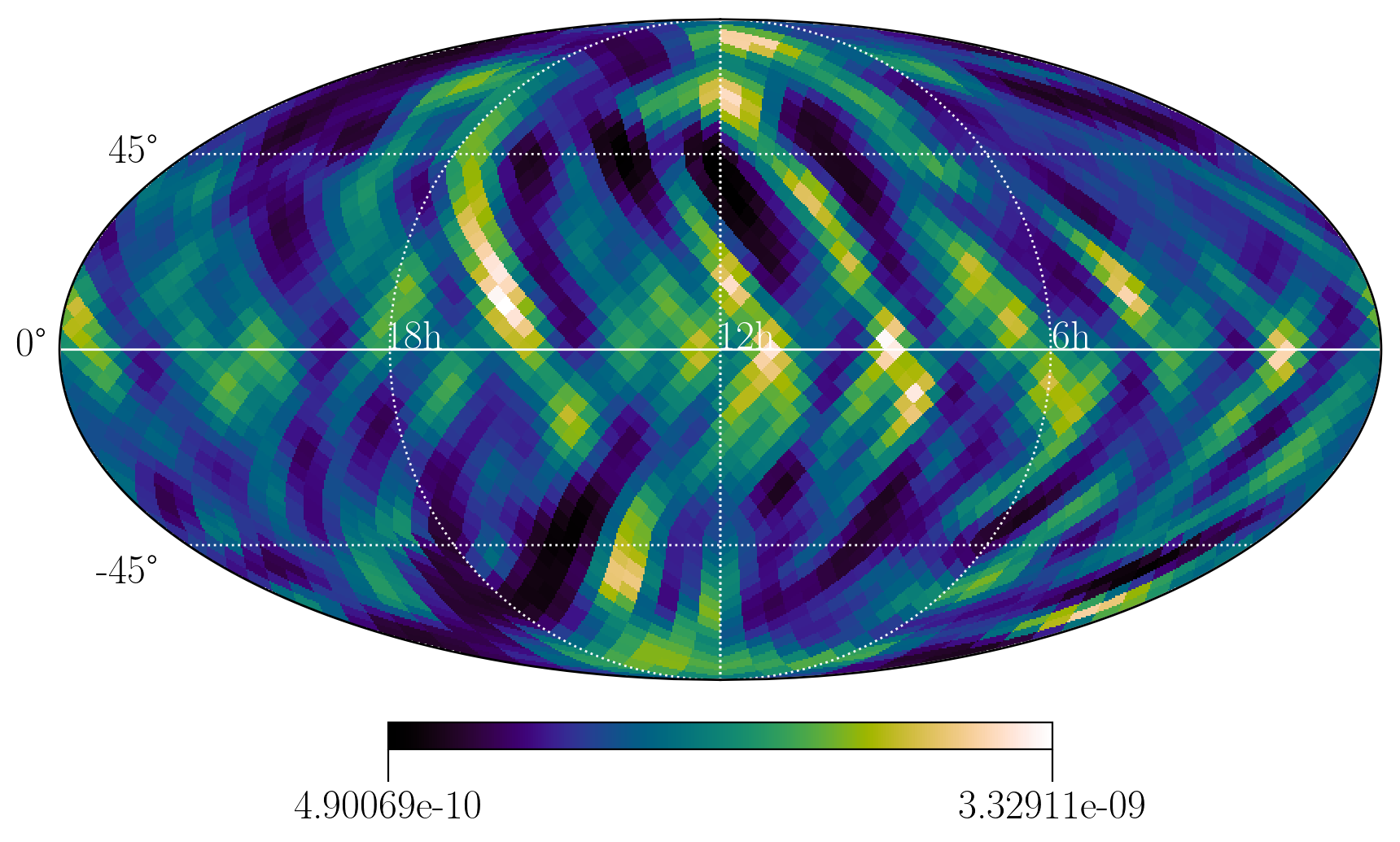}\\
\includegraphics[width = 0.3\textwidth]{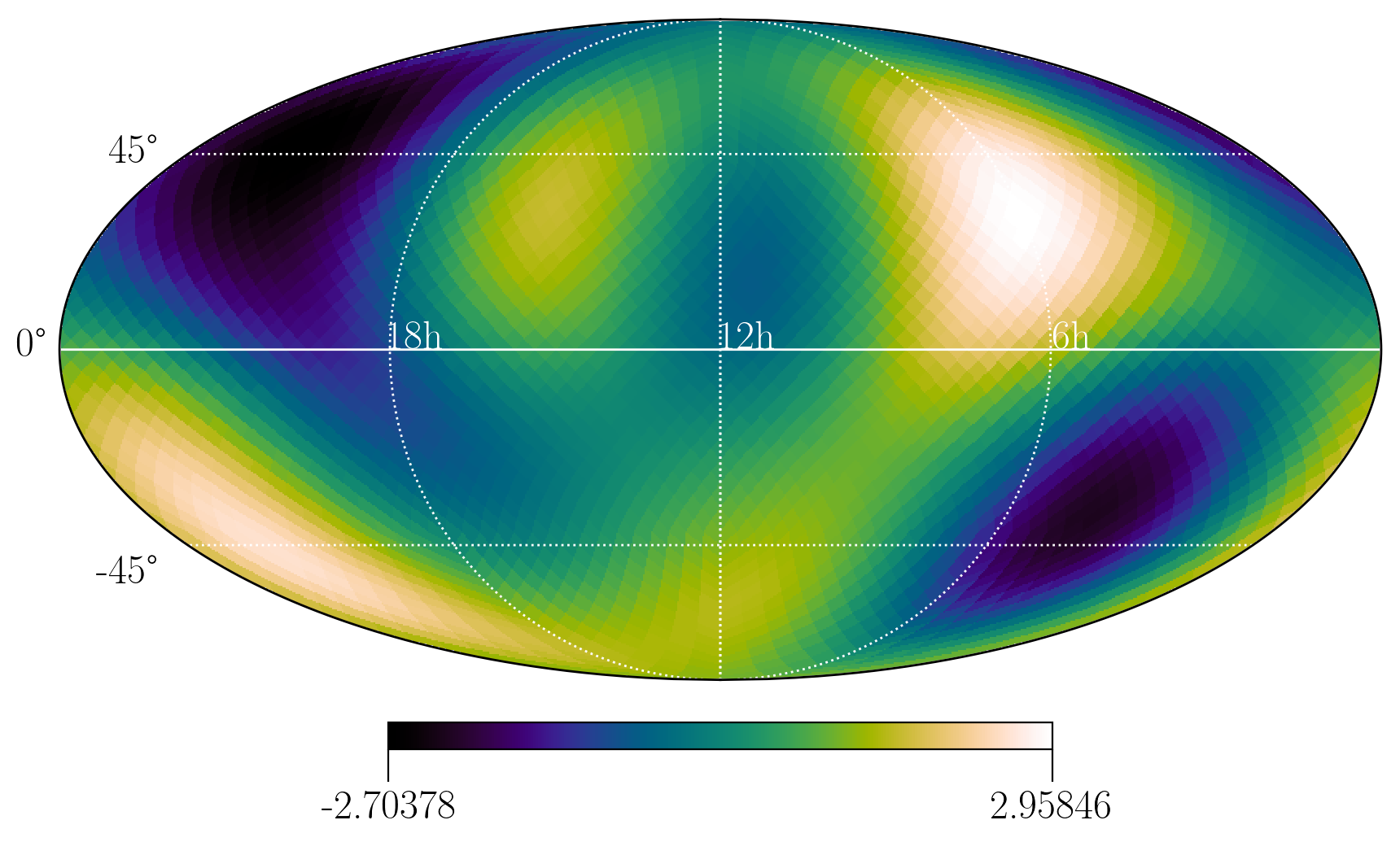} 
\includegraphics[width = 0.3\textwidth]{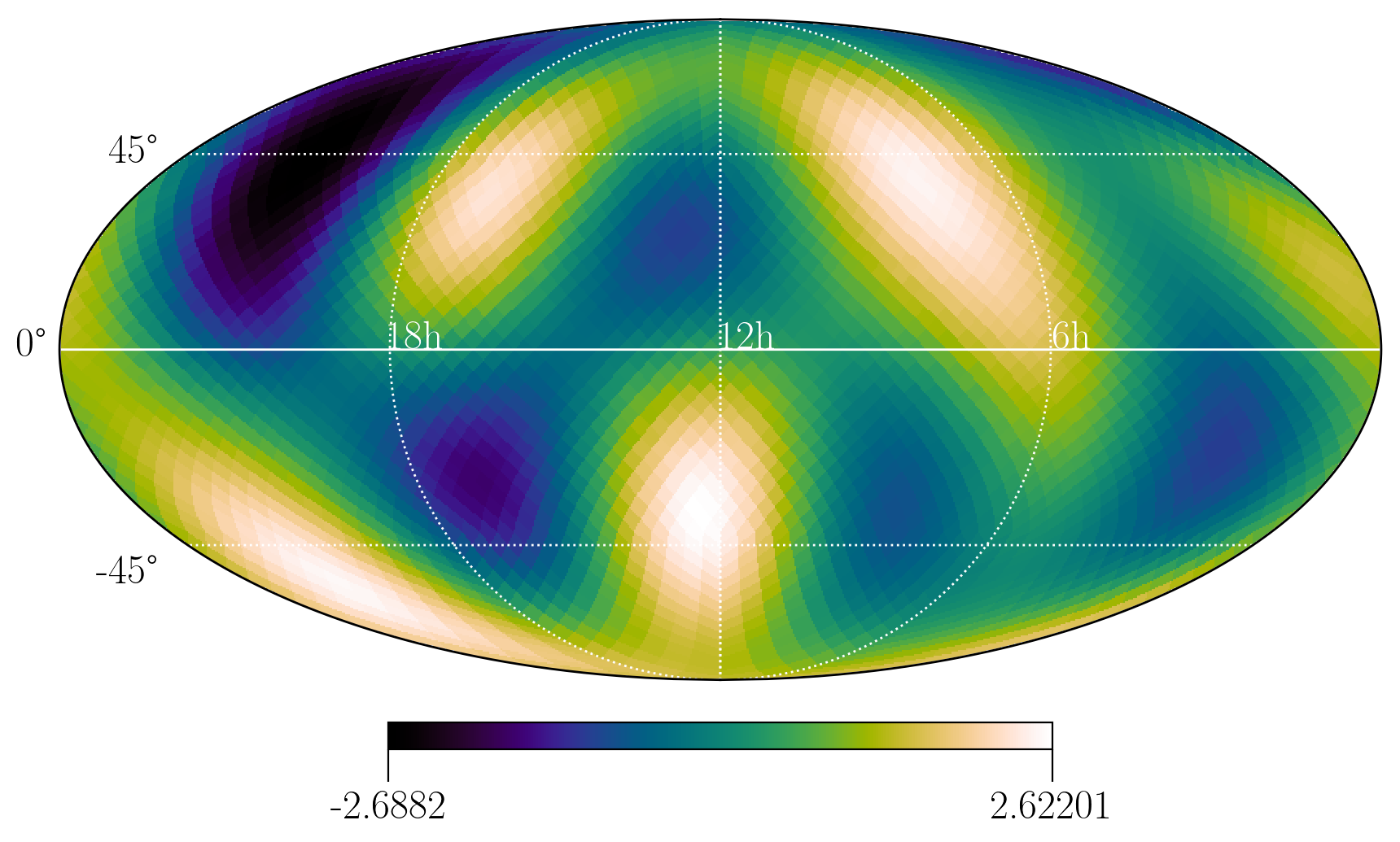} 
\includegraphics[width = 0.3\textwidth]{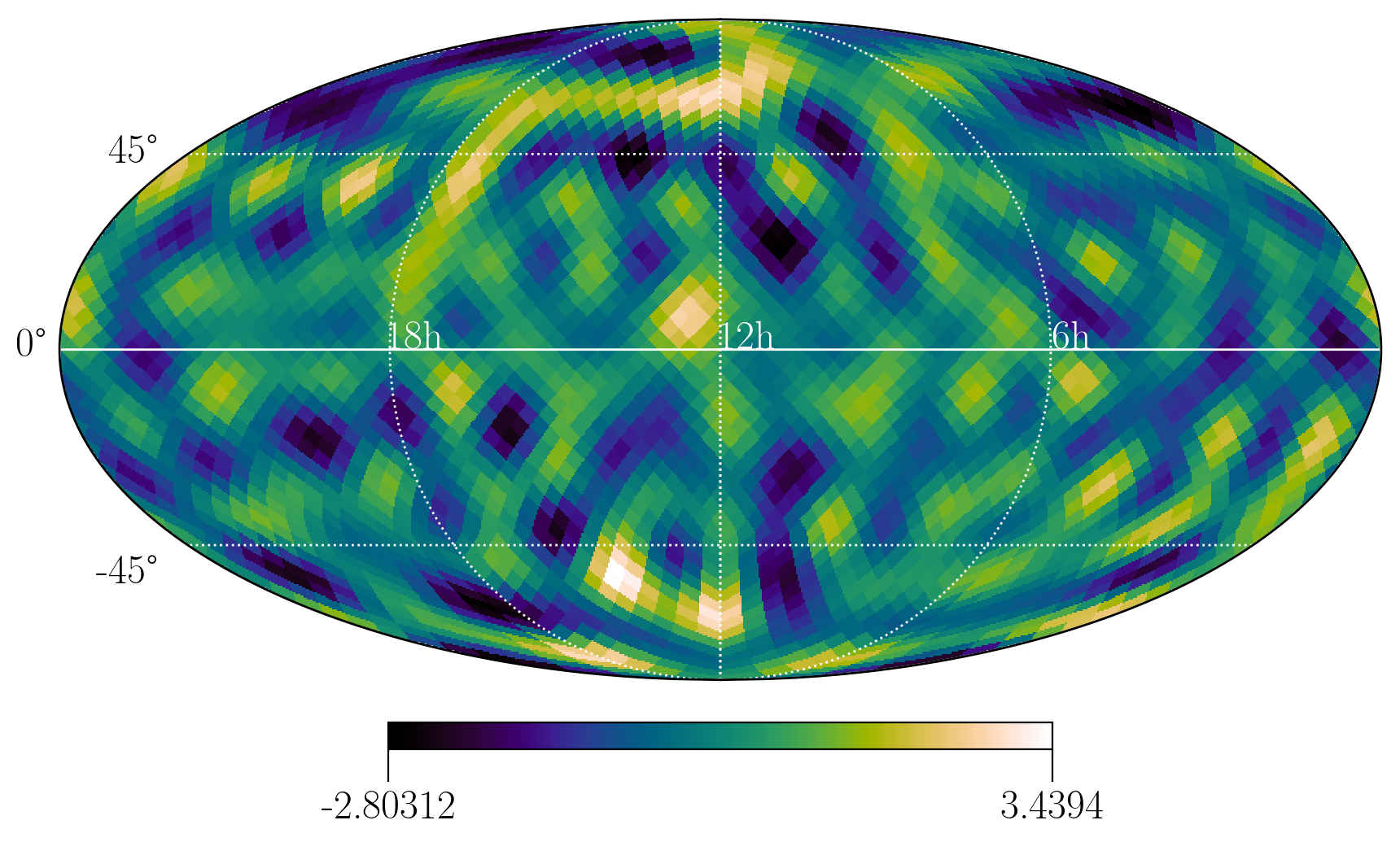} 
 \end{tabular}
 \caption{Broadband radiometer results with O1-O2 folded data for the power spectrum $H(f)$ with spectral index $\alpha=[0,2/3,3]$, from left to right. The first to fifth rows show results of the search performed in the pixel basis with the {\tt PyStoch} pipeline, while the sixth row shows results with folded data analyzed in the spherical basis but with the conventional pipeline. First row: dirty SNR sky maps using the definition of SNR in Eq.~(\ref{eq:pixel_SNR}) only considering the diagonal elements of the covariance matrix. Second row: the clean SNR sky maps obtained with norm regularization with $\kappa_N=[45.37,32.6,15]$ for $\alpha=[0,2/3,3]$, respectively. The third to fifth rows show the upper limit sky maps with $95\%$ confidence on GW flux $\mathcal{F}(\mathbf{\hat{\Omega}})$ (erg cm$^{-2}$ s$^{-1}$ Hz$^{-1}$ sr$^{-1}$) using conventional method, i.e., using diag(NCVM) of the dirty map (third row), the likelihood function formed from the clean map and its diag(NCVM) as in Eq.~(\ref{eq:norm_gaussian_lkhd}) (fourth row), and the likelihood function formed from the clean map and its approximated full covariance matrix as in Eq.~(\ref{eq:UL_lkhd_L4}) (fifth row). While the maps may look visually different, the results from these analyses are all consistent, as was confirmed by our injection study. Sixth row: SNR sky maps using spherical harmonic decomposition techniques. The consistency among the sky maps produced from both the pixel-based and spherical-harmonic-decomposition-based methods, when compared with the LVC O1-O2~\cite{O2paper} (updated) results, further validates the method and data set we used for the analysis. All the maps are represented as a color bar plot on a Mollweide projection of the sky in ecliptic coordinates.}
 \label{fig:bbr_maps}  
\end{figure*}

We apply the above analysis procedure to the folded data from the first two observing runs of Advanced LIGO’s H1 and L1 detectors. Our results are summarized in Fig.~\ref{fig:bbr_maps} and Table~\ref{table:BBR_results}. The top row of Fig.~\ref{fig:bbr_maps} shows the sky maps of dirty SNR values obtained by integrating over frequency range $20-500$ Hz. These sky-maps have $\rho_{\mbox{max}}=2.53, 2.58, 3.12$ for spectral indices $\alpha=0,2/3,3,$ respectively. The values of $\rho_{\mbox{max}}$ differ from the LVC O1-O2 results~\cite{O2paper} because of the cumulative effect of two main differences, namely, we use an updated list of notched frequencies~\cite{O3paper} and the LVC analysis used a Cartesian grid of sky locations with a pixel area of 1 square-degree, while we use the \hpx grid. Since the folding procedure takes advantage of sidereal day symmetry, it was recommended in ~\citet{Folding} to divide data into time segments having duration as a multiples of 52 s; otherwise, the mid-segments align differently, which may cause a subpercent difference. Note that these differences are small (a few percent fractional rms difference in SNR) and do not imply any inaccuracy or loss of precision.

We compute the full covariance matrix for each power-law spectral shape (see Fig.~\ref{fig:fisher_matrix}). We use them to obtain clean maps by norm-regularized deconvolution with $\kappa_N=[45.37,32.6,15]$ for $\alpha=[0,2/3,3]$ respectively (second row of Fig.~\ref{fig:bbr_maps}). Note that there is no unique choice for the condition number. We chose the condition number such that it is near the minimum of the NMSE (or bias) vs condition number plot [Fig.~\ref{fig:inj_4_a3}]. However, since higher condition numbers increase the noise (or variance) as seen in the figure, we used values of the condition number which are on the lower side. The final choice was made by performing the injection study to find the effect of condition number on significance estimation. For example, for $\alpha=3$, while applying norm regularization we tried $\kappa_N=7,15,50$. We see that $\kappa_N=15$ provides a reasonable choice considering all the above aspects.

We estimate the $p$-value using two methods: (i) maximum SNR distribution using noise simulations obtained with full covariance matrix, and (ii) likelihood $\mathcal{L}_4$ introduced in Eq.~(\ref{eq:p_L4_lkhd}). The false-alarm probability obtained from these calculations (shown in Table~\ref{table:BBR_results}) is consistent with the absence of a signal. Since these results are consistent with the expected Gaussian noise, we set an upper limit on GW power flux $\mathcal{F}(\mathbf{\hat{\Omega}})$ with power spectrum $H(f)$ calculated with a 25 Hz reference frequency integrated over the 20-500 Hz frequency band. The third to fifth rows of Fig.~\ref{fig:bbr_maps} show the $95\%$ upper limit maps with three likelihoods for comparison purposes: (i) the conventional method, (ii) with the clean map and its variances, and (iii) the clean map and its (approximated) NCVM. The upper limit values corresponding to each spectral shape are also summarized in Table~\ref{table:BBR_results}. While in certain cases the maps may look quite different, the upper limits are consistent within statistical errors. 
 
We also analyze the same O1-O2 folded data by considering the spherical harmonic (SpH) decomposition of the GW power on the sky. It was shown recently in~\citet{pystoch_SpH} that one can accurately transform the pixel-basis results to the SpH basis accurately using {\tt PyStoch}~\cite{pystoch_SpH}. However, to further validate the folded data sets, we compare the SNR sky maps obtained using the folded data with the published results~\cite{O2paper}. The results (see Refs.~\cite{Eric,pystoch_SpH,O2paper} for details regarding the formalism) are shown in the sixth row of Fig.~\ref{fig:bbr_maps}. As is evident from this figure, the results are matching with the LVC O1-O2~\cite{O2paper} results having a fractional rms difference less than $3\%$.

\section{Conclusion}
\label{sec:conclusion}
The search for an anisotropic gravitational-wave background and setting interesting upper limits on astrophysical and cosmological backgrounds play an important role in the current and future SGWB searches. We used folded data and the {\tt PyStoch} pipeline to search for the evidence of SGWBs from the first two observing runs of the Advanced LIGO detectors. We have used the full covariance matrix in the SGWB analysis in the pixel basis to analyze the data from the first two observing runs of the Advanced LIGO detectors. Earlier the noise covariance matrix was evaluated at $N_{\mathrm{side}}=8$ for the same data set~\cite{renzini1,renzini2} and at $N_{\mathrm{side}}=16$ for one day worth of simulated data~\cite{Radiometer}. Since the computation cost scales as $N_{\mathrm{side}}^4$, it was challenging to compute the matrix at $N_{\mathrm{side}}=16$ for the usable data for a full observation run (e.g., O1, which lasted for $\sim$4 months) without folding.

Since no evidence was found for a SGWB signal, we have set an upper limit on the GW flux in every direction on the sky. Though the upper limits are different for different schemes, they are consistent with the previously reported results by the LVC.

To incorporate the full noise covariance matrix in the analysis, we explore different schemes for regularization, significance estimation, and the Likelihood functions and study their performances. We carried out an extensive injection study to show that the upper limits obtained using the diagonals of the covariance matrix are close to those obtained using the full covariance matrix, and both the conservative $95\%$ upper limits satisfy the primary criteria that the upper limits are higher than the injected signals at least in $95\%$ of the pixels. Thus, the primary message from this study is that the approximate analysis published by the LVC is accurate enough.

This has an important implication for performing an all-sky-all-frequency (ASAF)~\cite{Folding,PyStoch,ASAF_2015,ASAF_2018} extension of the pixel-based radiometer analysis presented in this paper. In contrast to the broadband analysis, which integrates over a wide frequency range, the ASAF search has a much better possibility of detecting persistent narrow band sources, as the broadband search adds noise from all other frequency bands. Accounting for the full noise covariance matrix in the ASAF search will not only be a big computational challenge but these may be even more ill-conditioned, requiring more aggressive regularization causing hence larger bias. Our study indicates that using the ASAF dirty map with the diagonal components of the corresponding Fisher matrix, as was used for the broadband search by the LVC, will provide adequate accuracy. Even though the narrow band and broadband Fisher matrices will have different condition numbers, since the LVC approach does not involve an inversion of the Fisher matrix or deconvolution, we believe the outcome will be very similar. Nevertheless, further studies focused on ASAF, along the lines presented in this paper, may be necessary to ascertain the robustness of this claim, which may also lead to recipes that can provide more stringent upper limits without becoming computationally unfeasible. The use of analytical formula to calculate the $p$-value can speed up the calculation of the $p$-value for each frequency and each pixel instead of using noise simulations. In practice, the estimated $p$-value can also be biased. This may require determining a detection or follow-up significance threshold for the specific choice of the estimator and regularization schemes, if applicable.

In this paper, we limited our analysis to the pixel basis, though similar detailed studies may also be necessary for the search in the spherical-harmonic basis in order to assess the accuracy and possibilities to put more stringent upper limits on the GW power flux. Incorporating the bias in the likelihood can certainly help. Also, more methods can be explored to use a true clean-map covariance matrix in the likelihood. In general, incorporating the full noise covariance matrix in the analysis and regularized deconvolution are challenging problems, which require extensive studies specific to the application. While we have studied several possibilities here based on the commonly available literature, there may be more exciting possibilities even for the pixel basis. Such rigorous studies will be essential to claim a detection with enough confidence.

\begin{acknowledgments}
This work significantly benefitted from the interactions with the Stochastic Working Group of the LIGO-Virgo-KAGRA Scientific Collaboration. We acknowledge the use of Inter-University Centre for Astronomy and Astrophysics (IUCAA) and Caltech LDAS clusters for the computational/numerical work. D.A. acknowledges IUCAA, India for the funding support and expresses thanks to Dipankar Bhattacharya for useful discussion about the SVD regularization. J.S. acknowledges the support by Japan Society for the Promotion of Science (JSPS) KAKENHI Grant No. JP17H06361 and expresses thanks to Hideyuki Tagoshi for the helpful discussion. A.A. acknowledges support by Instituto Nazionale di Fisica Nucleare (INFN) Pisa and European Gravitational Observatory (EGO) and wants to thank Giancarlo Cella for his support. S.M. acknowledges support from the Department of Science and Technology (DST), India, provided under the Swarna Jayanti Fellowships scheme. This material is based upon work supported by LIGO Laboratory which is a major facility fully funded by the National Science Foundation (NSF). This research has  also made use of data obtained from the Gravitational Wave Open Science Center (gw-openscience.org), a service of LIGO Laboratory, the LIGO Scientific Collaboration, the Virgo Collaboration, and KAGRA. LIGO Laboratory and Advanced LIGO are funded by the United States National Science Foundation (NSF) as well as the Science and Technology Facilities Council (STFC) of the United Kingdom, the Max-Planck-Society (MPS), and the State of Niedersachsen/Germany for support of the construction of Advanced LIGO and construction and operation of the GEO600 detector. Additional support for Advanced LIGO was provided by the Australian Research Council. Virgo is funded, through the European Gravitational Observatory (EGO), by the French Centre National de Recherche Scientifique (CNRS), the Italian Istituto Nazionale di Fisica Nucleare (INFN) and the Dutch Nikhef, with contributions by institutions from Belgium, Germany, Greece, Hungary, Ireland, Japan, Monaco, Poland, Portugal, Spain. The construction and operation of KAGRA are funded by Ministry of Education, Culture, Sports, Science and Technology (MEXT), and Japan Society for the Promotion of Science (JSPS), National Research Foundation (NRF) and Ministry of Science and ICT (MSIT) in Korea, Academia Sinica (AS) and the Ministry of Science and Technology (MoST) in Taiwan. This article has a LIGO document number LIGO-P2000499. Parts of the results in this work make use of the color maps in the {\tt CMasher} package~\cite{cmasher_cmap}.
\end{acknowledgments}

\appendix
\section{Injection study results for $\alpha=2/3,0$ }\label{sec:inj_a23_0}

\begin{figure*}
\centering
\includegraphics[width = 0.3\textwidth]{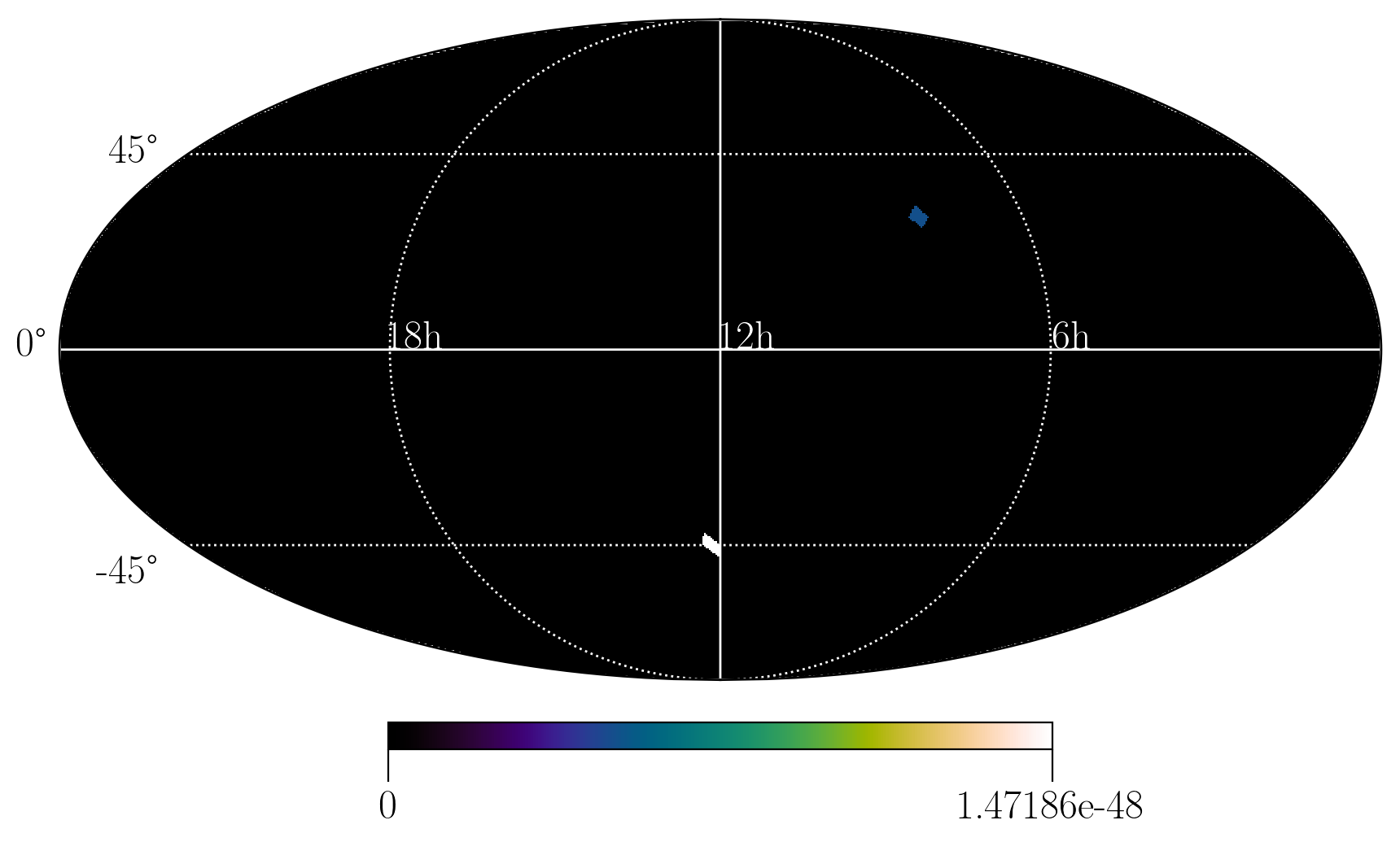} 
\includegraphics[width = 0.3\textwidth]{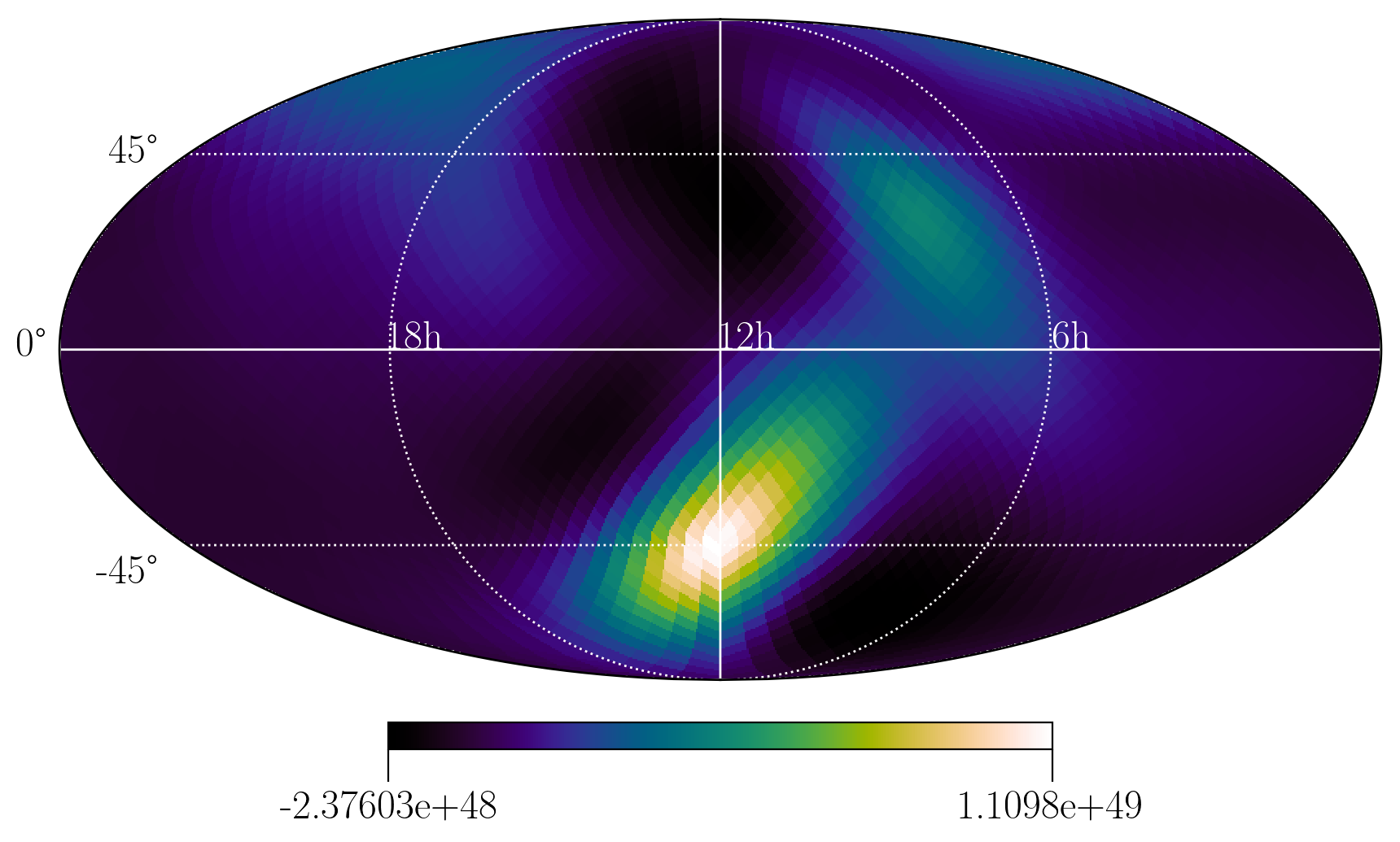} 
\includegraphics[width = 0.3\textwidth]{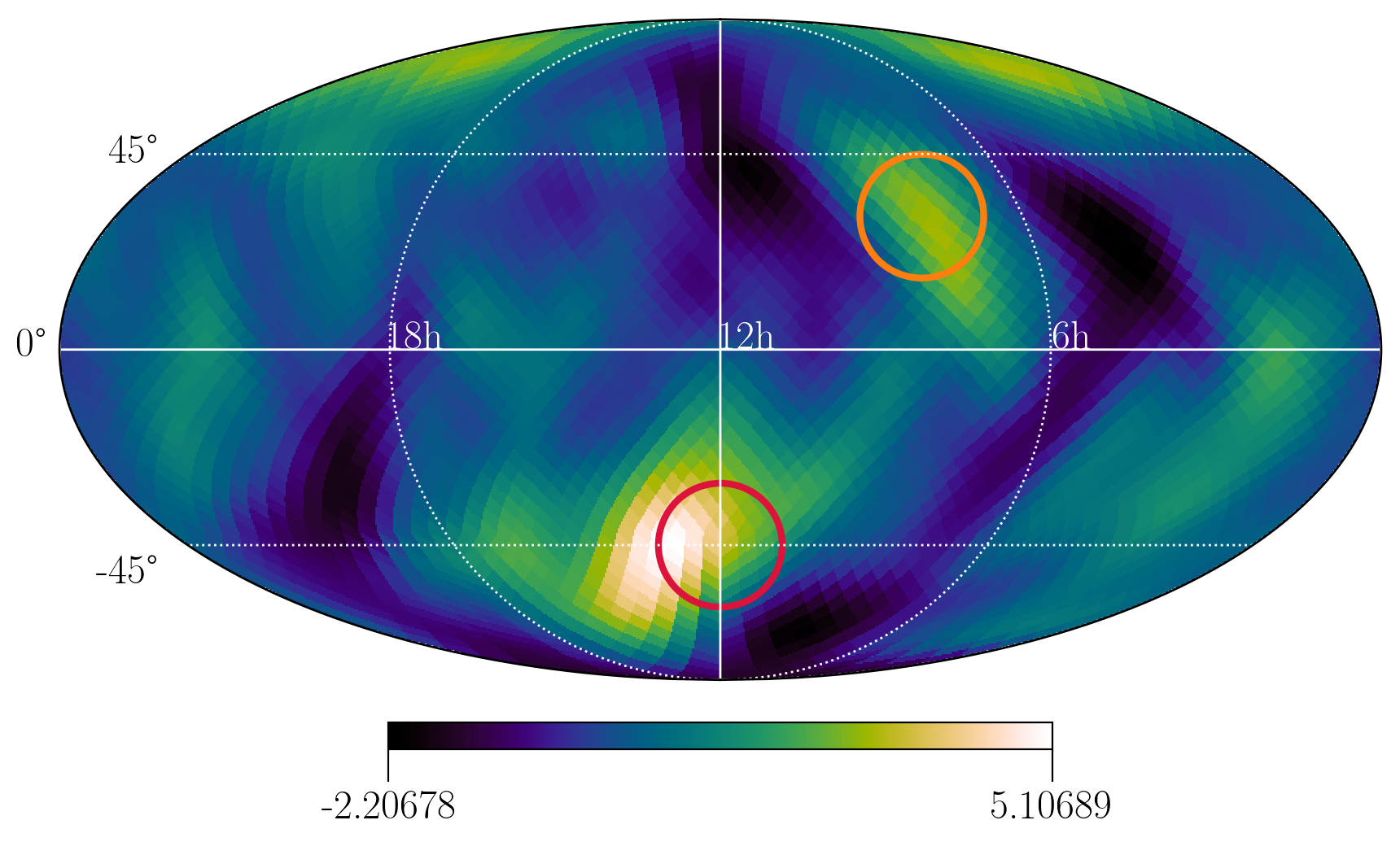} \\
\vspace{0.3cm}
\includegraphics[width = 0.3\textwidth]{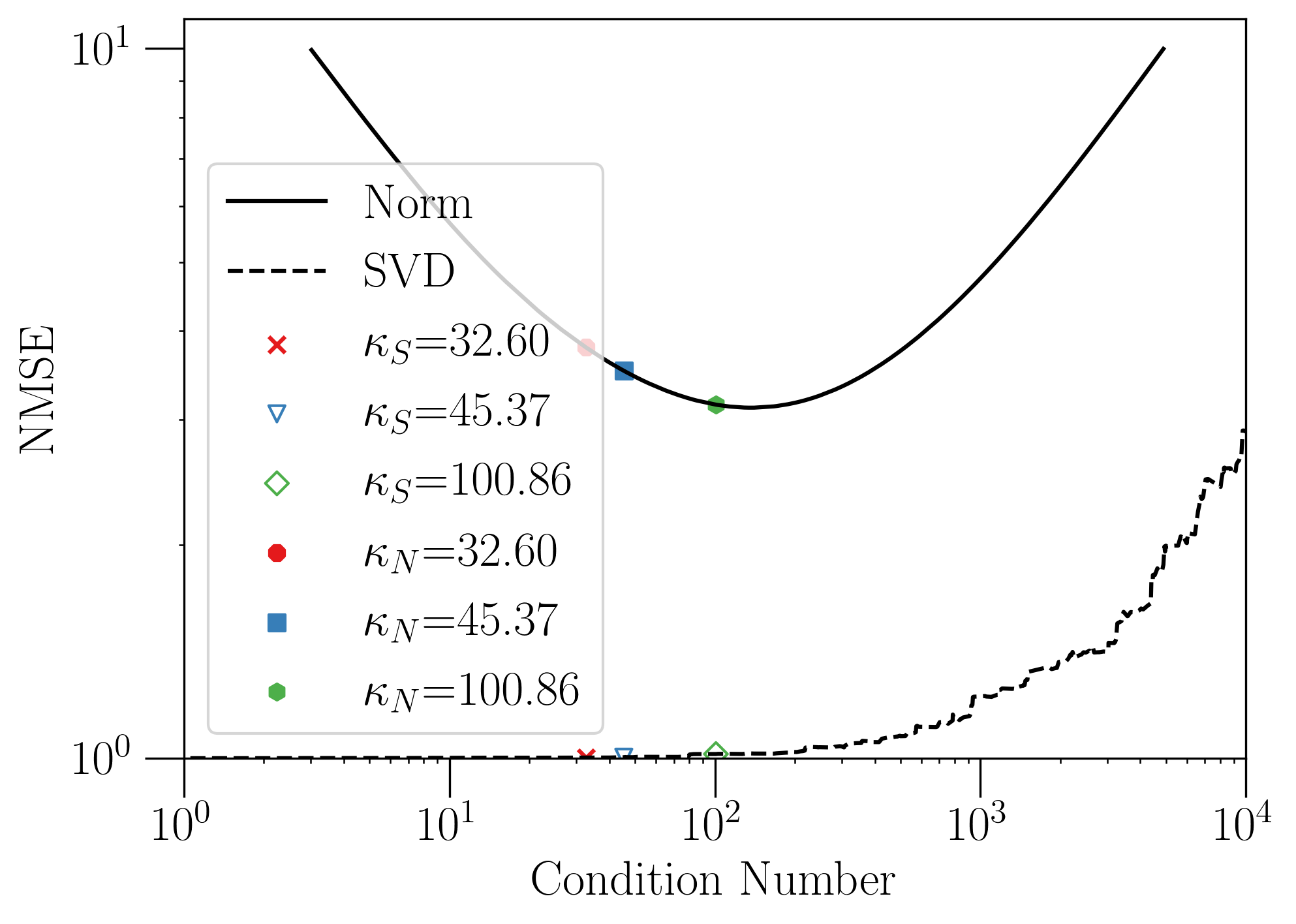}
\includegraphics[width = 0.3\textwidth]{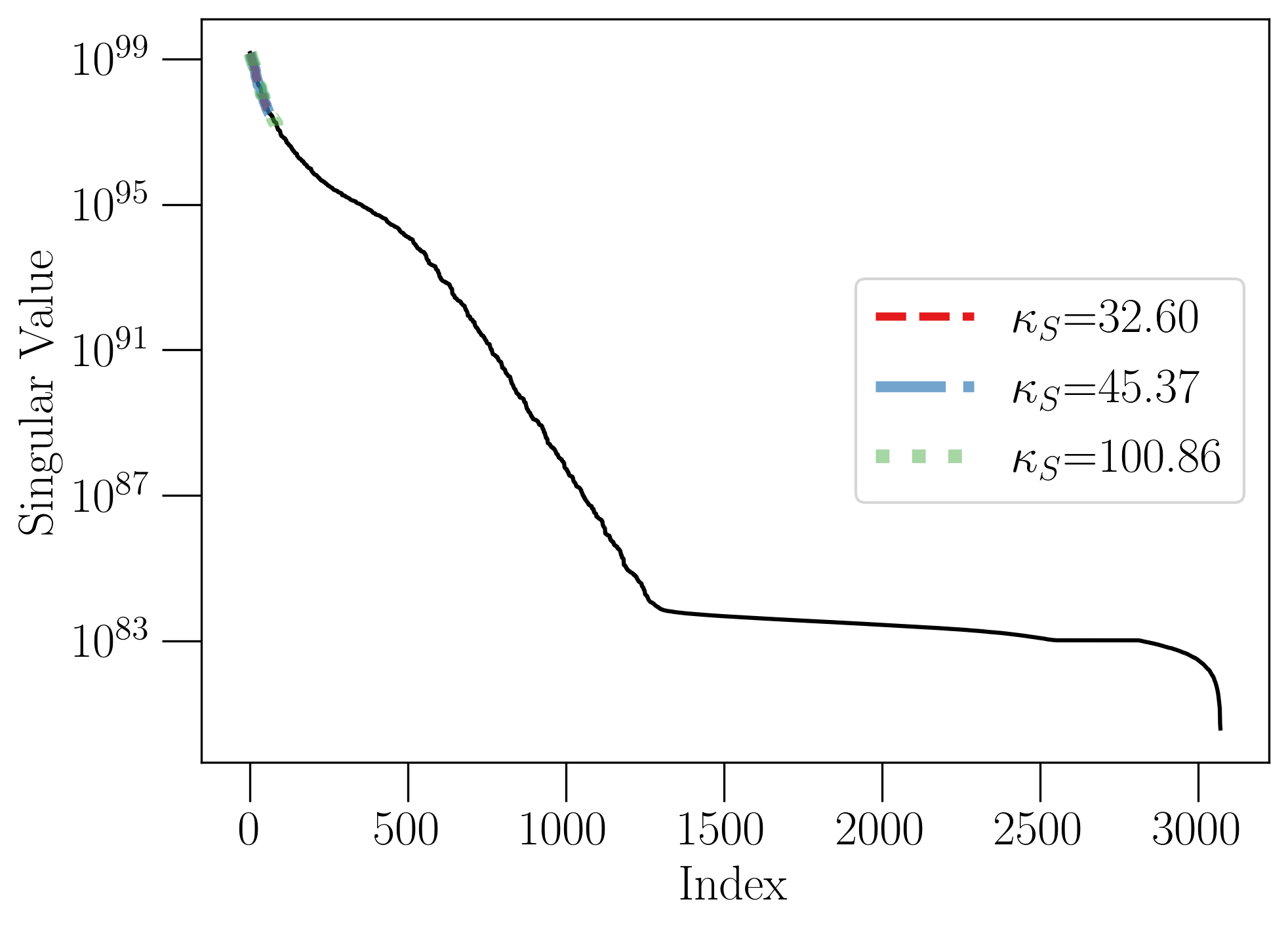}
\includegraphics[width = 0.3\textwidth]{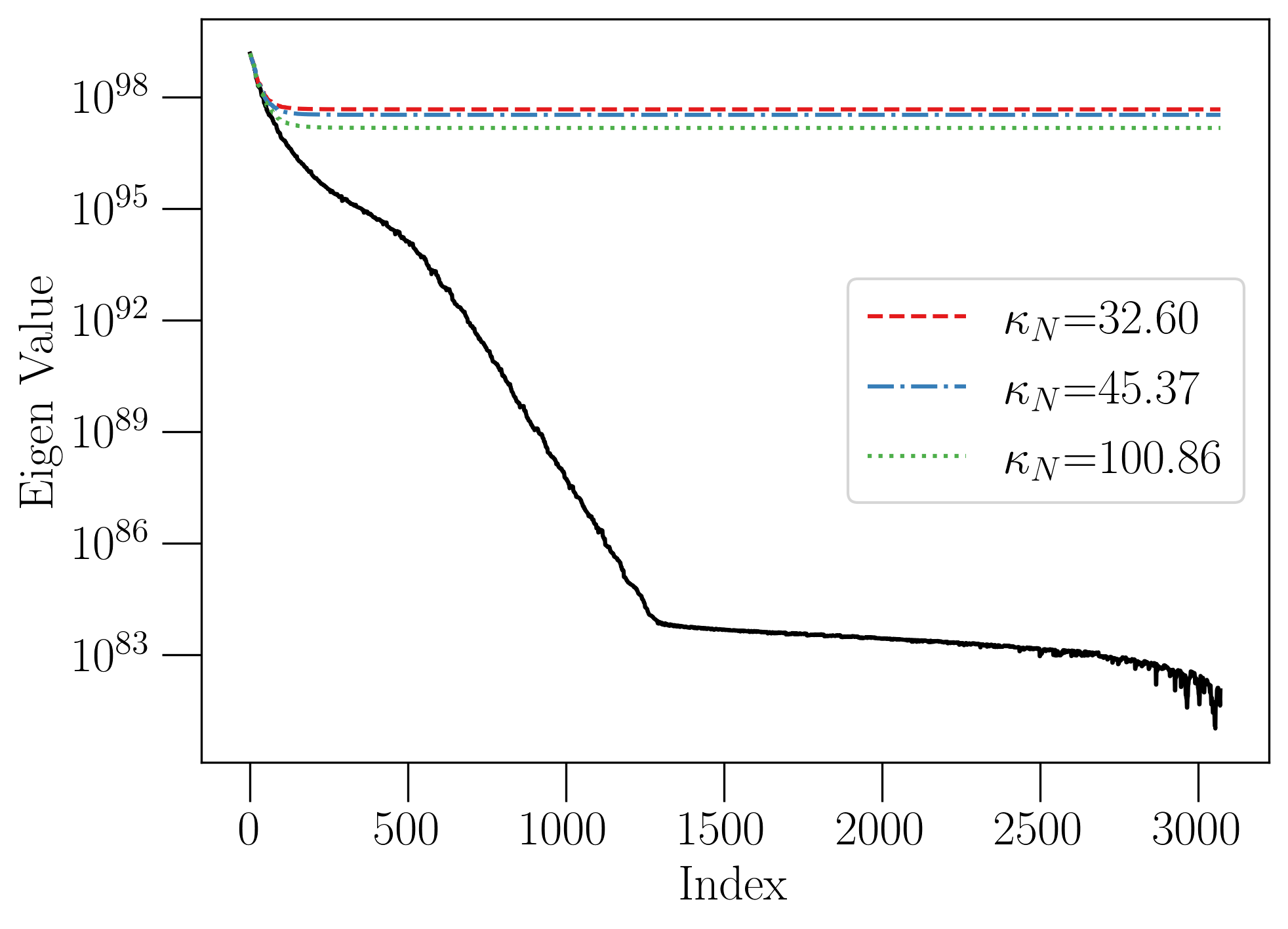} \\
\vspace{0.3cm}
\includegraphics[width = 0.3\textwidth]{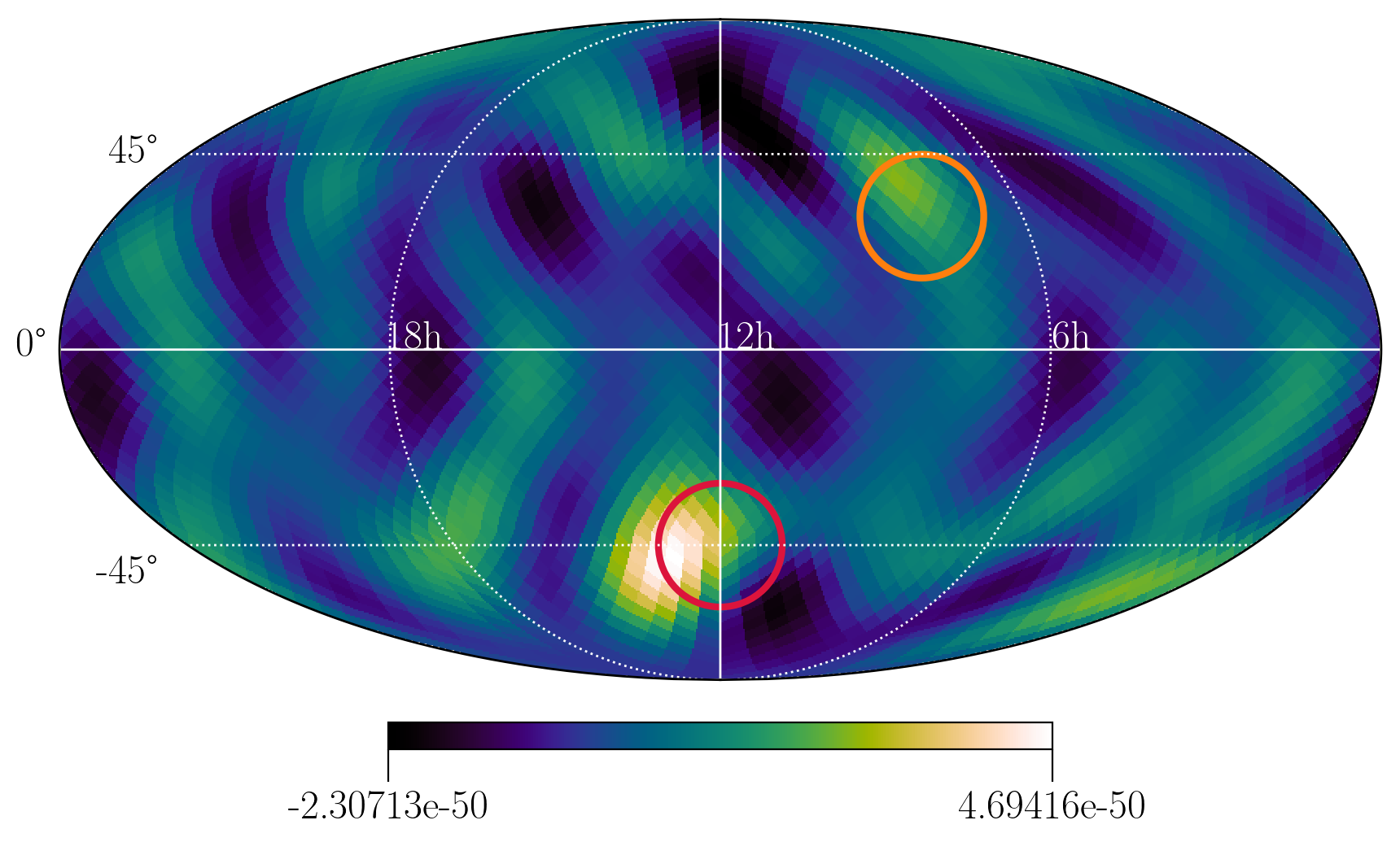} 
\includegraphics[width = 0.3\textwidth]{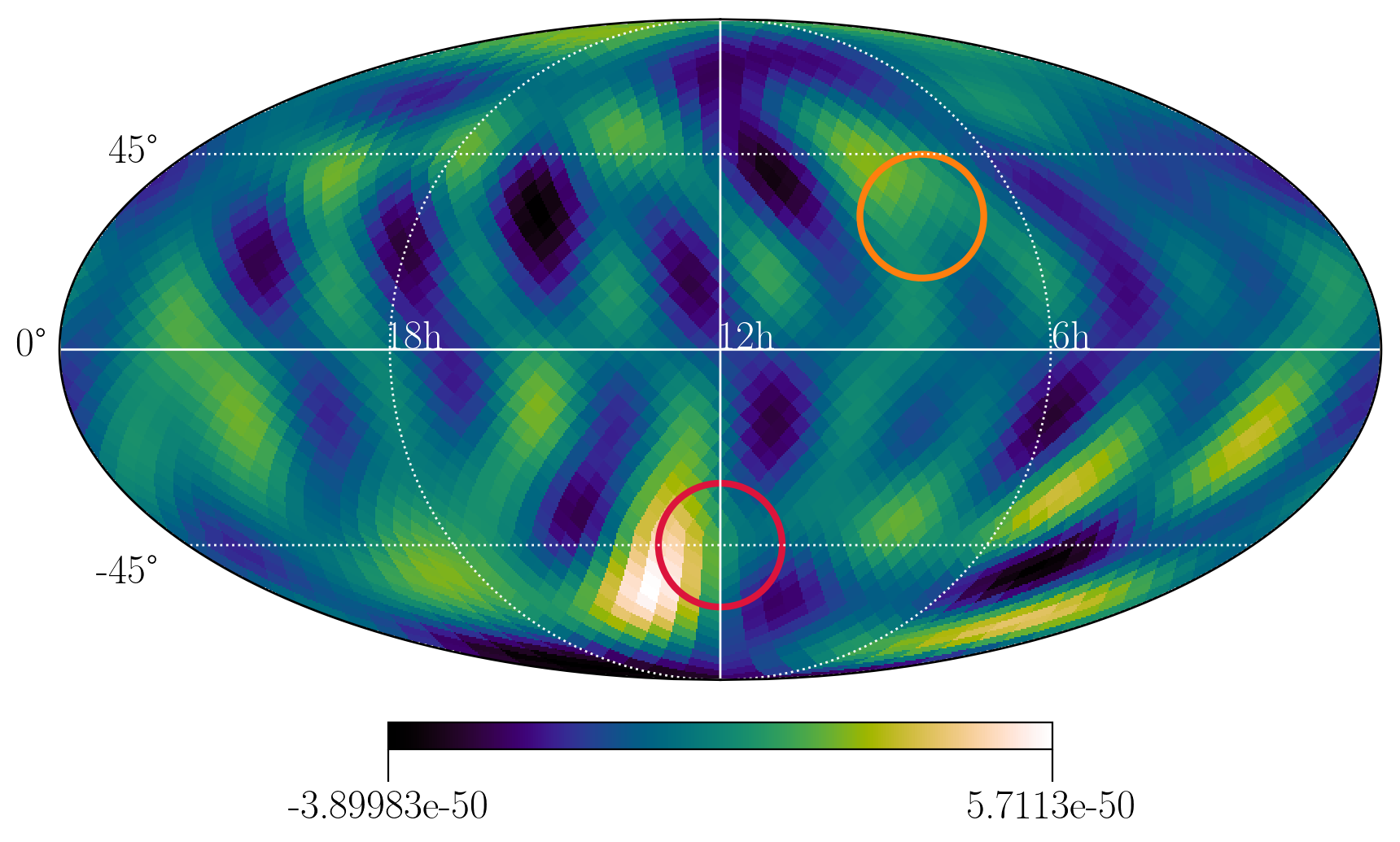} 
\includegraphics[width = 0.3\textwidth]{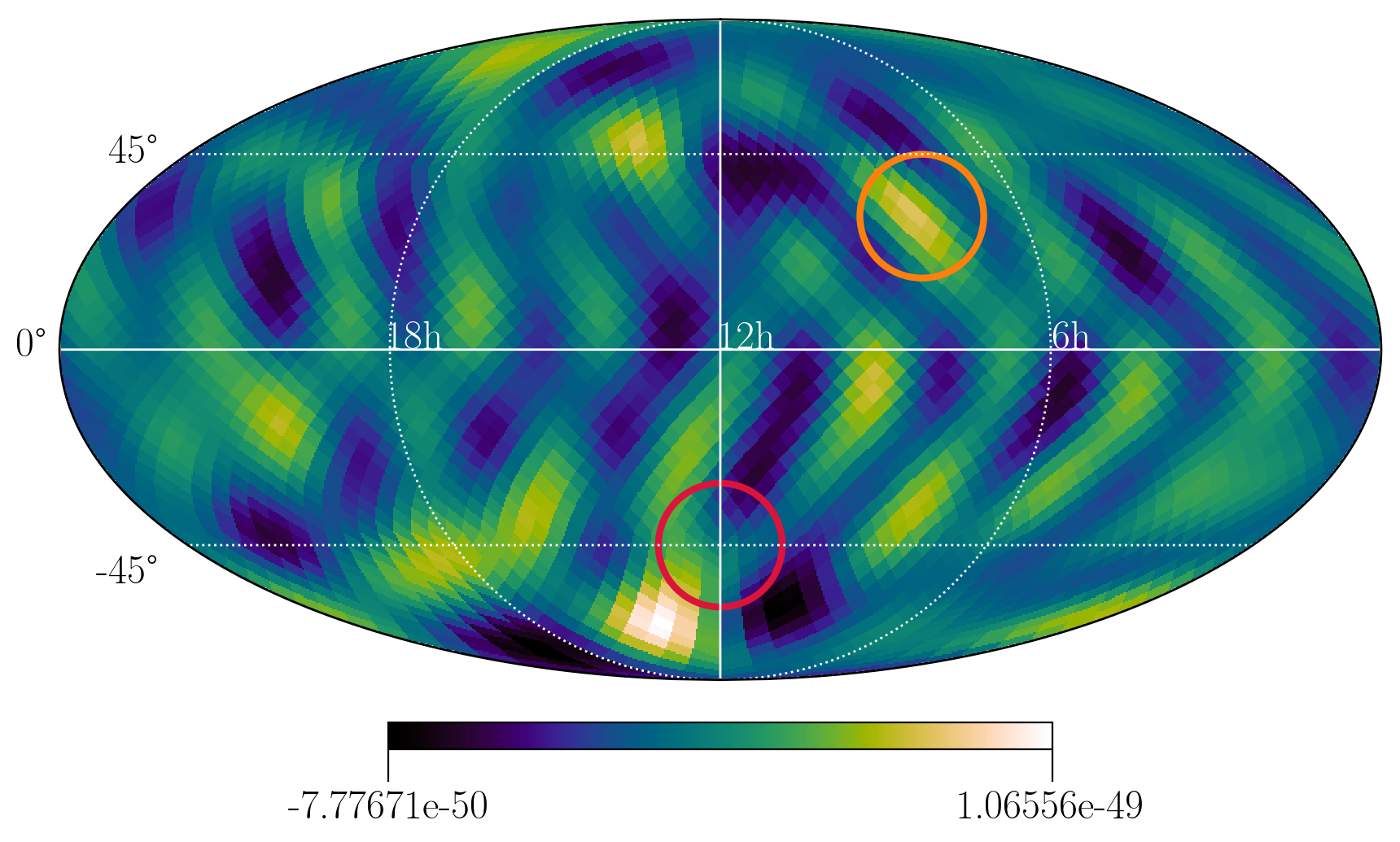} \\
\vspace{0.3cm}
\includegraphics[width = 0.3\textwidth]{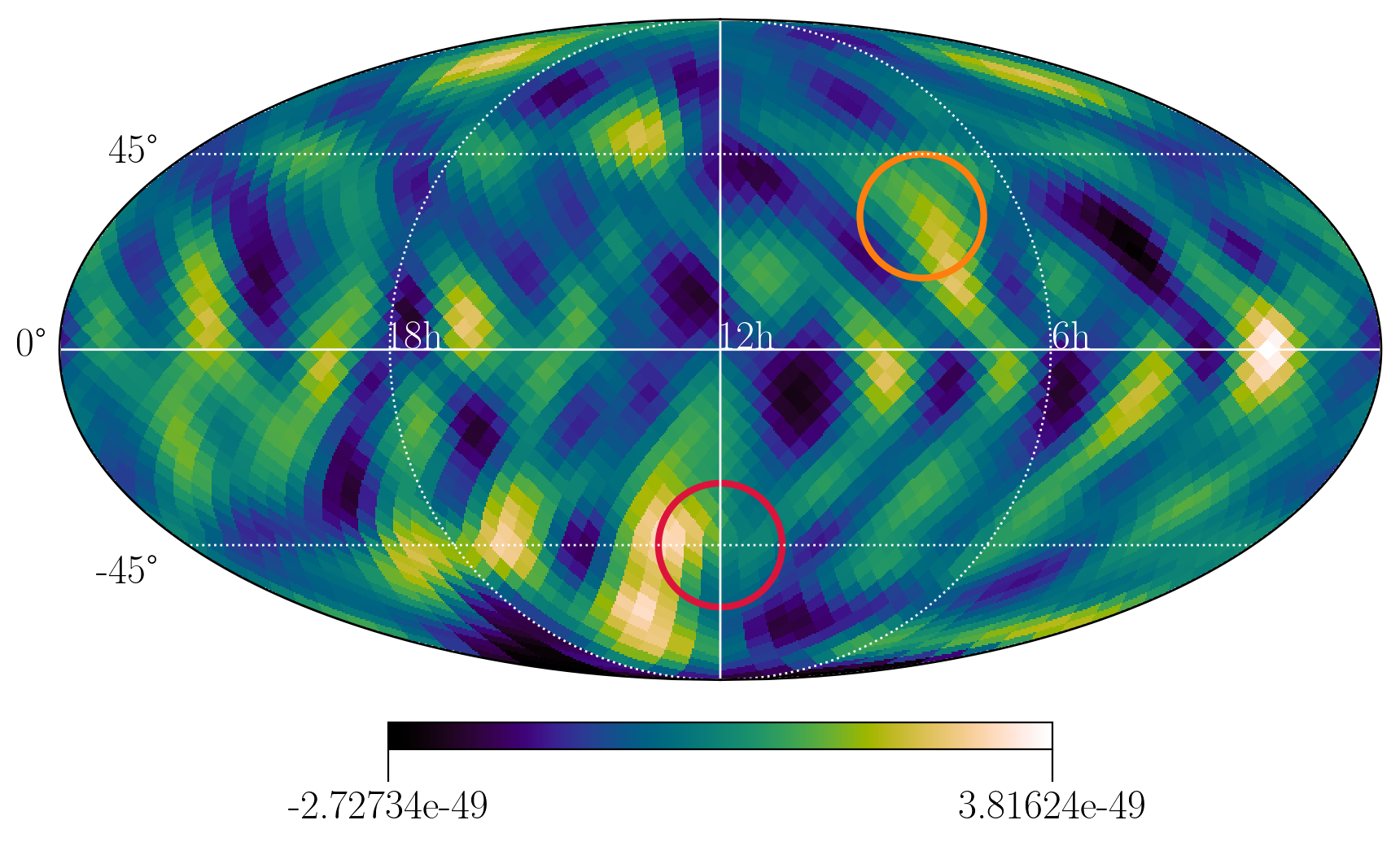} 
\includegraphics[width = 0.3\textwidth]{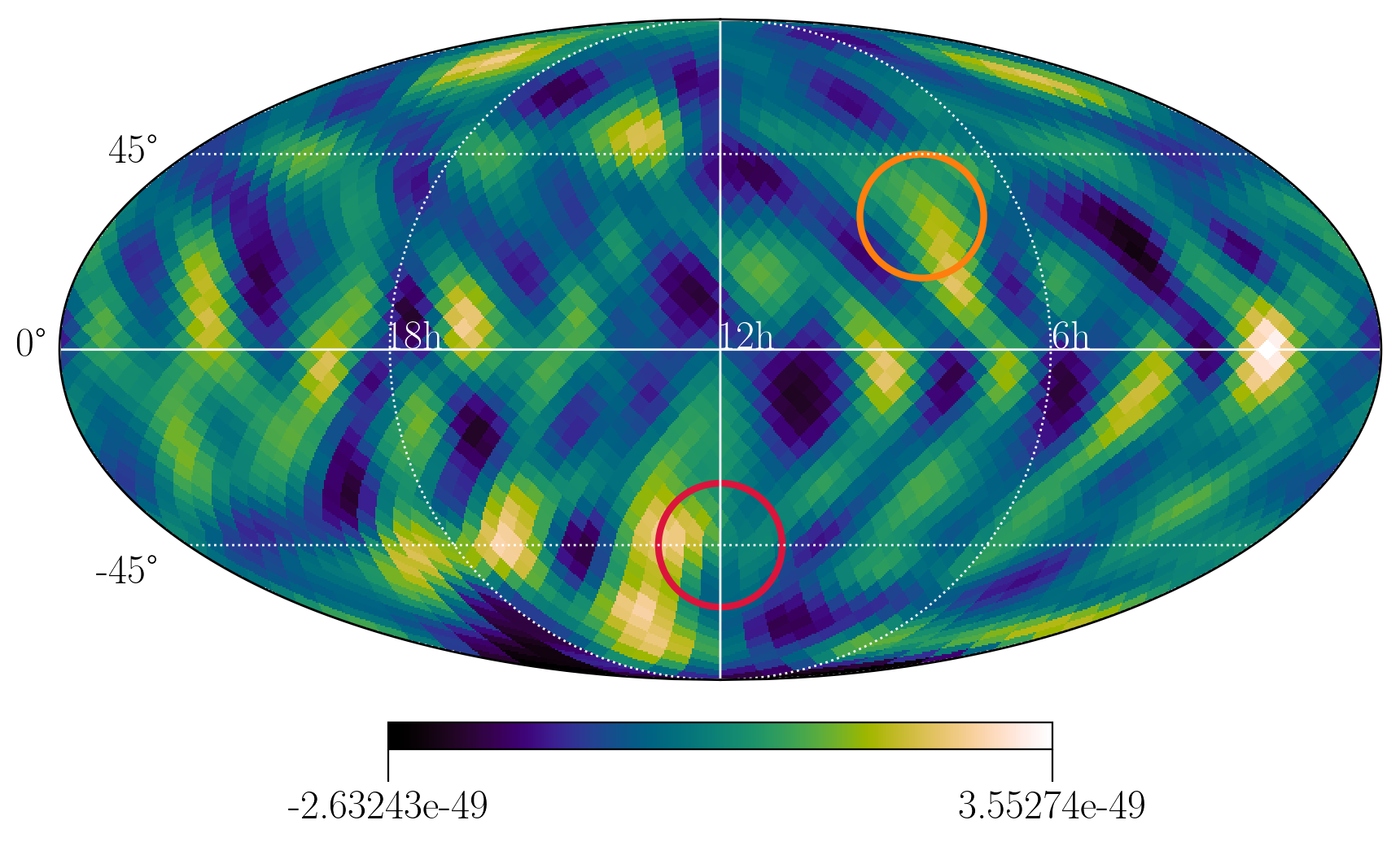} 
\includegraphics[width = 0.3\textwidth]{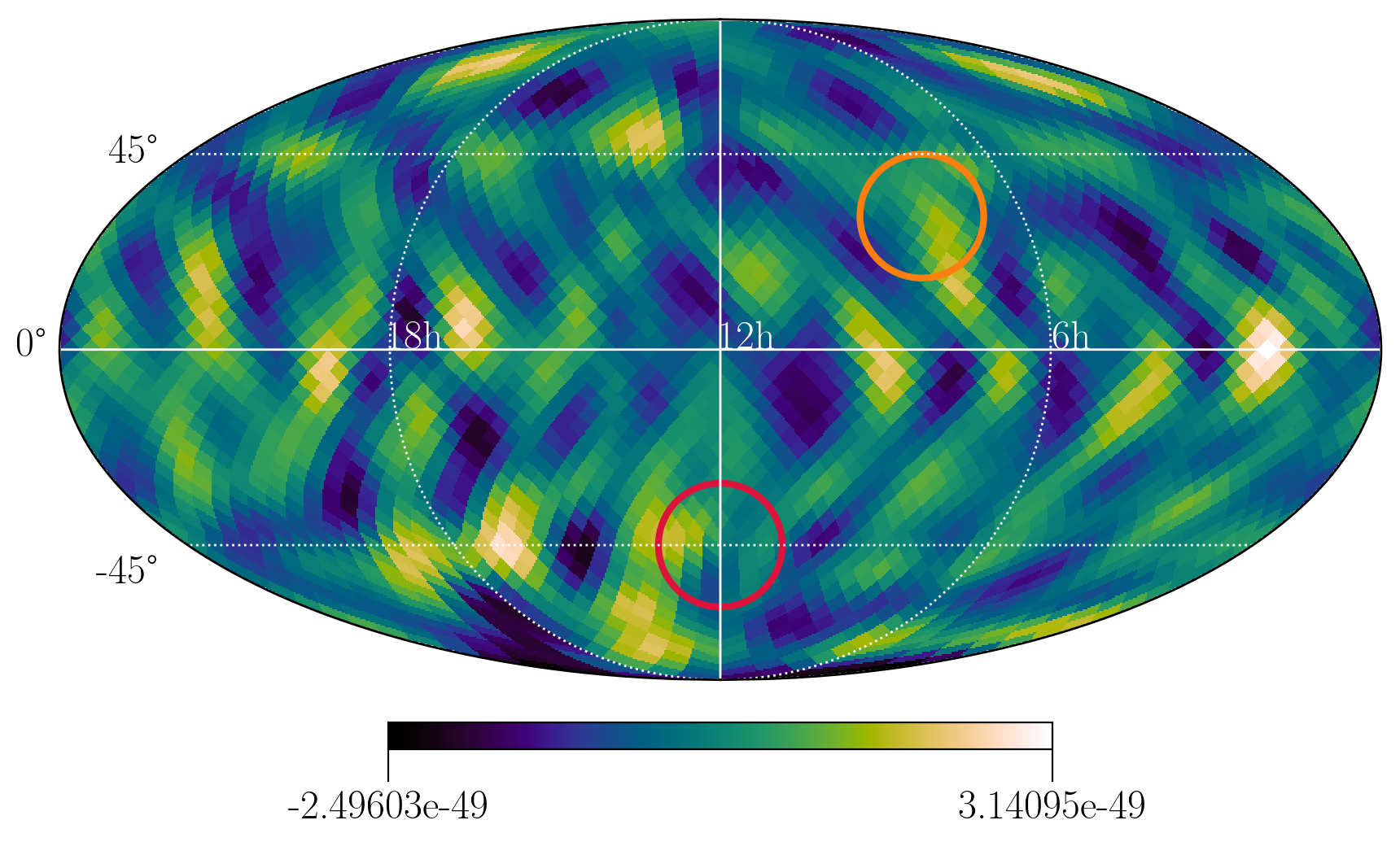} \\

\caption{Results of the injection study carried out to understand the effect of regularization recipes on deconvolution for $\alpha=2/3$ point source injections. The first row, from left to right, shows the injected source power map, the source map convolved with $\mathbf{\Gamma}$ without noise, and SNR dirty map with noise. In the second row, the leftmost plot shows the variation of NMSE with target condition number ($\kappa_S$ or $\kappa_N$). We have chosen three target condition numbers to show their effect on the recovery of injections. These condition numbers are marked in condition number-NMSE plot. The middle plot shows the singular value spectrum of $\mathbf{\Gamma}$ along with $\mathbf{\Gamma'}_S$ with the chosen target condition number. The rightmost plot shows the eigenvalue spectrum of $\mathbf{\Gamma}$ along with $\mathbf{\Gamma'}_N$ regularized with the chosen target condition number. The third row shows the ``scaled" clean power map with SVD regularization with $\kappa_S=[32.6,45.4,100.9]$ from left to right. The fourth row shows ``scaled" clean power map with norm regularization with $\kappa_N=[32.6,45.4,100.9]$ from left to right. The quantitative results are summarized in Table~\ref{table:Injection study}. All the maps are represented as a color bar plot on a Mollweide projection of the sky in ecliptic coordinates.}
\label{fig:inj_4_a23}
\end{figure*}

\begin{figure*}
\centering
\includegraphics[width = 0.3\textwidth]{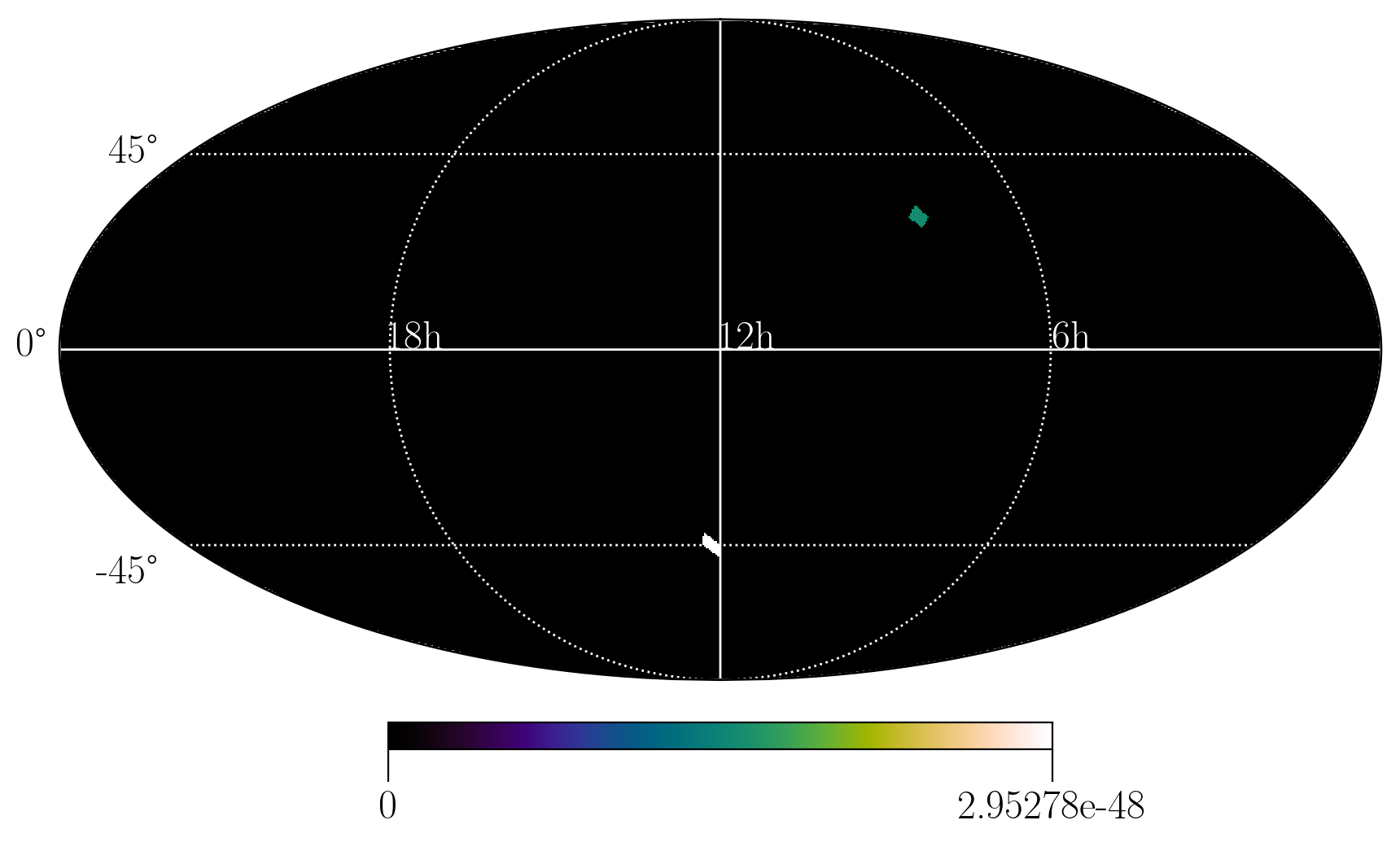} 
\includegraphics[width = 0.3\textwidth]{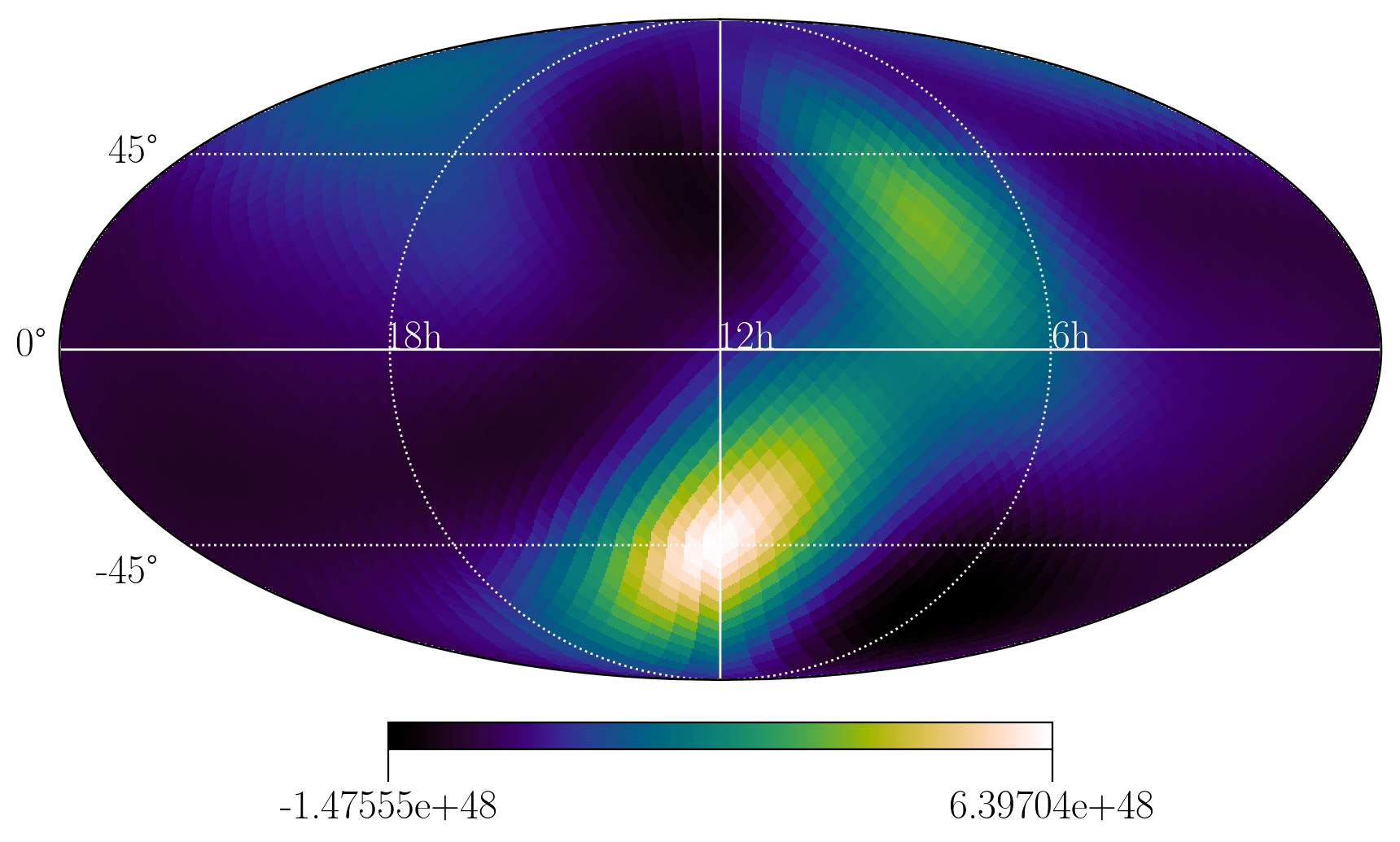} 
\includegraphics[width = 0.3\textwidth]{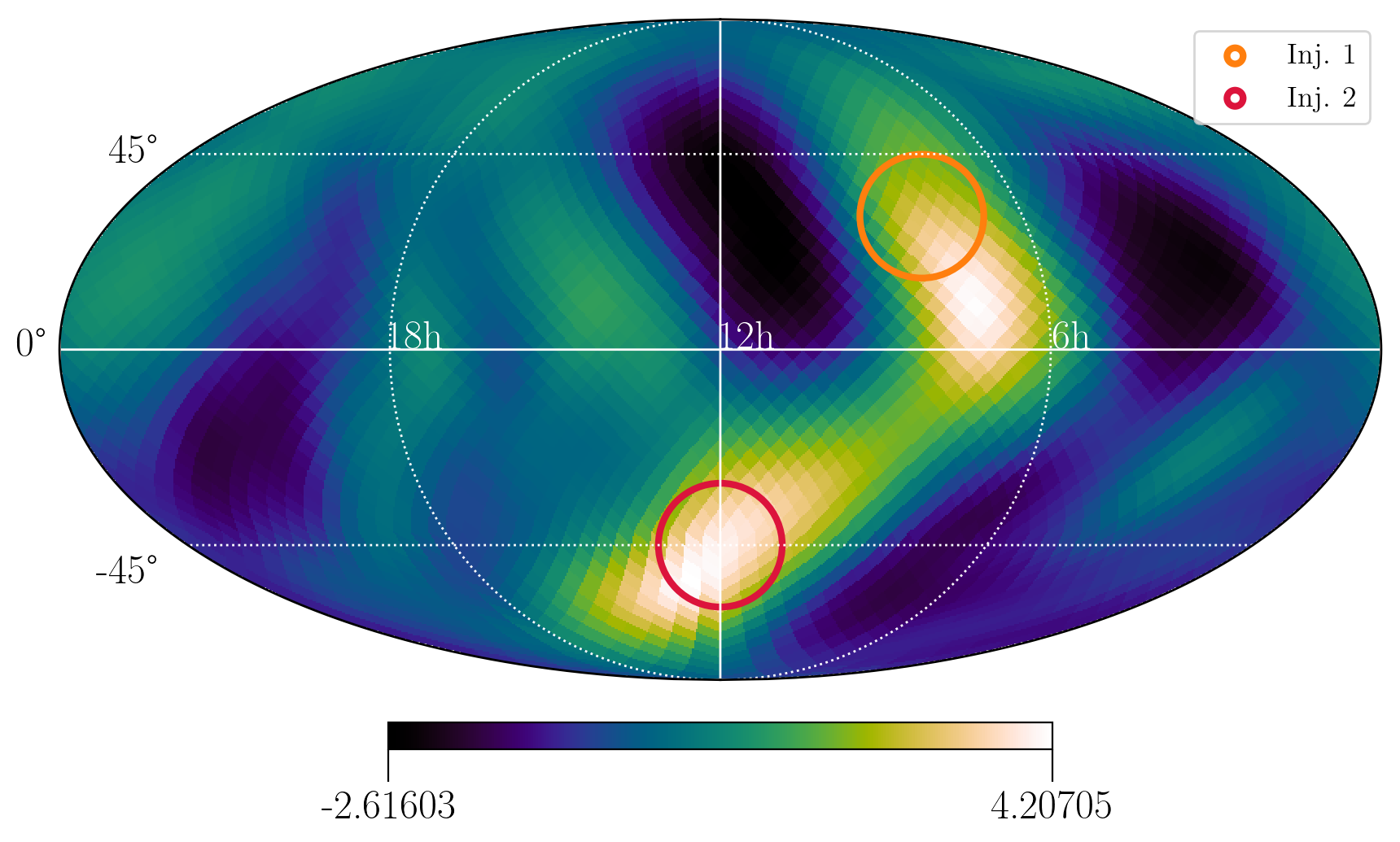} \\
\vspace{0.3cm}
\includegraphics[width = 0.3\textwidth]{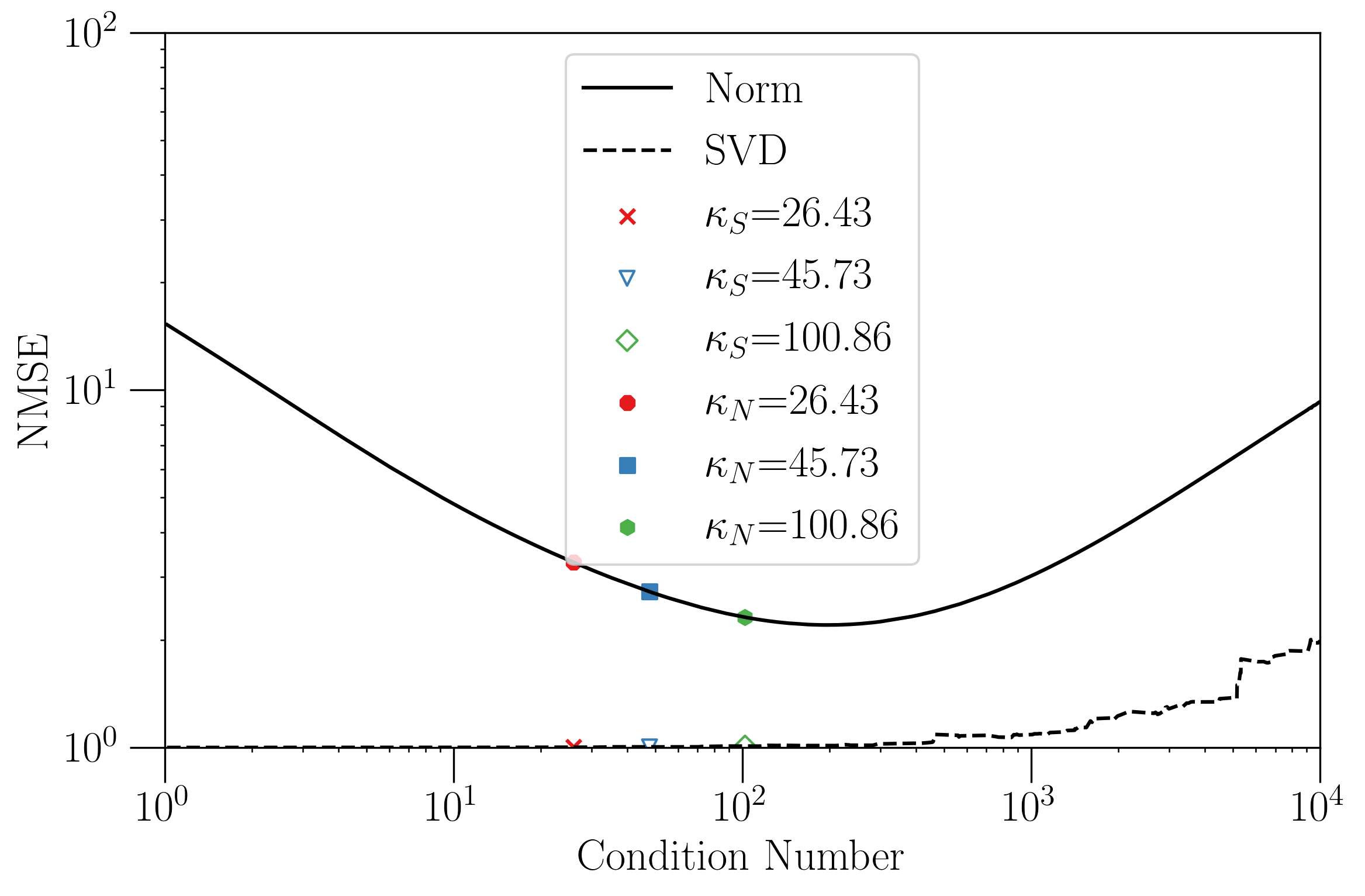}
\includegraphics[width = 0.3\textwidth]{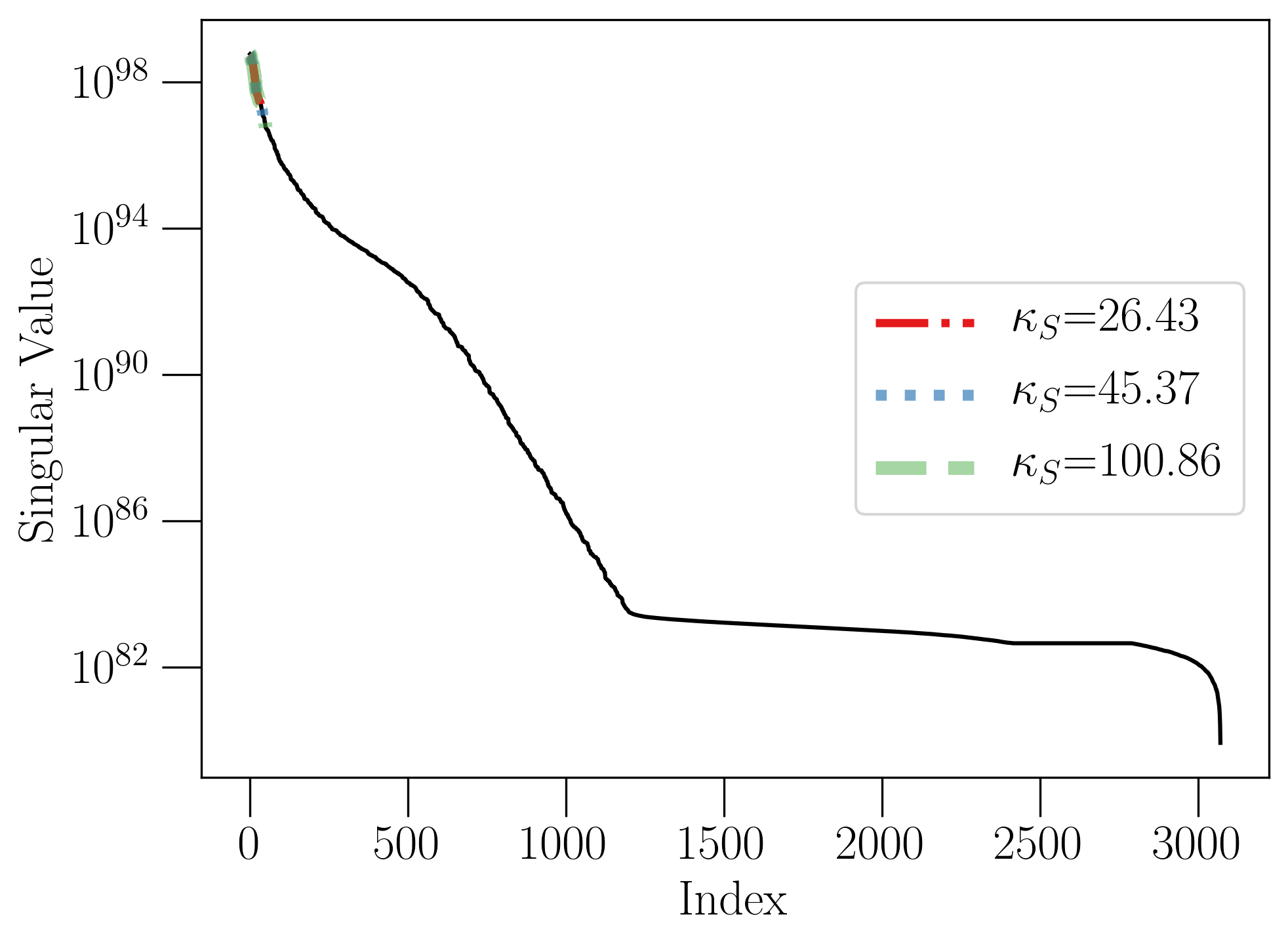}
\includegraphics[width = 0.3\textwidth]{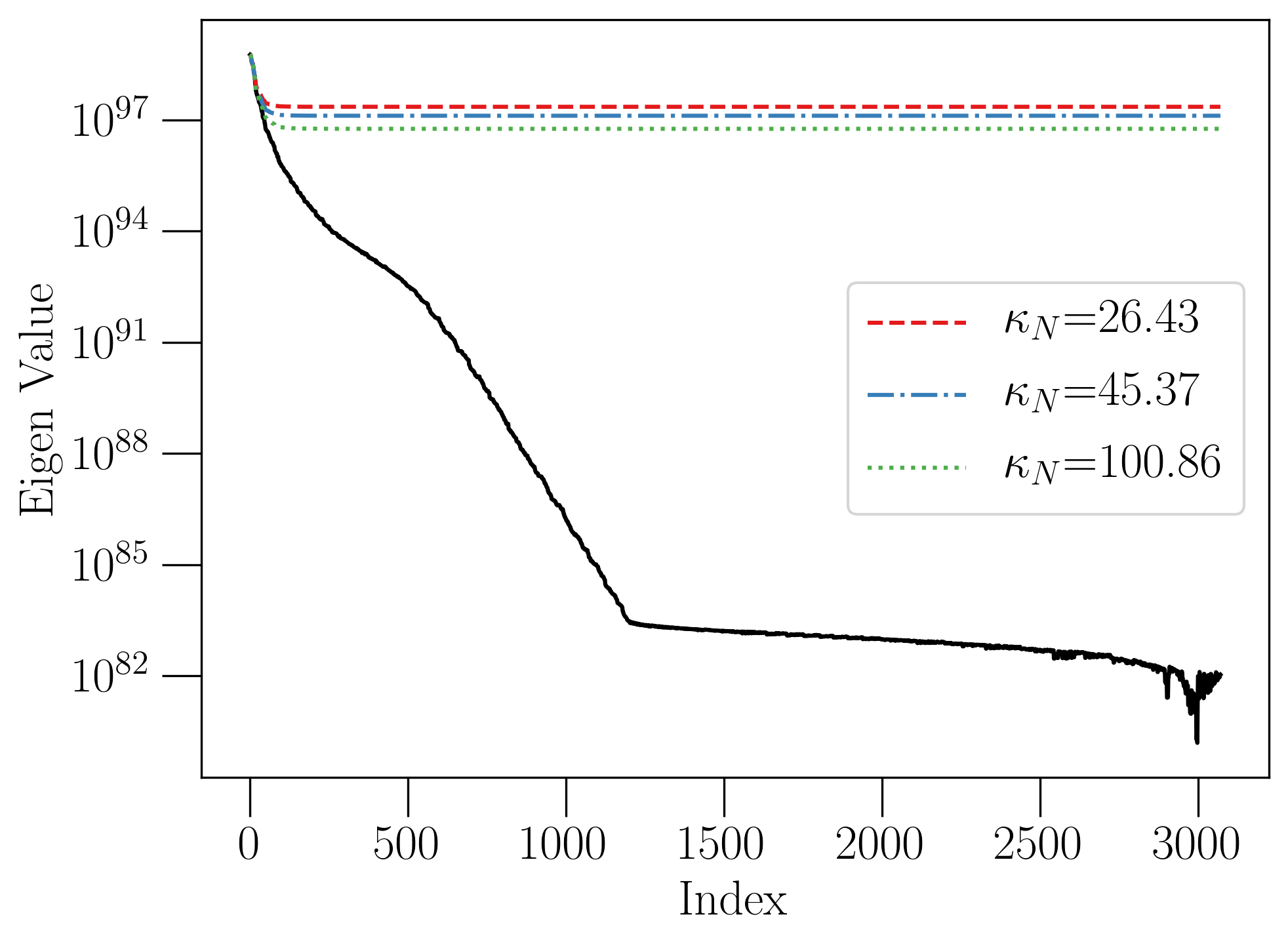} \\
\vspace{0.3cm}
\includegraphics[width = 0.3\textwidth]{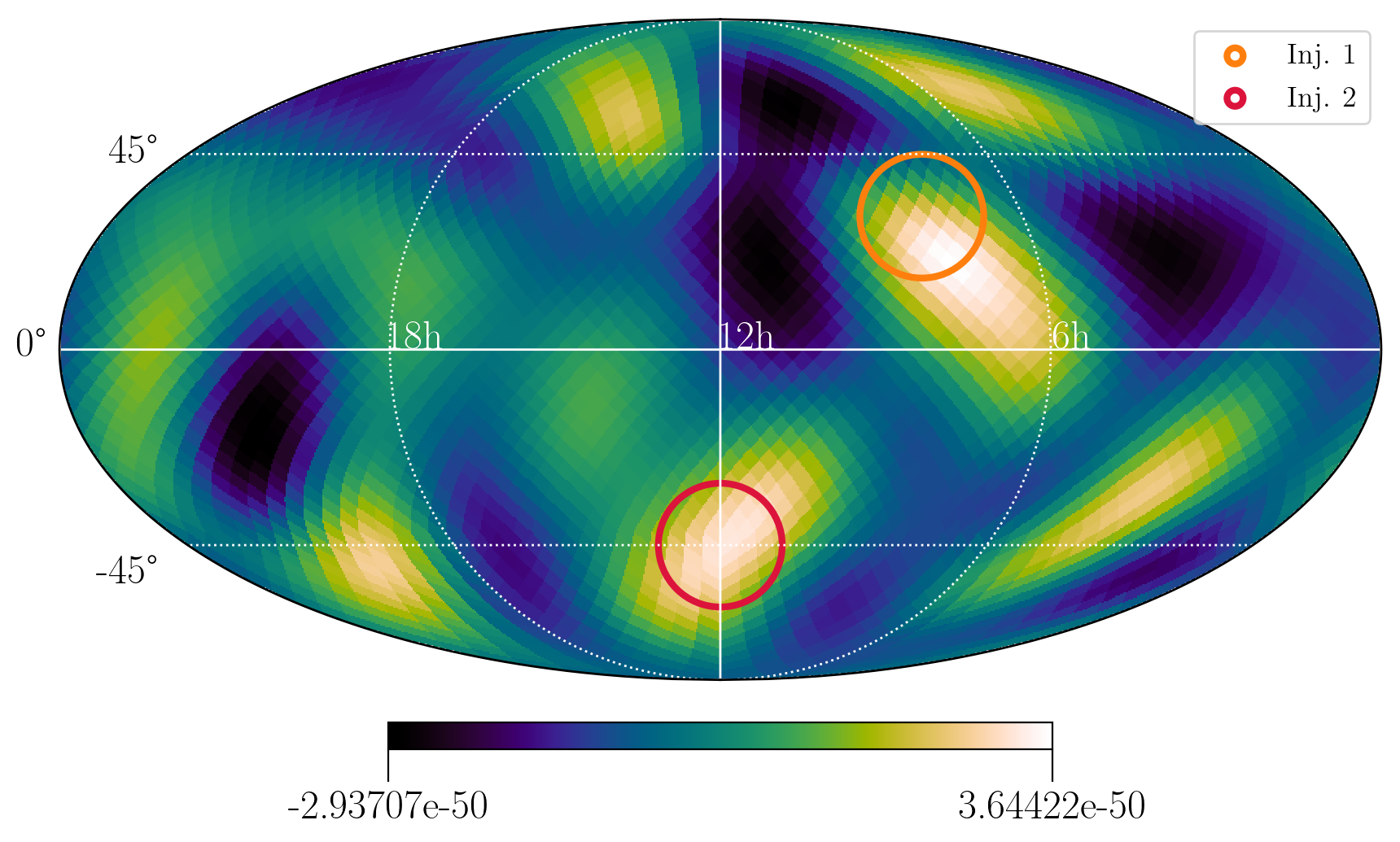} 
\includegraphics[width = 0.3\textwidth]{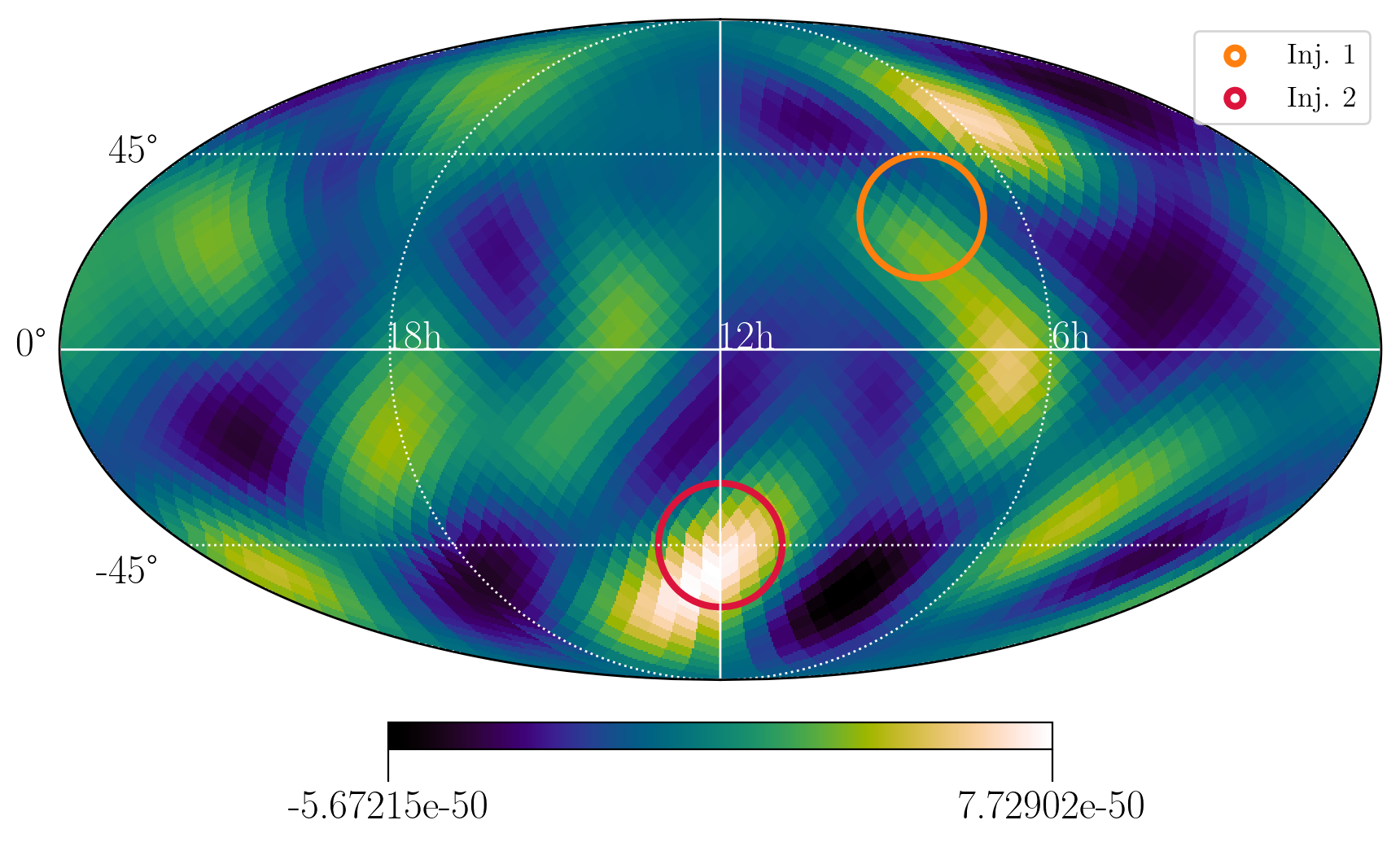} 
\includegraphics[width = 0.3\textwidth]{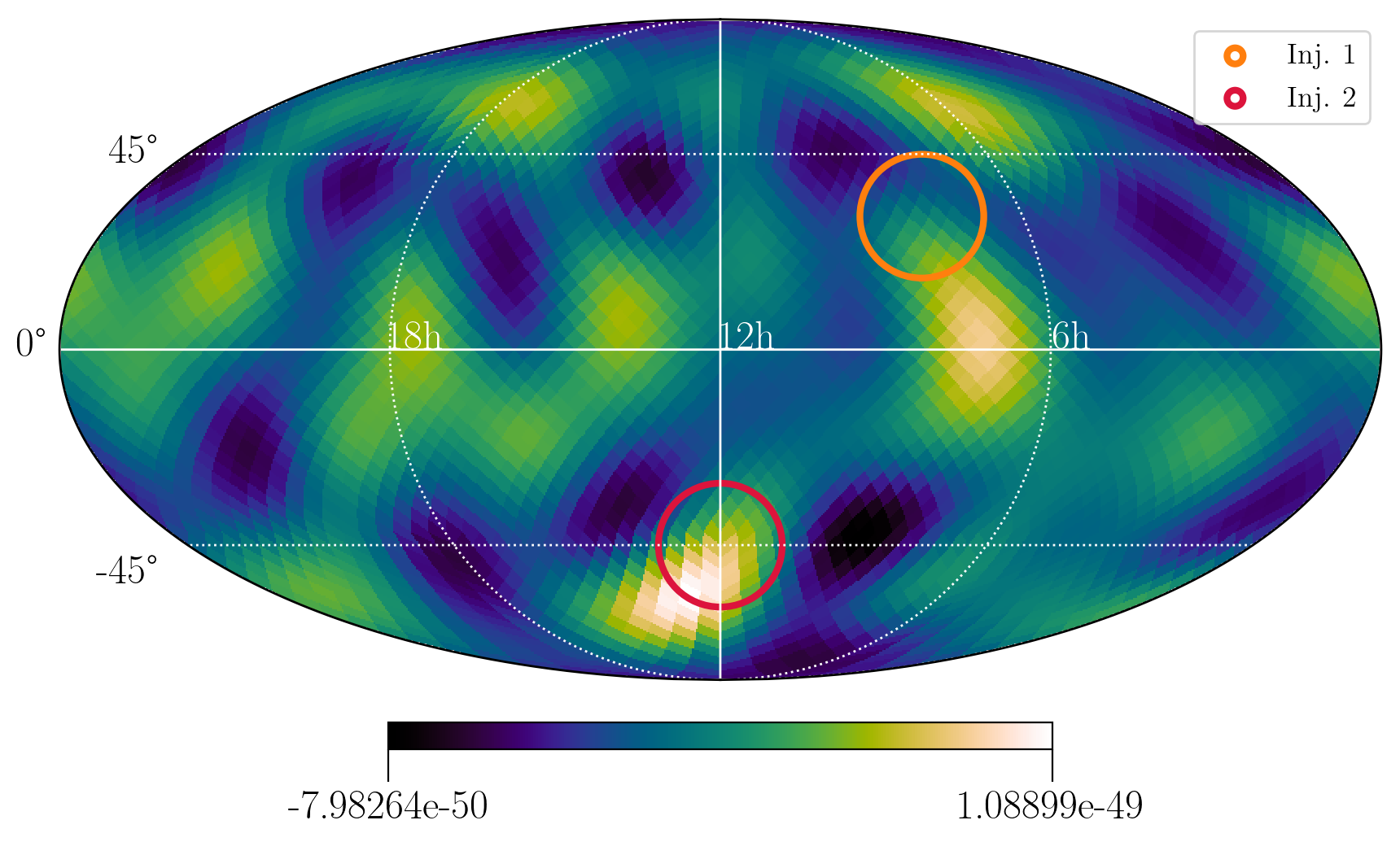} \\
\vspace{0.3cm}
\includegraphics[width = 0.3\textwidth]{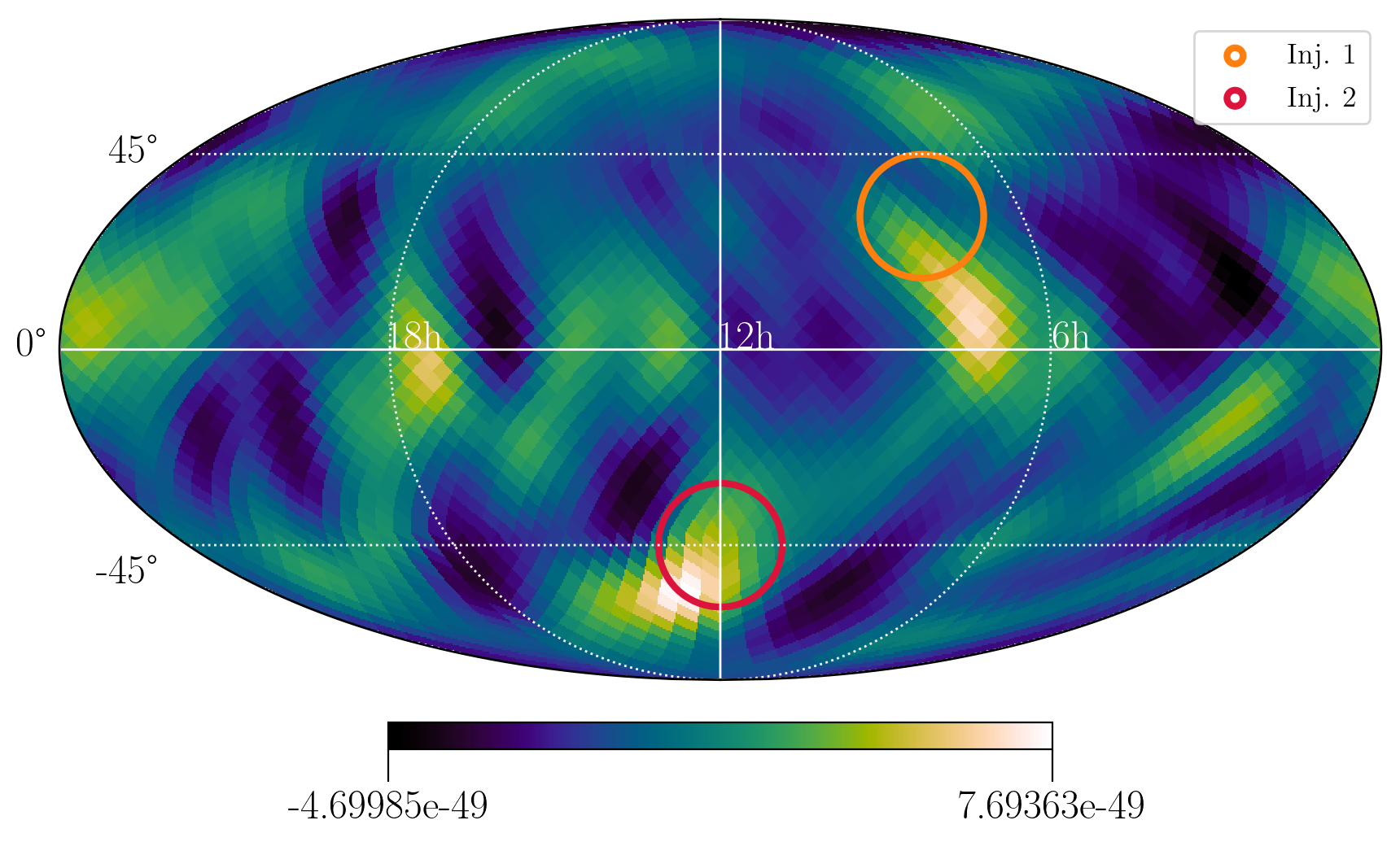} 
\includegraphics[width = 0.3\textwidth]{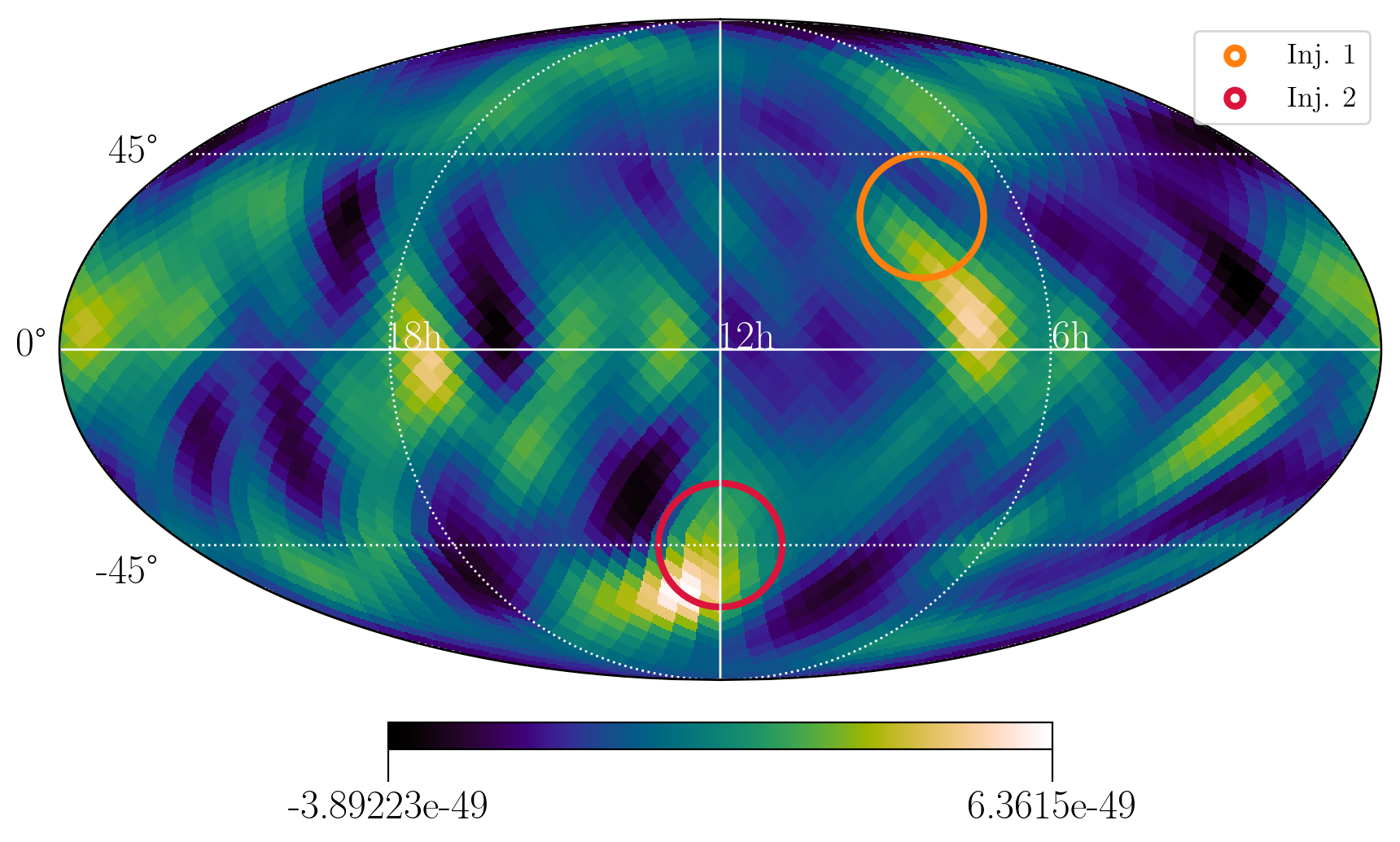} 
\includegraphics[width = 0.3\textwidth]{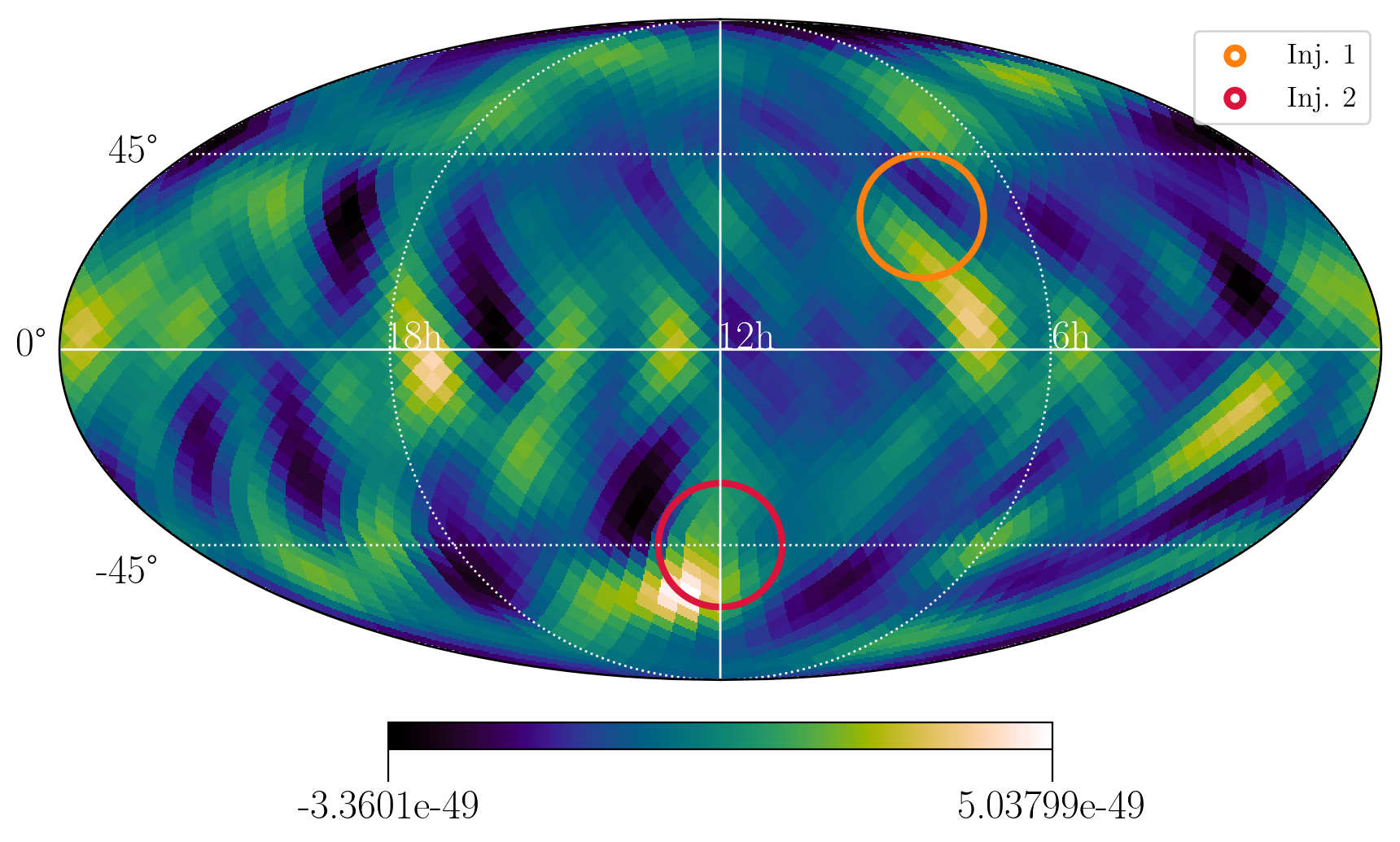} \\

\caption{Results of the injection study carried out to understand the effect of regularization recipes on deconvolution for $\alpha=0$ point source injections. The first row, from left to right, shows the injected source power map, the source map convolved with $\mathbf{\Gamma}$ without noise, and SNR dirty map with noise. In the second row, the leftmost plot shows the variation of NMSE with target condition number($\kappa_S$ or $\kappa_N$). We have chosen three condition numbers to show their effect on the recovery of injections. These condition numbers are marked in the condition number-NMSE plot. The middle plot shows the singular value spectrum of $\mathbf{\Gamma}$ along with $\mathbf{\Gamma'}_S$ with the chosen target condition number. The rightmost plot shows the eigenvalue spectrum of $\mathbf{\Gamma}$ along with $\mathbf{\Gamma'}_N$ regularized with the chosen target condition number. The third row shows the ``scaled" clean power map with SVD regularization with $\kappa_S=[26.4,45.4,102]$ from left to right. The fourth row shows the ``scaled" clean power map with Norm regularization with $\kappa_N=[26.4,45.4,100.9]$ from left to right. The quantitative results are summarized in Table~\ref{table:Injection study}. All the maps are represented as a color bar plot on a Mollweide projection of the sky in ecliptic coordinates.}
\label{fig:inj_4_a0}
\end{figure*}

The results of the injection study performed to understand the behavior of regularization recipes for point sources with a power law of spectral indices $\alpha=2/3,0$ are shown in Fig.~\ref{fig:inj_4_a23} $\&$~\ref{fig:inj_4_a0}. This is similar to the $\alpha=3$ case shown in Fig.~\ref{fig:inj_4_a3}. The quantitative results are summarized in Table~\ref{table:Injection study}. The plots are arranged similarly as was for the $\alpha=3$ case in Fig.~\ref{fig:inj_4_a3}. In the case of $\alpha=2/3,0$, the target condition number-NMSE plot suggests that the norm-regularized clean maps are noisier than SVD regularized clean maps. This is also observed by comparing the third and fourth rows of Figs.~\ref{fig:inj_4_a23} $\&$ \ref{fig:inj_4_a0}. In the case of SVD regularization, we have discarded noisy modes, while in norm regularization, the weights of the noisy modes are reduced, but they can still affect the results. These different ways of modifying modes while regularizing might cause differences in NMSE (or the clean map) for a particular target condition number.

\bibliography{ref.bib}

\end{document}